\documentclass[twocolumn]{aastex62}

\newcommand{\gtsim}{\protect\raisebox{-0.5ex}{$\:\stackrel{\textstyle >}{\sim}\:$}}
\newcommand{\ltsim}{\protect\raisebox{-0.5ex}{$\:\stackrel{\textstyle <}{\sim}\:$}}
\usepackage{amsmath}
\usepackage{longtable}

\received{September 13, 2018}
\accepted{March 12, 2019}
\submitjournal{ApJ}

\shorttitle{Mass accretion rate in LH\,95}
\shortauthors{Biazzo, Beccari, De Marchi \& Panagia}

\begin{document}

\title{Photometric determination of the mass accretion rates of pre-main sequence stars. \\
VI. The case of LH\,95 in the Large Magellanic Cloud
\footnote{Based on observations made with the NASA/ESA Hubble Space Telescope, obtained from the Data Archive at the Space Telescope 
Science Institute, which is operated by the Association of Universities for Research in Astronomy, Inc., under NASA contract 
NAS\,5-26555.}} 

\correspondingauthor{Katia Biazzo}
\email{katia.biazzo@inaf.it}

\author[0000-0002-1892-2180]{Katia Biazzo}
\affil{INAF - Osservatorio Astrofisico di Catania\\
Via Santa Sofia 78 \\
I-95123 Catania, Italy}

\author{Giacomo Beccari}
\affiliation{European Southern Observatory \\
Karl-Schwarzschild-Str. 2 \\
85748 Garching, Germany}

\author{Guido De Marchi}
\affiliation{European Space Research and Technology Centre \\
Keplerlaan 1\\
2200 AG Noordwijk, Netherlands}

\author{Nino Panagia}
\affiliation{Space Telescope Science Institute \\
3700 San Martin Drive\\
Baltimore MD 21218, USA}
\affiliation{Supernova Limited\\
OYV \#131, Northsound Rd.\\
Virgin Gorda VG1150, Virgin Islands, UK}

\begin{abstract}

We report on the accretion properties of low-mass stars in the LH\,95 association within the Large Magellanic Cloud (LMC). Using non-contemporaneous wide-band optical 
and narrow-band H$\alpha$ photometry obtained with the {\it Hubble Space Telescope}, we identify 245 low-mass pre-main sequence (PMS) candidates showing H$\alpha$ excess emission 
above the $4\sigma$ level. We derive their physical parameters, including effective temperatures, luminosities, masses ($M_\star$), ages, accretion luminosities, and mass accretion 
rates ($\dot M_{\rm acc}$). We identify two different stellar populations: younger than $\sim$8\,Myr with median $\dot M_{\rm acc} \sim 5.4 \times 10^{-8} M_\odot$yr$^{-1}$  
(and $M_\star \sim 0.15-1.8\,M_\odot$) and older than $\sim$8\,Myr with median $\dot M_{\rm acc} \sim 4.8 \times 10^{-9} M_\odot$yr$^{-1}$ (and $M_\star \sim 0.6-1.2\,M_\odot$). 
We find that the younger PMS candidates are assembled in groups around Be stars, while older PMS candidates are uniformly distributed within the region without evidence of clustering. 
We find that $\dot M_{\rm acc}$ in LH\,95 decreases with time more slowly than what is observed in Galactic star-forming regions (SFRs). This agrees with the recent interpretation according to which higher metallicity limits the accretion process both in rate and duration due to higher radiation pressure. The $\dot M_{\rm acc}$-$M_\star$ 
relationship shows different behaviour at different ages, becoming progressively steeper at older ages, indicating that the effects of mass and age on $\dot M_{\rm acc}$ cannot be treated independently. 
With the aim to identify reliable correlations between mass, age, and $\dot M_{\rm acc}$, we used for our PMS candidates a multivariate linear regression fit between these parameters. 
The comparison between our results with those obtained in other SFRs of our Galaxy and the Magellanic Clouds confirms the importance of the metallicity for the study of the 
$\dot M_{\rm acc}$ evolution in clusters with different environmental conditions. 

\end{abstract}


\section{Introduction} 
\label{sec:intro}

In the current star formation paradigm of the magnetospheric accretion scenario, a central low-mass star grows in mass over time through 
accretion of material from a circumstellar disk of dust and gas funneled by stellar magnetic field, assumed to be mostly dipolar (e.g., 
\citealt{camenzind1990, konigl1991}). The accretion disk is then truncated by the stellar magnetosphere at a few stellar radii (see a 
recent review on accretion onto pre-main sequence - PMS - stars by \citealt{hartmannetal2016}). Reliable measurements of the rate at which 
mass from circumstellar disk is transferred onto the central PMS star is therefore important for understanding the 
evolution of both the star and its disk, and for tracing possible planetary formation and subsequent evolution. In particular, the study of how 
the mass accretion rate changes with time as a star approaches the main sequence, how it depends on the mass of the forming star, and how it is 
affected by the metallicity and density of the parent cloud or by the proximity of early-type stars are of particular interest.

Ground-based studies of Galactic nearby star-forming regions show a decrease of the mass accretion rate ($\dot M_{\rm acc}$) with time, from 
$\sim 9\times10^{-8}\,M_\odot$\,yr$^{-1}$ at $\sim$1 Myr to $\sim 6\times10^{-10}\,M_\odot$\,yr$^{-1}$ at $\sim$10 Myr with a power law 
$\dot M_{\rm acc}^{-1.2}$ (e.g., \citealt{sicilia-aguilaretal2010}). This behaviour is in line with the expected evolution of viscous disks 
(\citealt{hartmannetal1998, rosottietal2017, muldersetal2017}), but the spread of the data may exceed 2\,dex at any given age. Such a scatter 
can be explained in part by the wide mass range covered by the observations, since the mass accretion rate depends also on the mass $M_\star$ 
of the forming star. While during the last twenty years several authors discussed about the steepness of the one-power $\dot M_{\rm acc}-M_\star$ 
relation (see, e.g., \citealt{alcalaetal2014}, and references therein), recently, observations suggested that two different exponents for this relation, 
at different mass regimes, can better describe the data than a single power-law (\citealt{fangetal2013, manaraetal2017, alcalaetal2017}), and 
this behaviour resembles theoretical predictions (\citealt{vorobyovbasu2009}). In particular, for $M_\star > 0.2 M_\odot$ mass accretion rate 
was found to scale with the power of $\sim 1.3-1.4$ of the stellar mass. These latter works claim the importance of modelling self-gravity 
of the disks in the early evolution of the more massive systems, but also of other physical processes, such as
photo-evaporation and planet formation during young stellar objects lifetime may lead to disk dissipation on different timescales depending 
on stellar mass (see \citealt{alcalaetal2017}, and references therein).

While potential systematic errors may contribute to the present uncertainty in the $\dot M_{\rm acc}$-$M_\star$ relation, one of the main 
limitations of the ground-based observations comes from the relatively paucity of available measurements. Indeed, most of the results so far obtained are 
based on the mass accretion rates of some hundred stars, all located in nearby Galactic star forming regions (SFRs), covering a limited range of ages 
($\sim 0.5-20$\,Myr) and with essentially solar metallicity (see, e.g., \citealt{muzerolleetal1998, herczeghillenbrand2008, antoniuccietal2011, 
rigliacoetal2011, biazzoetal2012, costiganetal2012, fangetal2013, alcalaetal2014}). The origin of this limitation can be found in the methods used to 
measure $\dot M_{\rm acc}$. Indeed, current techniques based on medium-high resolution single-object spectroscopy or low-medium resolution 
multi-object spectroscopy (or UV photometry) of nearby regions limit the number of objects and the distance to the regions that can be 
observed (because, e.g., of crowding). To partially overcome these limitations, \cite{demarchietal2010} have developed and tested to the SN\,1987A field 
a new method to reliably measure the mass accretion rate from photometry. This method, successfully applied to several regions of the 
Large Magellanic Cloud (\citealt{spezzietal2012,demarchietal2017}), the Small Magellanic Cloud (\citealt{demarchietal2011,demarchietal2013}), 
and the Milky Way (\citealt{beccarietal2010,beccarietal2015,zeidleretal2016}), combines wide-band $V$ and $I$ photometry with narrow-band 
H$\alpha$ imaging to identify all stars with significant H$\alpha$ excess emission and to derive from it the accretion luminosity ($L_{\rm acc}$) 
and hence $\dot M_{\rm acc}$ for many hundreds of objects all at once.

Here, we apply the same method to the young association LH\,95. This region, first identified by \cite{luckehodge1970}, is one of the stellar 
aggregates located north of the Large Magellanic Cloud (LMC) at $\alpha \sim 5^{\rm h}37^{\rm m}04\fs32$ and 
$\delta \sim -66^\circ22'00^{\prime\prime}.7$ in $J2000$. This group was recognized by \cite{kontizasetal1994} as an association rather poor in 
total number of stars and low in density ($\sim 0.05-0.07\,M_\odot$ pc$^{-3}$). It is embedded in the bright \ion{H}{2} region LH$\alpha$ 
120/N\,64C (\citealt{henize1956}), in an area situated to the north-east of the super-bubble LMC\,4. \cite{lucke1974} detected four early-type 
stars ($B-V \sim 0$) with $13<V<16$ mag, while \cite{kontizasetal1994} counted in it 15 blue stars and estimated for the region a mean age of 
$2\pm1\times10^7$ yr.

Most recently, \cite{gouliermisetal2002} determined a diameter slightly highly than about $2^{\prime}$ for LH\,95. From the $R-H\alpha$ 
color index versus the color index $B-V$, they identified a central cluster of four Be stars which strongly determine the \ion{H}{2} emissivity 
in an area of $\sim 4^{\prime}.1 \times 5^{\prime}.3$. They estimated a reddening of $0.1-0.2$\,mag in $B-V$ color and an age as young as $\sim 8$\,Myr 
within the cluster and older than $\sim$50 Myr in the field. They discuss about the possibility that LH\,95 is not a large mass segregated system, 
but rather a small young system. Studies on the initial mass function from high-mass ($\sim 70$\,$M_\odot$) down to sub-solar 
($\sim 0.4$\,$M_\odot$) regime were led by \cite{darioetal2009, darioetal2012}, while claims about the possibility of age spread within the PMS 
stars in the associations were reported by \cite{darioetal2010}.

Thanks to the high angular resolution and wide field achievable with the instruments on board of the {\it Hubble Space Telescope} (HST), several 
LMC associations cointaining PMS candidates were identified (\citealt{gilmozzietal1994, romaniello1998, panagiaetal2000, romanielloetal2006, 
gouliermisetal2006, gouliermisetal2007, demarchietal2010, demarchietal2017, spezzietal2012}). The spatial distribution of PMS stars within LMC 
associations shows the existence of significant substructures, as in the case of Galactic OB associations. Moreover, the locations of the 
detected low-mass PMS stars on color-color diagrams are found to be in excellent agreement with those of T\,Tauri stars with $\ltsim 2$\,$M_{\odot}$ 
in young associations of the Milky Way (\citealt{bricenoetal2007}). \cite{gouliermisetal2007} analyzed the stellar content of LH\,95 finding for 
the association a mass distribution from bright OB stars ($\sim 7$\,$M_{\odot}$) to faint red PMS stars ($\sim 0.3$\,$M_{\odot}$). They found that the 
PMS members of this association seem to be clustered in stellar sub-groups containing also a few early-type stars.

In the present paper, we study the LH\,95 association taking advantage of the HST photometry in three bands. In Section\,\ref{sec:obs}, we describe 
the photometric observations, while in Section\,\ref{sec:analysis} we address the identification of PMS candidates via their color excess, the measurement 
of the H$\alpha$ luminosity, and the derivation of the stellar parameters. Section\,\ref{sec:accretion} shows how $L_{\rm acc}$ and $\dot M_{\rm acc}$ is 
obtained from the H$\alpha$ luminosity. A general discussion about our results is provided in Section\,\ref{sec:discussion}, while summary and 
conclusions are drawn in Section\,\ref{sec:summ_Concl}. The Appendix\,\ref{sec:appendixA} provides a discussion of how stellar mass, age, mass 
accretion rate change using different evolutionary tracks, while Appendix\,\ref{sec:appendixB} lists the stellar and accretion parameters of the 
selected low-mass PMS candidates.

\section{Photometric observations} 
\label{sec:obs} 

\begin{figure*}  
\begin{center}
\vspace{-2.5cm}
\includegraphics[width=19cm]{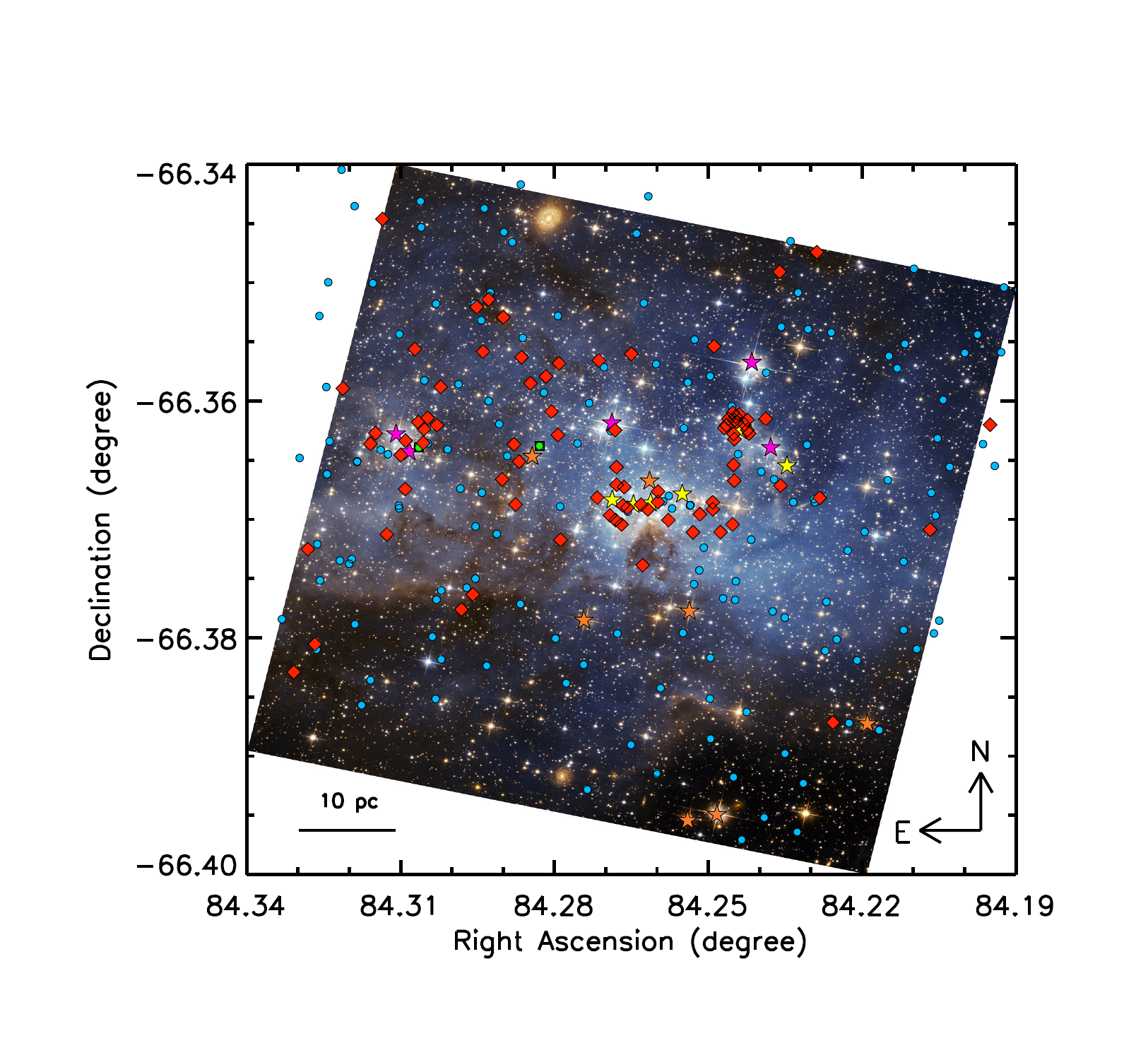}
\vspace{-2cm} 
\hspace{-1.8cm} 
\caption{Color-composite image from ACS/WFC observations in the $F555W$ and $F814W$ filters of LH\,95. Stars showing H$\alpha$ excess are displayed with 
filled dots and diamonds, the latter representing the younger pre-main sequence candidates; green squared circles are the two bright early-type targets excluded 
from our analysis (see Sect.\,\ref{sec:iden} for the selection criteria of PMS candidates). Star symbols mark the 
position of B stars identified by \cite{gouliermisetal2002} (yellow), OB stars analyzed by \cite{darioetal2012} (purple), probable massive young
targets selected from the $2MASS$ catalogue (\citealt{cutrietal2003}) with $J-H<0.8$ and $J<15$ mag (orange). These early-type stars are listed in 
Table\,\ref{tab:early-type}. North is up and East to the left. The field covers an area of about $0^\circ.15 \times 0^\circ.06$. [Picture credits: NASA, 
ESA, and the Hubble Heritage Team (STScI/AURA)-ESA/Hubble Collaboration; Acknowledgment: D. Gouliermis (Max Planck Institute for Astronomy, Heidelberg).]}
\label{fig:LH95_image} 
\end{center} 
\end{figure*}

The data were collected as part of HST programs \#10566 (PI: Gouliermis), \#12872 (PI: Da Rio), and \#13009 (PI: De Marchi). The LH\,95 region was 
observed with the Wide-Field Channel (WFC) of the Advanced Camera for Surveys (ACS) in the narrow-band filter $F658N$ (centered on the H$\alpha$ line) 
and in the wide-band filters $F555W$ and $F814W$, equivalent to the standard Johnson $V$ and $I$ bands, respectively. A log of the ACS/WFC observations is 
given in Table\,\ref{tab:log}. A color-composite image of the region in the filters $F555W$ and $F814W$ is shown in Fig.\,\ref{fig:LH95_image}. 
These $F555W$ and $F814W$ band observations are among the deepest ever taken toward the LMC (see \citealt{gouliermisetal2007, darioetal2009}) and will
allow us to explore photometrically the accretion properties for these resolved extragalactic low-mass PMS stars.

The entire data-set was reduced using the package {\sc daophotii}~\citep[][]{ste87}. We used more than 30 well sampled and not saturated sources 
to model the point spread function (PSF) of all the $F555W$ and $F814W$ images. We then used the task {\sc daophotii/montage} to stack all the 
$F555W$ and $F814W$ images together in order to produce a deep image cleaned from cosmic rays and detector imperfections. We used the stacked 
image to create a master list of sources. We accepted all objects identified at 5$\sigma$ above the background. The master list was used to 
perform accurate PSF fitting photometry on each single frame using the task {\sc allframe}~\citep[][]{ste94}. In order to retain a source in our final catalogue we 
require that the object is detected in at least 3 out of 5 images both in the $F555W$ and $F814W$ bands. The average of the magnitudes measured 
in each individual frame was adopted as stellar magnitude while the standard deviation was adopted as the associated photometric error.

Given the low level of stellar crowding affecting the images taken with the $F658N$ filter, we used aperture photometry to measure the $m_{658}$ 
magnitude of the stars. Aperture photometry was obtained with {\sc daophotii} using 5 deep drizzled images. These images are available for download 
as part of the high level scientific products of the archive of the Space Telescope Science Institute (STScI). 
We used the catalogue of sources measured in $F555W$ and $F814W$ as master list of stars. We registered the position of the stars listed in this catalogue 
on the astrometric system of the drizzled images with an overall accuracy of 0\farcs05. Once again the average of the magnitudes measured in at least 
3 out of 5 frames for every master list object was adopted as the stellar $m_{658}$ magnitude in the final catalogue, while we took the resulting standard 
deviation around the mean for each object as the associated photometric uncertainty. 

The final catalog contains 24515 objects with measured $m_{555}$ and $m_{814}$ magnitudes, of which 21512 have also a measure of $m_{658}$. This is expected 
as the images acquired with the $F658N$ narrow filter are slightly shallower with respect to the broad-band images. We emphasize here that this is not an issue 
since we are here mostly interested in the identification of H$\alpha$ excess emitters, hence stars with high flux in the $F658N$ band.

The $m_{555}$, $m_{814}$, and $m_{658}$ magnitudes were calibrated in the VEGAMAG system following the recipe in~\citet{si05} and using the most recent 
zero point values available through the ACS Zeropoint Calculator (\citealt{ryonetal2018}). As for the extinction law, we adopted the one derived specifically 
in the field of LH\,95 by \cite{darioetal2009}. The reddening distribution obtained by the same authors does not show evidence of patchy nature 
of the absorption, unlike other LMC (e.g., \citealt{demarchietal2017}) or nearby SFRs (see, e.g., \citealt{luhman2007, hillenbrandetal2013}).
Therefore, we applied a uniform reddening correction for all stars in the sample (details will be discussed in Sect.\,\ref{sec:iden}).

\begin{table}  
\caption{Summary of the observations.}
\label{tab:log} 
\small 
\begin{center} 
\begin{tabular}{lcccc} 
\hline
\hline 
Program ID &   Date  & UT         & Filter & $T_{\rm exp}$\\ 
(\#)  &  (yy-mm-dd)  & (hh:mm:ss) &  & (s) \\ 
\hline
10566 &  2006-03-02  & 20:53:03 & $F555W$ & 5200 \\ 
10566 &  2006-03-06  & 20:59:47 & $F814W$ & 5200 \\ 
12872 &  2013-05-26  & 18:47:27 & $F658N$ & 2718 \\ 
12872 &  2013-05-28  & 03:07:15 & $F658N$ & 2718 \\ 
12872 &  2013-05-29  & 03:02:02 & $F658N$ & 2862 \\ 
12872 &  2013-05-30  & 01:20:39 & $F658N$ & 2864 \\ 
13009 &  2013-05-26  & 21:58:25 & $F658N$ & 2753 \\ 
\hline
\end{tabular} 
\end{center} 
\normalsize 
\end{table}

\begin{table}  
\caption{Characteristics of the early-type stars identified in the field of our observations. Right ascension and declination are given in J2000. } 
\label{tab:early-type} 
\tiny
\begin{center} 
\begin{tabular}{lcccc} 
\hline 
\hline 
Name & Right Ascension   & Declination  &  $V/J$    \\ 
     &  (hh:mm:ss)       & (\degr:$^\prime$:\arcsec)     & (mag) \\ 
\hline
\multicolumn{4}{c}{Sample of Be stars by \cite{gouliermisetal2002}$^\ast$ }\\ 
\hline 
ID114$^a$  & 05:37:05.38 & $-$66:21:59.13   & 16.16$^b$ \\ 
ID124$^a$  & 05:37:04.38 & $-$66:22:00.53   & 16.24$^b$  \\ 
ID157$^a$  & 05:36:59.33 & $-$66:21:37.54   & 16.48$^b$ \\ 
ID239$^a$  & 05:36:57.14 & $-$66:21:48.53   & 16.87$^b$  \\ 
ID410$^a$  & 05:37:03.57 & $-$66:22:00.03   & 17.33$^b$   \\ 
ID1388$^a$ & 05:37:02.08 & $-$66:21:57.29   & 18.53$^b$  \\ 
\hline 
\multicolumn{4}{c}{Sample of massive stars by
\cite{darioetal2012}$^{\ast\ast}$ }\\ 
\hline 
SK-66 170 & 05:36:58.945 & $-$66:21:16.130 & 13.17$^b$  \\ 
SK-66 172 & 05:37:05.553 & $-$66:21:34.950 & 13.58$^b$ \\ 
SK-66 174 & 05:37:15.723 & $-$66:21:38.355 & 13.74$^b$ \\ 
ID18$^a$  & 05:36:58.007 & $-$66:21:42.613 & 14.53$^b$   \\ 
ID85$^a$  & 05:37:15.129 & $-$66:21:44.304 & 15.58$^b$   \\ 
\hline 
\multicolumn{4}{c}{Sample of $2MASS$ stars by \cite{cutrietal2003} }\\
\hline 
2MASS\,J05370037$-$6623410 & 05:37:00.371 & $-$66:23:41.03 & 13.254$^c$ \\ 
2MASS\,J05365327$-$6623109 & 05:36:53.274 & $-$66:23:10.98 & 14.846$^c$ \\  
2MASS\,J05370664$-$6622387 & 05:37:06.645 & $-$66:22:38.73 & 14.212$^c$ \\ 
2MASS\,J05370166$-$6622355 & 05:37:01.665 & $-$66:22:35.55 & 14.653$^c$ \\ 
2MASS\,J05370175$-$6623421 & 05:37:01.753 & $-$66:23:42.15 & 14.554$^c$ \\ 
2MASS\,J05370915$-$6621455 & 05:37:09.151 & $-$66:21:45.59 & 14.813$^c$ \\ 
2MASS\,J05370361$-$6621532 & 05:37:03.619 & $-$66:21:53.22 & 14.848$^c$ \\ 
\hline 
\end{tabular}
\end{center} 
$^\ast$ These objects were classified as Be stars by \cite{gouliermisetal2002} through $BVR$ photometry; $^{\ast\ast}$ These stars were classified as 
B0.2IIIp, O2III(f*)+OB, O7.5III(f), O6.5V, and B1.5III, respectively, by \cite{darioetal2012} using high-resolution spectroscopy.\\ 
$^a$ Catalogue ID number of \cite{gouliermisetal2002}. $^b$ Magnitudes in $V$ band. $^c$ Magnitudes in $J$ band. 
\normalsize 
\end{table}

\section{Data Analysis} 
\label{sec:analysis}
In order to measure the H$\alpha$ luminosity, $L_{\rm H\alpha}$, we need a solid estimate of the stellar continuum in the H$\alpha$ band, i.e. without 
the contribution of the emission from the background. This is important for deriving accretion luminosity and mass accretion rate, whose measurement will 
be discussed in the following sub-sections.

\subsection{Identification of PMS candidates}
\label{sec:iden}

With the aim to identify PMS candidates within LH\,95, we followed the method described and first tested in \cite{demarchietal2010} and 
then applied in a series of papers (\citealt{beccarietal2010, beccarietal2015, demarchietal2011, demarchietal2013, demarchietal2017, spezzietal2012, 
zeidleretal2016}). This method relies on the detection of H$\alpha$ excess emission in low-mass star-forming stars (see, e.g., \citealt{konigl1991}).

\begin{figure}[h]
\hspace{-1cm}\includegraphics[width=10cm]{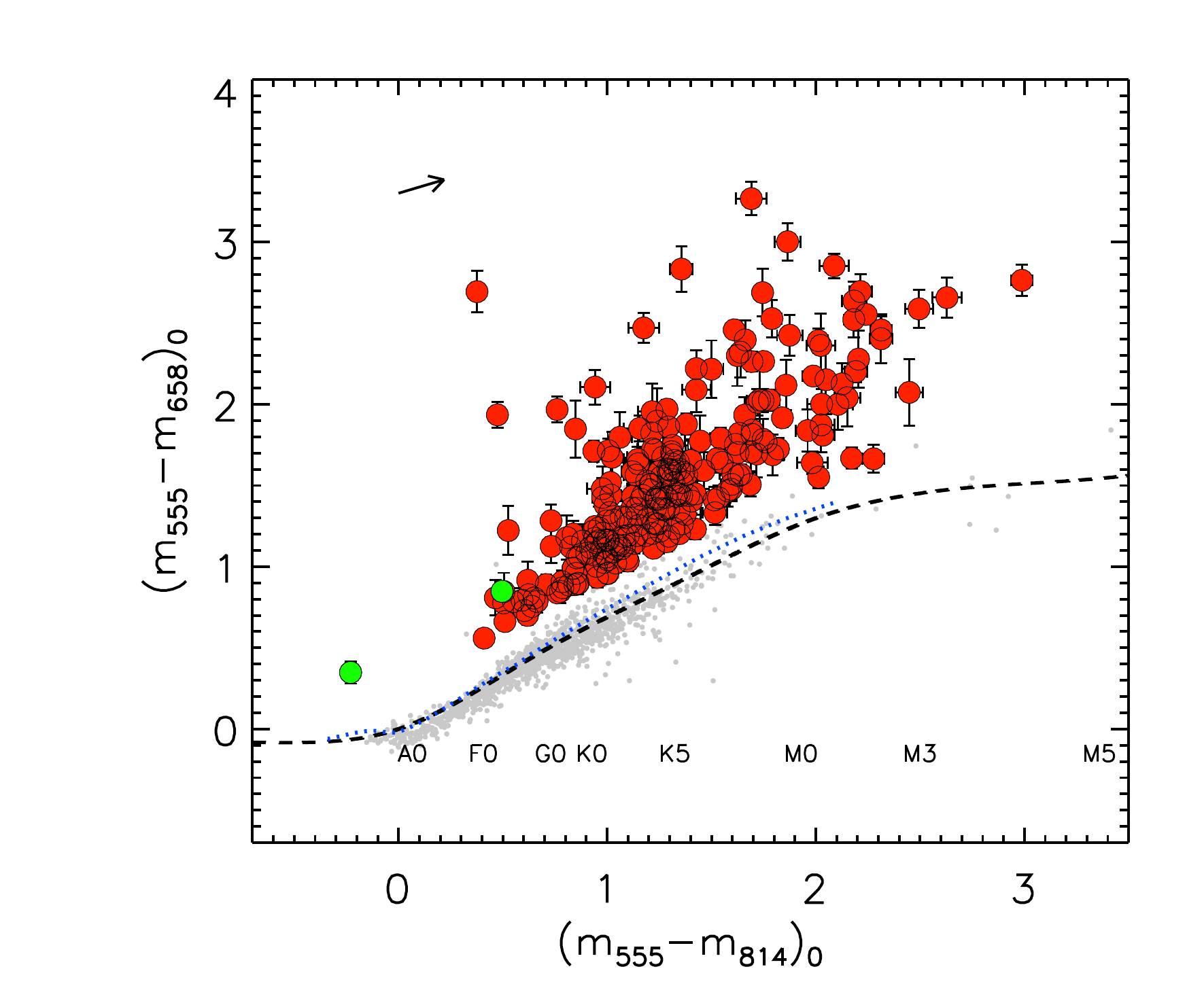}
\caption{De-reddened color-color diagram of the selected 1541 stars in the field of LH\,95. All magnitudes are already corrected for both the extinction 
contribution of our Galaxy and of LH\,95. The dashed line represents the running median photospheric $(m_{555}-m_{814})_0$ color for the 1294 stars with small 
($<0.05$\,mag) photometric uncertainties in all three bands (grey small dots). The dotted line shows the model atmospheres of \cite{besselletal1998} 
computed for the three ACS/WFC filters. The arrow displays the reddening vector of $E(m_{555}-m_{814})^{\rm LH\,95}=0.2$ and 
$E(m_{555}-m_{658})^{\rm LH\,95}=0.09$ for the adopted LH\,95 extinction law. A total of 247 objects with $(m_{555}-m_{658})_0$ excess larger than 
$4\sigma$ are indicated with large red dots. Green dots mark the position of the two brighest ($\ltsim 22$\,mag in the $F555W$ band) targets with H$\alpha$ 
excess emission. Error bars are also shown. Spectral types as in \cite{pecautmamajek2013} are marked in the bottom of the plot.} 
\label{fig:VI_VHalpha} 
\end{figure}

\begin{figure}  
\hspace{-1cm}\includegraphics[width=10cm]{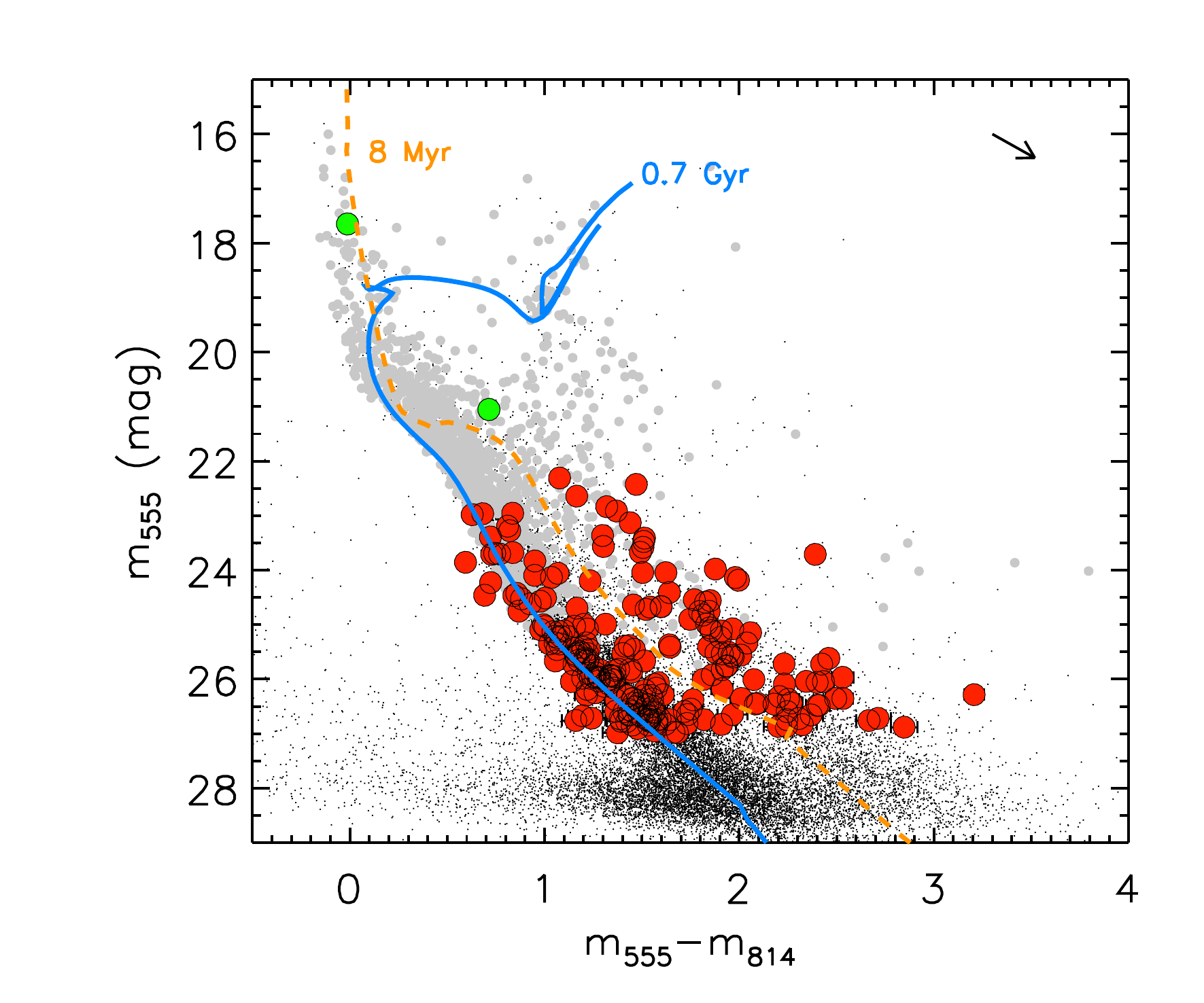}
\caption{Color-magnitude diagram of the field of LH\,95 (24515 sources). All magnitudes are already corrected for the extinction law of our Galaxy. As in 
Fig.\,\ref{fig:VI_VHalpha}, green and red big dots represent the 247 PMS star candidates with H$\alpha$ excess emission at the $4\sigma$ level with $m_{555}$ 
magnitude brighter and fainter than 22\,mag, respectively, while small grey dots are, as in Fig.\,\ref{fig:VI_VHalpha}, the stars with photometric uncertainties 
$<0.05$\,mag in all three filters (see text). Error bars are also shown, but they are within the symbol size in most cases. The arrow is the reddening 
vector of $E(m_{555}-m_{814})^{\rm LH\,95}=0.2$ and $A_{555}^{\rm LH\,95}=0.44$ mag applicable to LH\,95. Solid and dashed lines show the theoretical isochrones from 
\cite{bressanetal2012} for ages of 0.7 Gyr and 8 Myr, respectively, metallicity $Z=0.007$, and a distance modulus $(m_V-M_V)_0=18.55$. These models include only 
the absorption due to our Galaxy along the line of sight (i.e. $E(m_{555}-m_{814})^{\rm MW}=0.1$ and $A_{555}^{\rm MW}=0.22$ mag).} 
\label{fig:CMD} 
\end{figure}

We selected from our photometric catalogue in the $F555W$, $F658N$, and $F814W$ bands all those stars whose photometric uncertainties 
$\delta_{555}$, $\delta_{658}$, and $\delta_{814}$ in each individual band do not exceed 0.05\,mag. A total of 1294 stars satisfy this condition 
(grey small dots in the color-color diagram of Fig.\,\ref{fig:VI_VHalpha}), out of 24515 sources in the complete catalogue (black little dots in 
the color-magnitude diagram of Fig.\,\ref{fig:CMD}). These stars are typically old main sequence (MS) and do not have appreciable H$\alpha$ excess; 
they define the reference with respect to which one should look for excess emission in the $(m_{555}-m_{658})_0$ color at given $(m_{555}-m_{814})_0$ 
color (see the running median represented by the dashed line in Fig.\,\ref{fig:VI_VHalpha}). For comparison, this reference sequence is in good 
agreement with the theoretical color relationship in the same filters obtained using the \cite{besselletal1998} model atmospheres for MS stars with 
effective temperature $3500 \le T_{\rm eff} \le 40000$\,K, surface gravity $\log g=4.5$, and metallicity index $[M/H] \simeq -0.5$ 
(\citealt{coluccietal2012}), appropriate for the LMC (dotted line in the same figure). The root mean square (rms) deviation between the model and 
the data amounts to $\sim$0.03\,mag and is dominated by the systematic departure around $(m_{555}-m_{814})_0 \sim 1.5$ most likely due to the coarse 
sampling of our data. As pointed out by \cite{demarchietal2017}, even before correction for reddening, such a kind of color-color diagram provides a 
robust identification of stars with H$\alpha$ excess, since in these bands the reddening vector runs almost parallel to the median photospheric colors 
of non-accreting objects, and, moreover, our targets do not have a known patchy nature of the interstellar absorption (see \citealt{darioetal2009}). 

To select the most probable accretors, after the exclusion of the 1294 stars taken as reference, we first selected the targets with $\delta_{555}<0.1$\,mag, 
$\delta_{658}<0.3$\,mag, and $\delta_{814}<0.1$\,mag, namely 5155 objects. Then, we retained those whose de-reddened $(m_{555}-m_{658})_0$ color exceeds the 
local average by at least four times the individual combined photometric uncertainty $\delta_3$ in the color in the three bands $F555W$, $F658N$, and $F814W$, 
where 
\begin{equation} 
\label{eq:delta3} 
\delta_3=\sqrt{\frac{ \delta_{555}^2 + \delta_{658}^2 + \delta_{814}^2}{3}}\,, 
\end{equation} 
with $\delta_{555}$, $\delta_{658}$, and $\delta_{814}$ being the photometric uncertainties in each individual band. It should be noted that, for the 
selected 5155 sources, $\delta_3$ is dominated by the uncertainty on the H$\alpha$ magnitude, which is on average around 0.15\,mag, while the median value 
of the uncertainty in the other two bands is $<\delta_{555}> \sim 0.03$\,mag and $<\delta_{814}> \sim 0.02$\,mag, respectively. In the end, a total of 
247 stars satisfy the condition that we have set and they must be regarded as having bona fide H$\alpha$ excess above the 4$\sigma$ level 
(big dots in Fig.\,\ref{fig:VI_VHalpha}). This allows us to select most probable PMS candidates even when the uncertainty in the $F658N$ 
band is not negligible. Indeed, as we will show in Sect.\,\ref{sec:equiv_linelum}, our selection in H$\alpha$ excess emission translates directly into 
H$\alpha$ equivalent widths typical of accretors, allowing us to safetely remove possible contaminants from our sample, such as chromospherically active stars 
(see, e.g., \citealt{whitebasri2003, biazzoetal2007, frascaetal2008, beccarietal2015}). Other classes of objects whose spectra might present H$\alpha$ emission 
are interacting binaries, but typically they are very rare in Local Group galaxies (e.g., \citealt{dobbieetal2014}) and their intrinsic colors are bluer with respect to the main sequence 
(see, e.g., \citealt{beccarietal2014}, and references therein).

The $m_{555}$ magnitude versus the $m_{555}-m_{814}$ color of the detected sources is shown in Fig.\,\ref{fig:CMD}. From this 
color-magnitude diagram, the two targets with $m_{555} < 22$\,mag are bright objects with H$\alpha$ excess, that we exclude from our following analysis as 
we are searching for low-mass PMS candidates\footnote{Even though we could derive the parameters of these two objects from their colors, we prefer to 
focus only on low-mass objects. In fact, without spectroscopy the age of these early-type stars would be rather uncertain, because they could be both 
PMS and post-MS stars, making the comparison with the lower-mass PMS objects of interest here more difficult.}. \cite{gouliermisetal2007} found in the CMD 
a pronounced turnoff at $V \sim 22.5$\,mag and a red clump at $V \sim 19$\,mag and $V-I \sim 1.2$. The red clump (RC) and 
old MS population are best matched by a $0.7$ Gyr isochrone taken from the Padova-Trieste Stellar Evolution Code (PARSEC; see \citealt{bressanetal2012}). 
Stars with H$\alpha$ excess are shown in red and they define two distinct groups, nicely separated by a 8 Myr old PMS isochrone from the same authors. 
The theoretical isochrones of \cite{bressanetal2012} are shown, respectively, with solid blue and orange dashed lines in Fig.\,\ref{fig:CMD}, where the distance 
of $51.4\pm1.2$\,kpc (\citealt{panagia1999}), corresponding to a distance modulus $(m_V-M_V)_0=18.55$, and a metallicity of $Z=0.007$, typical of young LMC stars 
(e.g., \citealt{coluccietal2012}) were adopted. These isochrones includes both the effects of the Milky Way intervening absorption along the line of sight 
and the LH\,95 mean absorption within the field. According to \cite{fitzpatricksavage1984}, the former amounts to 
$A_{555}^{\rm MW}=0.22$\,mag and $E(m_{555}-m_{814})^{\rm MW}=0.1$, while for the LH\,95 field we considered the following mean extincion values 
(\citealt{darioetal2009}): $R_{555}=A_{555}/E(m_{555}-m_{814}) \simeq 2.18$ and $R_{814}=A_{814}/E(m_{555}-m_{814}) \simeq 1.18$. 

To determine the presence and extent of possible differential extinction, \cite{darioetal2009, darioetal2012} analyzed the position of the upper main sequence (UMS) 
stars in the CMD. In particular, after subtracting field stars, they compared the CMD position of the UMS objects with that expected according to grids of evolutionary models. 
They concluded that there is not a significant level of differential extinction for the UMS stars in the LH\,95 field. This conclusion is very relevant to our investigation because 
young PMS objects and UMS stars share the same spatial distribution (see, e.g., \citealt{demarchietal2011, demarchietal2013}). Those authors also provide the total optical 
extinction and reddening toward LH\,95, namely, $A_{555}^{\rm tot}=0.6$\,mag and $E(m_{555}-m_{814})^{\rm tot} \sim 0.3$, respectively; 
therefore the reddening within LH\,95 is $E(m_{555}-m_{814})^{\rm LH\,95} \sim 0.2$. Finally, considering the relation $A_{658}/A_{555} \simeq 0.8$ 
(\citealt{rodrigoetal2012}) and the adopted Galactic and LMC extinction laws, the total extinction in the H$\alpha$ band toward LH\,95 is $A_{658}=0.48$\,mag.

Looking at the color-magnitude diagram shown in Fig.\,\ref{fig:CMD}, most stars with $m_{555}-m_{814} \gtsim 0.6$ and $m_{555} \gtsim 22$\,mag could be old MS, 
or PMS objects, or red giants. This is why it is important to search for PMS objects analyzing the H$\alpha$ excess emission as signature of accretion (and 
therefore of youth).

\subsection{From H$\alpha$ color excess to line luminosity and equivalent width} 
\label{sec:equiv_linelum}

As pointed out by \cite{demarchietal2010}, the contribution of the H$\alpha$ line to the $m_{555}$ magnitude is completely negligible and therefore the 
magnitude $\Delta H\alpha$ corresponding to the excess emission is:

\begin{equation} 
\label{eq:delta_halpha} 
\Delta H\alpha = (m_{555}-m_{658})^{\rm obs} - (m_{555}-m_{658})^{\rm ref} \,, 
\end{equation}
\noindent{where the superscript ``obs'' refers to the observations and ``ref'' to the reference sequence at each $m_{555}-m_{814}$ color (dashed line in 
Fig.\,\ref{fig:VI_VHalpha}). Once $\Delta H\alpha$ is derived in this way, the H$\alpha$ emission line luminosity $L_{\rm H\alpha}$ can be immediately obtained 
from the photometric zero point ($ZP$), the absolute sensitivity of the instrumental setup (in this case the $F658N$ of the ACS/WFC), and the distance to the sources. 
We took the $F658N$ photometric properties of the instrument at the exact observing dates from \cite{ryonetal2018}, namely the inverse sensitivity 
$PHOTFLAM=1.967 \times 10^{-18}$\,erg\,cm$^{-2}$\,s$^{-1}$\,\AA, and the zero-point in VEGAmag $ZP=22.383$\,mag. Considering the rectangular width of the 
$F658N$ filter $RECTW=74.98$\,\AA\,and assuming a distance to SN\,1987A of $51.4\pm1.2$ kpc (\citealt{panagia1999}), we derived a median value of the H$\alpha$ 
luminosity of the 245 low-mass objects with H$\alpha$ excess of $\sim 1.2 \times 10^{31}$\,erg\,s$^{-1}$ or $\sim 0.3 \times 10^{-2}$\,$L_\odot$. This value 
is lower than that measured by \cite{demarchietal2010} in the SN\,1987A field ($\sim 10^{-2}$\,$L_\odot$) and that found by \cite{demarchietal2017} in the 
30\,Doradus Nebula ($\sim 3 \times 10^{-2}$\,$L_\odot$). This difference is not surprising because our observations include stars with greater $m_{555}-m_{814}$ 
colors, and therefore lower masses}.

The total uncertainty on our $L_{\rm H\alpha}$ measurements is typically $\sim 16\%$ and is dominated by the inaccuracy on the H$\alpha$ magnitude, the 
uncertainty on the distance and on the instrumental setup accounting for $\sim 5\%$ and $\sim 3\%$, respectively (see also \citealt{demarchietal2010}). 
Extinction indeed does not have any influence in the $V-H_{\rm H\alpha}$ color excess, and therefore in the $L_{\rm H\alpha}$ uncertainty, both because the 
reddening vector in Fig.\,\ref{fig:VI_VHalpha} is substantially parallel to the reference template (dashed line) and also because our targets do not seem to 
have differential reddening, as indicated, for instance, by the relatively compact RC in the CMD.

Following \cite{demarchietal2010}, the difference between the observed H$\alpha$ magnitude ($m_{658}$) and the level of the H$\alpha$ continuum ($m_{658}^{\rm c}$) 
provides a direct measure of $EW_{\rm H\alpha}$. In particular, since the line width is narrow compared to the width of the 
filter, the line profile falls completely within the filter bandpass. If we assume that the stars defining the reference template have no H$\alpha$ absorption 
features, their $m_{555} - m_{658}$ index would correspond to the color of the pure continuum. Therefore, $EW_{\rm H\alpha}$ is given by the following relationship:

\begin{equation}
\begin{split}
EW_{\rm H\alpha} & = RECTW \times [1 - 10^{-0.4 \times (m_{658} - m_{658}^{\rm c})}] \\
  & = RECTW \times [1 - 10^{-0.4 \times \Delta H\alpha}] \,, 
\end{split}
\end{equation}
\noindent{with $RECTW$ the rectangular width of the $F658N$ filter. As for $L_{\rm H\alpha}$, the statistical uncertainty on $EW_{\rm H\alpha}$, typically 
$\sim 6$\%, is dominated by the uncertainty on the H$\alpha$ photometry. The validity of this method was also independently tested and then confirmed by 
\cite{barentsenetal2011, barentsenetal2013}, who considered both photometric and spectroscopic observations for T Tauri stars in the Galactic 
NGC\,2264 and IC\,1396 young regions. The authors found strong correlation between H$\alpha$ equivalent widths derived with both spectroscopic and 
photometric methods. This already suggests that the possible contribution of veiling due to accretion is greatly reduced by the subtraction of the continuum flux 
from the band flux. 

During the last thirty years, veiling was more and more accurately measured mainly thanks to spectroscopic observations of SFRs in our Galaxy (see, e.g., 
\citealt{hartiganetal1995, manaraetal2013, alcalaetal2014}, to cite a few works), but in general, its effects must be taken into account also for broad-band 
measurements. As discussed in detail in \cite{demarchietal2010}, any nebular 
continuum unrelated to the stellar photosphere (and therefore also the one associated with the accretion luminosity) will add to the intrinsic continuum of the 
stars, thereby affecting both the observed total level and the slope. This could alter the measured broadband colors of the source, thereby thwarting our attempts 
to infer the level of the continuum in the H$\alpha$ band from the observed $m_{658}$ and $m_{814}$ magnitudes, and ultimately also affecting the effective temperature and 
bolometric luminosities that we measure (see Section\,\ref{sec:Teff_Lstar}). 

Fortunately, the contribution of the nebular continuum appears to be insignificant for the stars in our sample. To prove this, \cite{demarchietal2010} assumed 
a fully ionized gas of pure H, considering only bound-free and free-free transitions and ignoring the contribution to the continuum from two-photon emission 
(\citealt{spitzergreenstein1951}). Using Osterbrock's (1989, chapter 4) tabulations, they find H$\alpha$ line intensity and H$\alpha$ continuum fluxes of the 
nebular gas for gas electron temperatures in the range 5000–20000 K (see their Table 2). The purely nebular $EW_{\rm H\alpha}$ ranges from 5000\,\AA\,to 
9000\,\AA, or more than 2 orders of magnitude higher than what we measure for our PMS objects (see Figure 4). We can, therefore, safely conclude that the 
nebular contribution to the continuum is negligible (less than 1\%).  

Furthermore, for gas temperatures in the range 5000–10000\,K the $m_{658}-m_{814}$ color of the nebular continuum varies from 1.4 to 0.5, spanning a range typical of 
G–K type stars. Thus, the effects of the nebular continuum on the $m_{658}-m_{814}$ color of PMS stars remains insignificant even for the objects with the highest $EW_{\rm H\alpha}$ 
in our sample. Therefore, for these objects also the relationships between $m_{658}-m_{814}$ and effective temperature and, in turn, the bolometric luminosity that we will derive in 
Section\,\ref{sec:Teff_Lstar} are not affected by the veiling introduced by the additional nebular continuum.}

\begin{figure}  
\hspace{-1cm}\includegraphics[width=10cm]{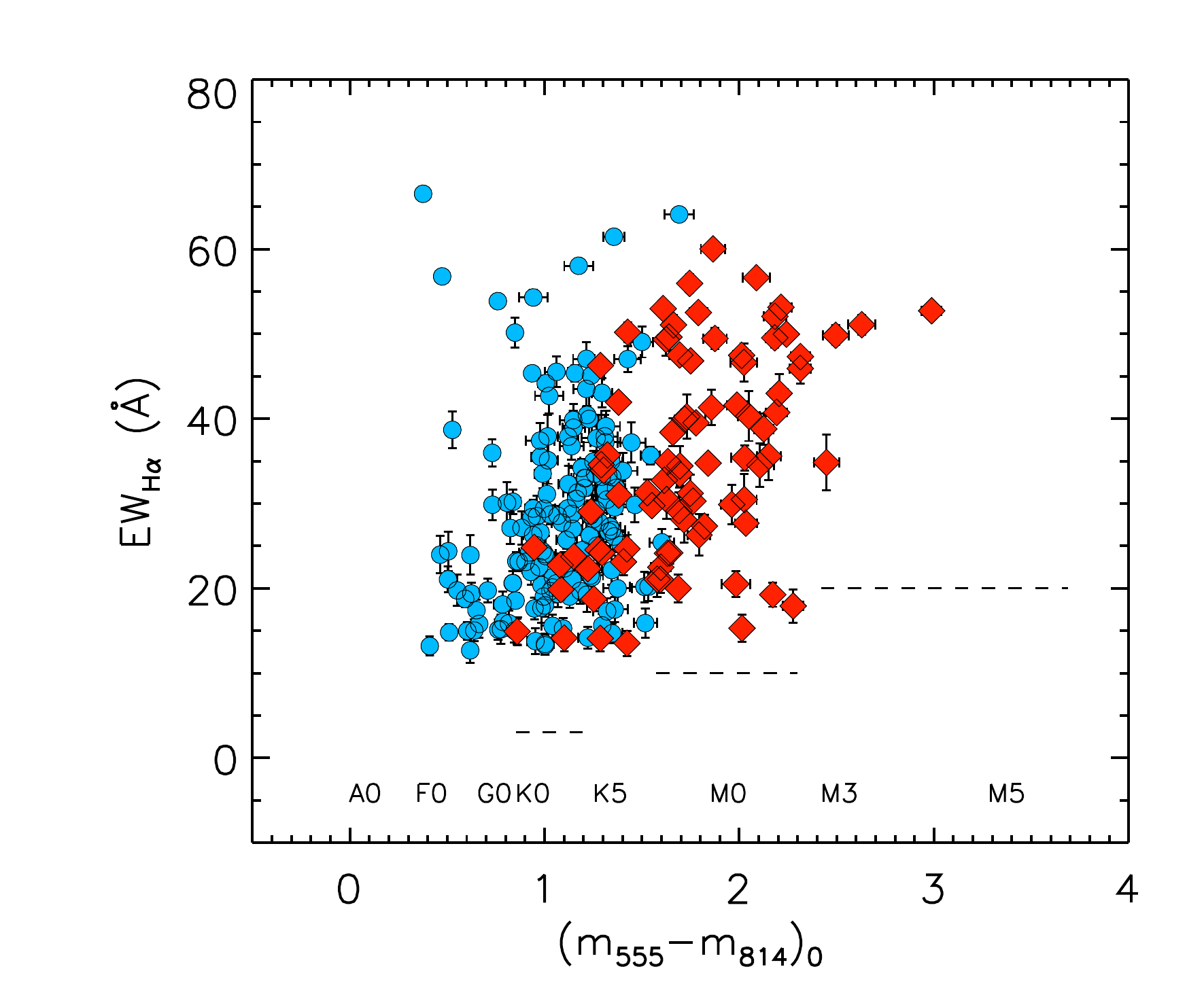}
\caption{H$\alpha$ equivalent width of the selected 245 low-mass PMS candidates in the field of LH\,95, as a function of the dereddened $m_{555}-m_{814}$ 
color. The red diamonds mark the position of the younger PMS candidates while the blue dots represent the older PMS stars in our sample with H$\alpha$ excess (see 
Section\,\ref{sec:astr_paramet} for details). Error bars are also shown. Spectral types as in \cite{pecautmamajek2013} are marked at the bottom of the plot. Dashed 
lines define the thresholds above which most probable accretors are positioned at given spectral types, according to \cite{whitebasri2003}.} 
\label{fig:ew_halpha} 
\end{figure}

The values of $EW_{\rm H\alpha}$ that we obtain in the field of LH\,95 are shown in Fig.\,\ref{fig:ew_halpha} as a function of the dereddened $m_{555}-m_{814}$ color 
for the 245 low-mass PMS candidates. All selected targets at the 4$\sigma$ level fall well above the threshold established by \cite{whitebasri2003} to identify 
probable accretors as a function of spectral type (see also \citealt{beccarietal2014}), thus confirming that our selection criteria of PMS candidates is cautious. Indeed, 
the sample that we selected must be considered as a lower limit to the number of objects with 
genuine H$\alpha$ excess. This means that most probably we are excluding from the sample the weakly accreting PMS stars, but completeness is not the aim of 
this work. Instead, we are interested in studying the properties of the mass accretion process in PMS stars and for this reason it is important to have a solid sample 
of candidates.

Values of $EW_{\rm H\alpha}$ for our sample range from $\sim 12$\,\AA\,to $\sim 70$\,\AA, with a median of 29\,\AA. These values are typical of 
PMS stars. It should be noted that, because of the width of the specific $F658N$ filter, $\Delta$H$\alpha$ includes small contributions due to the emission of the two 
forbidden [\ion{N}{2}] lines at $\lambda$6548\,\AA\,and $\lambda$6584\,\AA. \cite{demarchietal2010}, following a conservative approach, have estimated corrections 
of $\sim 0.98$, on average, for the ACS $F658N$ filter. This translates into a lower $EW_{\rm H\alpha}$ value by $\sim 0.2-1.4$\,\AA\,in the range characteristic 
of our targets, i.e. within the uncertainties of our measurements (see Table\,\ref{tab:param}).

\subsection{Astrophysical parameters of PMS candidates}
\label{sec:astr_paramet} Physical parameters of the PMS candidates identified in Section\,\ref{sec:iden}, i.e. their effective temperature, bolometric luminosity, 
mass, and age, were obtained as explained in the following two sub-sections.

\subsubsection{Effective temperature and bolometric luminosity}
\label{sec:Teff_Lstar}

We derived the effective temperature $T_{\rm eff}$ from the observed $m_{555}-m_{814}$ color, properly corrected for the total reddening, due to both our Galaxy 
and LH\,95, as explained in Section\,\ref{sec:iden}. The models of \cite{besselletal1998} with $3500\,{\rm K} \le T_{\rm eff} \le 40000$ K, $\log g =4.5$, and 
metallicity index of $[M/H]=-0.5$\,dex were used for the conversion from color to effective temperature, following the work by \cite{demarchietal2010} for ACS/WFC 
filters (see that paper for details). Since models of \cite{besselletal1998} are available only for $T_{\rm eff} > 3500$ K, for lower temperatures we decided to consider 
the $T_{\rm eff}$-$(V-I_{\rm C})$ calibration by \cite{pecautmamajek2013}\footnote{We verified that in the $T_{\rm eff}$ range in common, the two calibrations are 
in very good agreement. In particular, considering the parameter space of the present work, the agreement is within $\sim 0.03$\,mag and $\sim 70$\,K in color and 
$T_{\rm eff}$, respectively.}, assuming for simplicity that the calibrated $m_{555}$ and $m_{814}$ magnitudes coincide with $V$ and $I_{\rm C}$. 
The reason for using a different calibration at temperatures lower than those covered by the models of \cite{besselletal1998} is to avoid possibly larger uncertainties 
in $T_{\rm eff}$ due to arbitrary extrapolations, as the relationship between $T_{\rm eff}$ and $m_{555}-m_{814}$ is critical for very cool temperatures. 
The bolometric luminosity $L_\star$ was obtained from the $m_{555}$ magnitude corrected for the interstellar extinction (see Section\,\ref{sec:iden}), having adopted a 
distance to LH\,95 of 51.4\,kpc (\citealt{panagia1999}) and a bolometric solar magnitude $M_{\rm bol}^{\odot}=4.74$\,mag (\citealt{pecautmamajek2013}), and having 
applied at any $T_{\rm eff}$ the bolometric corrections of the latter authors.

\begin{figure}  
\hspace{-1cm}
\includegraphics[width=10cm]{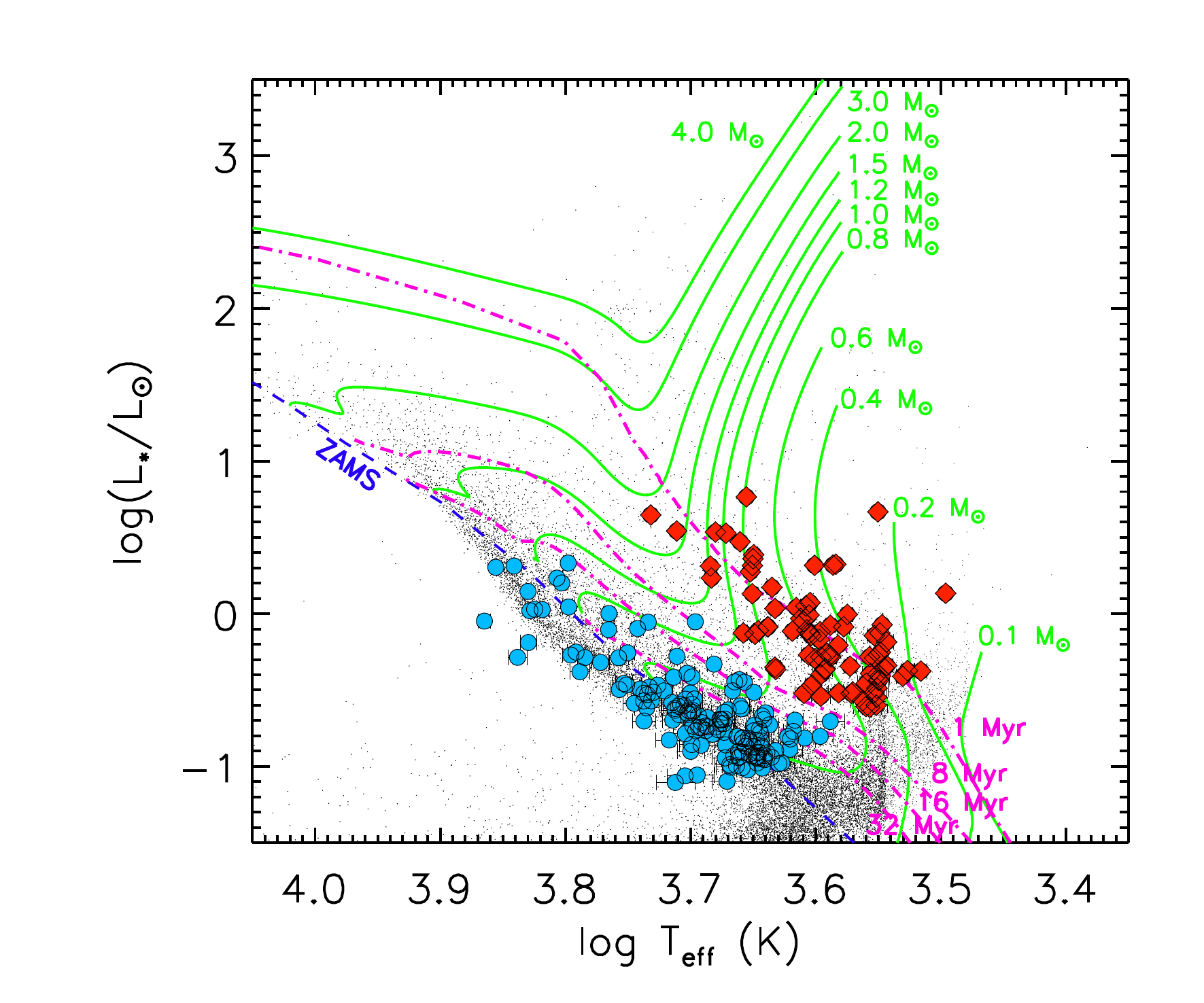} 
\caption{Location of the PMS candidates in the HR diagram (diamonds: younger PMS stars; dots: older PMS stars). Overimposed are the PARSEC 
evolutionary tracks (solid green lines) and theoretical isochrones (dot-dashed red lines) of \cite{bressanetal2012} for metallicity $Z=0.007$. The masses, in $M_\odot$, 
and the ages, in Myr, are indicated next to each track and isochrone, respectively. The position of the ZAMS is marked with a dashed line.}
\label{fig:HRdiagram} 
\end{figure}

The positions of the PMS candidates in the HR diagram are displayed in Fig.\,\ref{fig:HRdiagram}, where the $\pm 1 \sigma$ uncertainties on $T_{\rm eff}$ and 
$L_\star$ are also shown; in most cases, these errors are within the symbol size. They are mostly due to uncertainties in photometry and 
distance\footnote{Note that uncertainties on $T_{\rm eff}$ and $L_\star$ do not include, respectively, possible color/magnitude temporal variability 
due to the observations in the $F555W$ and $F814W$ filters, having been obtained 4 days apart (see Table\,\ref{tab:log}).}. As reference, 
we traced the PMS theoretical isochrones of \cite{bressanetal2012} for metallicity $Z=0.007$, as appropriate for the young populations of the LMC (e.g., 
\citealt{coluccietal2012}), and for ages of 1, 8, 16, 32 Myr from right to left (dot-dashed lines). Also shown are the representative PARSEC evolutionary tracks 
for masses of 0.1, 0.2, 0.4, 0.6, 0.8, 1.0, 1.2, 1.5, 2.0, 3.0, and 4.0 $M_\odot$ from the same \cite{bressanetal2012} models (solid lines). The dashed line in the 
same figure defines the Zero Age Main Sequence (ZAMS) by \cite{bressanetal2012}. From the stellar bolometric luminosity and effective temperature we also derived 
the stellar radius $R_\star$ assuming 5770\,K as effective temperature of the Sun. Typical mean uncertainties on $R_\star$ are around 5\% and include both 
uncertainties in $T_{\rm eff}$ and $L_\star$.

In the HR diagram of our PMS candidates, a bimodal distribution in age and $T_{\rm eff}$ seems to be evident, with a separation around 8 Myr. In particular, 
stars younger than 8\,Myr have a mean $T_{\rm eff} \sim 3965$\,K, while older targets have a mean $T_{\rm eff}$ of $\sim 4990$\,K. This apparent bimodality 
will be also evident in the accretion properties, thus proving that it is not caused by detection limits. This issue will be discussed in the following Sections.

\subsubsection{Mass and age} 
\label{sec:mass_age}

After having identified a population of PMS candidates in Sect.\,\ref{sec:iden} and derived their effective temperature and bolometric luminosity in the 
previous sub-section, it is important for our purposes to determine their mass and age from the HR diagram (see Fig.\,\ref{fig:HRdiagram}). We followed the approach 
originally discussed in \cite{romaniello1998} and most recently refined by \cite{demarchietal2011, demarchietal2013}. The method, without making assumptions on the 
properties of the population and on the pure basis of the measurement errors, provides the probability distribution for each individual star to have a given value 
of mass and age, with typical uncertainties of $\sim 5\%$ and $\sim 15\%$, respectively (see \citealt{demarchietal2017} for a thorough explanation of the procedure). 
In particular, we determined the most likely mass $M_\star$ of the 245 low-mass PMS candidates of LH\,95 by comparing the location of each object on the HR diagram with theoretical 
PMS evolutionary tracks. As for the latter, we adopted the already mentioned PARSEC tracks for $Z=0.007$ and available down to $M_\odot=0.09M_\odot$ 
(\citealt{bressanetal2012}).

Comparing masses computed from different evolutionary tracks is important for the determination of the uncertainty on the mass accretion rate and of its relationship 
with stellar mass and age. In order to assess how differences in the evolutionary models affect our results, in Appendix\,\ref{sec:appendixA} we compare mass, 
age, and mass accretion rate measurements obtained using the Pisa tracks from \cite{tognellietal2011}, available down to $M_\star=0.2M_\odot$, 
with those obtained using the PARSEC tracks. For an extensive discussion of the model-dependent age estimation in clusters, see the recent review by \cite{soderblometal2014}.

We determined the ages of individual objects by interpolating between the isochrones in the HR diagram. As already mentioned in Section\,\ref{sec:iden}, from 
Fig.\,\ref{fig:HRdiagram} it is evident that the PMS candidates appear to be distinct in two populations with a ``gap'' around 8\,Myr. We thus decided to divide 
the sample of selected PMS candidates in two sub-samples depending on their age: from now on we will indicate as {\it younger PMS candidates} those with 
age $t<8$\,Myr and {\it older PMS candidates} those with $t>8$\,Myr. With such an age difference, older PMS stars must belong to a previous generation with respect 
to the younger PMS objects, thus no spatial relationship (whether a correlation or anti-correlation) should be expected between the two types of objects, as indeed is 
evident in Fig.\,\ref{fig:LH95_image}. This will be discussed in more details in Sect.\,\ref{sec:spatial_Macc_Age}. The younger PMS candidates are about 35\% (85/245) of 
the total sample, while the older PMS candidates represent about 65\% (160/245).

\begin{figure}  
\hspace{0.5cm}
\includegraphics[width=8cm]{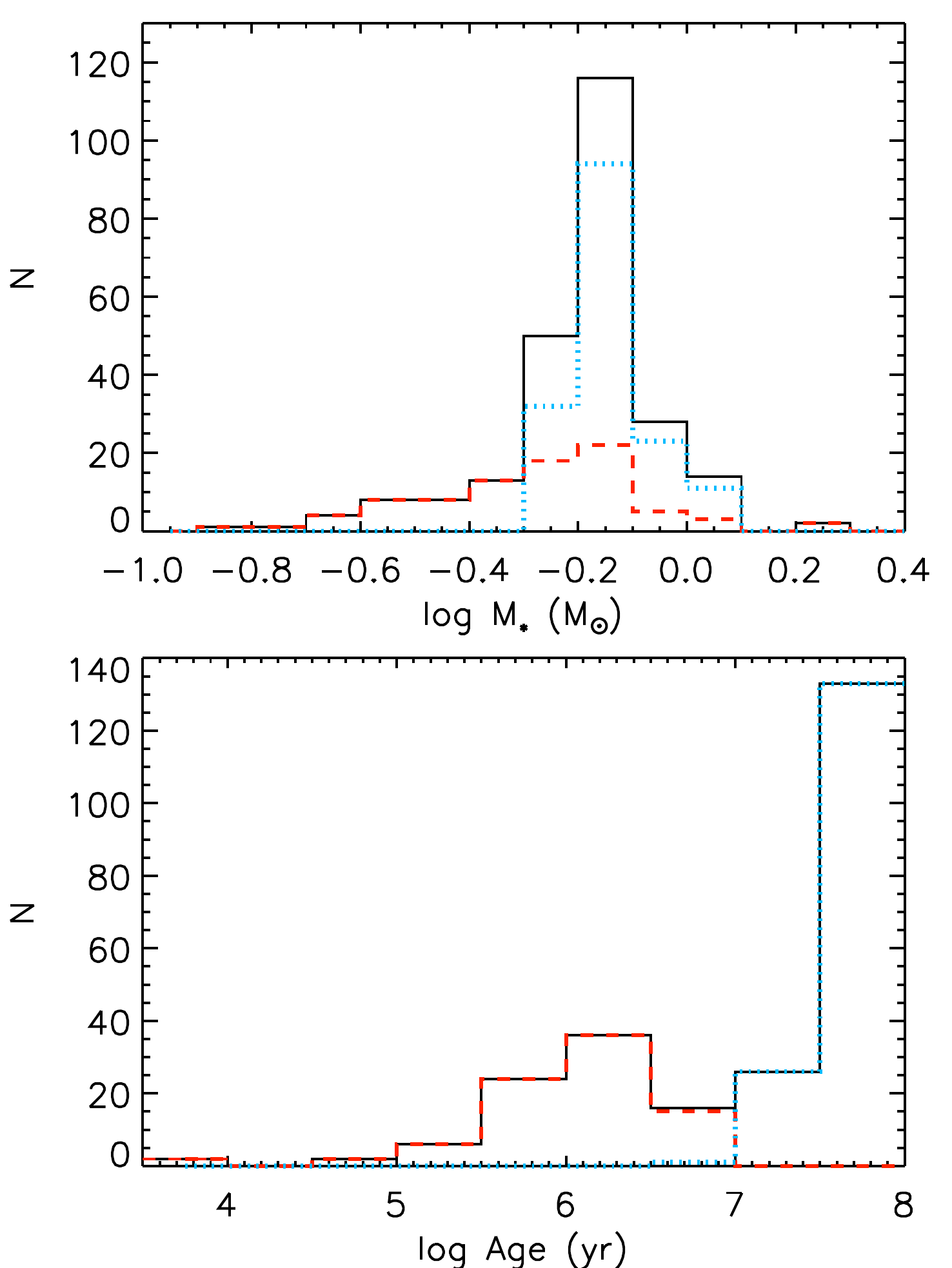} 
\caption{Histograms of the stellar mass ({\it upper panel}) and age ({\it lower panel}) for the 245 low-mass PMS candidates (solid lines). 
Dashed (red) and dotted (blue) lines correspond to the distributions of the younger and older populations, respectively.}
\label{fig:hist_mass_age} 
\end{figure}

The histograms with the mass and age distributions for the 245 low-mass PMS candidates are shown in Fig.\,\ref{fig:hist_mass_age}. Different line types correspond to younger 
(dashed lines) and older (dotted lines) PMS candidates. The mass distributions are peaked at similar values both for younger and older populations, but they show 
different ranges, with the younger population having wider range in mass than the older one. Not surprisingly, old PMS stars comprise many low-mass 
stars, since the more massive objects have already reached the MS. Inspecting the age distribution, a clear separation between younger 
and older PMS stars is evident. We highlight here that these measurements are not reliable to put constrains on the shape of the mass function or on the exact value 
of the star formation rate, since we are only considering PMS with H$\alpha$ excess emission at the 4$\sigma$ level at the time of the observations. Moreover, we are 
not taking into account photometric incompleteness, which is unavoidably more severe at lower masses. 
 
As already found by \cite{demarchietal2010, demarchietal2017} in other regions of the LMC, it is noteworthy that many of the PMS candidates in Fig.\,\ref{fig:HRdiagram} 
are close to the MS and would have been missed if no information on their H$\alpha$ excess had been available. Since we would expect to find very young objects 
above and to the right of the MS in the CMD, it is indeed customary to identify PMS objects by searching in that area of the CMD. However, this method of identification 
of PMS stars is not very reliable, because of the presence of older populations and possible age spreads in the same field, which prevent the true identification 
of PMS stars on the basis of the stellar effective temperatures and luminosities alone (see discussion in \citealt{demarchietal2010}). In fact, these older PMS stars 
were not detected by \cite{gouliermisetal2007} and were considered as field stars by the same authors.

In summary, we find evidence for a series of at least two star formation episodes, which correspond to two distinct stellar populations with different ages. We indeed 
identify a generation of younger PMS stars with ages ranging from $<1$\,Myr up to $\sim 7$\,Myr (median value $\sim 1$ Myr) and a generation of older 
PMS stars with ages of $\sim 10-60$\,Myr (median value $\sim 50$ Myr). Objects of this type are to be expected, also according to the evolutionary tracks. In fact, from 
the PARSEC tracks at metallicity $Z=0.007$, a star with mass of $\sim 0.7M_\odot$, i.e. around the peak histogram of our sample (see Fig.\,\ref{fig:hist_mass_age}), 
takes $\sim 50$\,Myr to reach the main sequence.

\section{Accretion properties}
\label{sec:accretion} 
\subsection{Accretion luminosity} 
\label{sec:Lacc}

In the magnetospheric accretion scenario, the accretion luminosity can be determined from the measurement of the reradiated energy from the circumstellar gas ionized 
and heated by the funnel flows (e.g., \citealt{hartmannetal1998}). The H$\alpha$ line, and hence its luminosity, generated in this process can be used as a diagnostics 
to derive the accretion luminosity. From the analysis of a set of $L_{\rm H\alpha}$ measurements of a group of T Tauri stars in Taurus-Auriga compiled by 
\cite{dahm2008}, \cite{demarchietal2010, demarchietal2013} found the following $L_{\rm acc}-L_{\rm H\alpha}$ relationship, that we adopt in this work:

\begin{equation} 
\label{eq:lacc_lhalpha} 
\log L_{\rm acc}/L_\odot = \log L_{\rm H\alpha}/L_\odot + (1.72 \pm 0.25)\,,
\end{equation}

\noindent{where the ratio $L_{\rm acc}$/$L_{\rm H\alpha}$ is linear. Recently, \cite{alcalaetal2017}, using X-shooter spectra of class II objects in the Galactic 
Lupus SFR, concluded that their relationships derived empirically between $L_{\rm acc}$ and the luminosity of several lines from UV to NIR are compatible with linear 
relationships.}

Taking into account the Eq.\,\ref{eq:lacc_lhalpha}, the median value of the accretion luminosity thus obtained for our sample of 245 low-mass PMS candidates is $\sim 0.17\,L_\odot$. 
The statistical uncertainty on $L_{\rm acc}$ is dominated by the quoted uncertainty of $\sim 16$\,\% on $L_{\rm H\alpha}$ mainly associated with the photometric 
error in the H$\alpha$ magnitude. There is also a systematic error due to the $L_{\rm acc}$-$L_{\rm H\alpha}$ relationship, but since we used the Eq.\,\ref{eq:lacc_lhalpha} 
for all PMS stars, this uncertainty will not prevent the comparison between different targets. Comparable uncertainties in the $L_{\rm acc}$-$L_{\rm H\alpha}$ 
relationship were obtained by \cite{alcalaetal2017} for the Lupus SFR.

\subsection{Mass accretion rate} 
\label{sec:Macc}

Once $L_{\rm acc}$ is known, the mass accretion rate $\dot{M}_{\rm acc}$ can be derived from the free-fall equation that links the luminosity released in the impact 
of the accretion flow with the rate of mass accretion according to the following relationship (see, e.g., \citealt{hartmann1998}):

\begin{equation} 
\label{eq:Macc} 
\dot{M}_{\rm acc} = \left(1 - \frac{R_{\star}}{R_{\rm in}} \right)^{-1} \frac{L_{\rm acc} R_{\star}}{G M_{\star}}\, 
\approx 1.25 \frac{L_{\rm acc} R_{\star}}{G M_{\star}}\,,
\end{equation}

\noindent{where $M_\star$ and $R_\star$ are the stellar mass and the photospheric radius, respectively, $R_{\rm in}$ is the inner radius of the accretion disk, and 
$G$ is the universal gravitational constant. $R_{\rm in}$ corresponds to the distance at which the disk is truncated, because of the stellar magnetosphere, and 
from which the disk gas is accreted and channeled by the magnetic field lines; therefore, its value is rather uncertain because it depends on how the accretion 
disk is coupled with the star. Following \cite{gullbringetal1998}, we assume $R_{\rm in}=5\,R_\star$ for all PMS stars.}

The median value of the distribution of mass accretion rates is $\sim 7.5 \times 10^{-9}$\,$M_\odot$\,yr$^{-1}$, with higher values for the younger PMS candidates 
($\sim 5.4 \times 10^{-8}$\,$M_\odot$\,yr$^{-1}$) and lower values for the older PMS candidates ($\sim 4.8 \times 10^{-9}$\,$M_\odot$\,yr$^{-1}$).

Concerning the statistical errors on $\dot{M}_{\rm acc}$, the first source of uncertainty is $L_{\rm H\alpha}$. With our selection criteria, the typical uncertainty 
on $L_{\rm H\alpha}$ is $\sim 16\%$ and is dominated by random errors. The other sources of uncertainty for $\dot{M}_{\rm acc}$ are the stellar mass and radius. The 
uncertainty on $R_\star$ is typically $\sim 5\%$, including the systematic uncertainty on the distance modulus. As for the mass, its determination is linked to the 
comparison of the location in the HR diagram with the evolutionary tracks. When we interpolate through the PMS evolutionary tracks to estimate the mass, the uncertainties on 
effective temperature and stellar luminosity imply an error of $\sim 7\%$ on $M_\star$. Combining all the sources of errors, statistical uncertainty on 
$\dot{M}_{\rm acc}$ is $\sim 18\%$\footnote{Our observations in $F555W$ and $F814W$ are not simultaneous to those in $F658N$ (see Table\,\ref{tab:log}), 
thus implying that part of the scatter in $\dot{M}_{\rm acc}$ could be also due to intrinsic stellar variability. At timescales of a few years, as in our case, variations 
may be up to $\sim 0.3$\,dex in $\log \dot{M}_{\rm acc}$ (see \citealt{costiganetal2014}). This is fully consistent with the observed 0.25\,dex dispersion around the 
average relation between mass accretion rate and age for stars of similar mass reported by \cite{demarchietal2011}. Intrinsic 
stellar variability of PMS candidates in several clusters of the Magellanic Clouds (including LH\,95) will be the subject of a forthcoming work (De Marchi et al., in prep.).}.

The systematic uncertainty on $\dot{M}_{\rm acc}$ is dominated by the knowledge of the ratio $L_{\rm acc}$/$L_{\rm H\alpha}$ reported in the 
Equation\,\ref{eq:lacc_lhalpha}. As already mentioned in Sect.\,\ref{sec:Lacc}, this ratio, even if uncertain by a factor of $\sim$2 due to the variations of the 
H$\alpha$ line intensity, is the same for all stars, therefore the comparison between different objects is not hampered by this uncertainty, as long as the statistical 
errors are small (see \citealt{demarchietal2010}).

As pointed out by \cite{demarchietal2010}, other sources of systematics errors on the derived $\dot{M}_{\rm acc}$ are due to theoretical evolutionary tracks and 
isochrones, reddening, H$\alpha$ emission generated by processes different from accretion, and contribution of nebular continuum to the photometric colors. Concerning 
the first source of errors, the main uncertainty on the derived mass and age comes from differences between models computed by different authors or from the use of 
models with metallicity that might not properly describe the stellar population under study. As shown in Appendix\,\ref{sec:appendixA}, if we for instance had used 
the \cite{tognellietal2011} PMS tracks instead of those of \cite{bressanetal2012} at the same metallicity, we would have obtained similar values of mass and age for 
PMS, to within 2\% and 6\% percent, with the largest discrepancy for $0.35 \ltsim M_\star  \ltsim 0.70\,M_\odot$ (see Appendix\,\ref{sec:appendixA}). Concerning the 
metallicity, had we used tracks with $Z$ lower by 30\%, the masses of our PMS objects would be systematically smaller by about 10\% and the ages younger by a negligible 
amount for the luminosity and temperature ranges typical of our targets. For what concerns the reddening, we followed \cite{demarchietal2010}, and concluded that 
underestimating the $E(m_{555}-m_{814})$ color excess by $\sim 0.2$\,mag would lead to a 30\% overestimate of $R_\star/M_\star$. This translates into the same 
overestimate of $\dot{M}_{\rm acc}$, which is smaller than the typical measurement uncertainties in the determination of the mass accretion rate. Finally, the 
possibility that processes different from accretion (e.g., chromospheric activity, H knots along the line of sight, ionization of H gas from nearby massive stars) 
or nebular continuum may alter the determination of $\dot{M}_{\rm acc}$ was addressed in detail in \cite{demarchietal2010}, with the conclusion that their contribution 
is negligible. Indeed, the contribution of the background emission was safely removed thanks to the fact that the $m_{658}$ magnitude of each star was determinated 
above the background calculated locally in an annulus of few arcseconds around the centroid of the star (see also Section\,\ref{sec:obs}).

\section{Discussion} 
\label{sec:discussion} 

We will now explore the distribution of the accreting PMS candidates we identified in Sect.\,\ref{sec:iden}, how $L_{\rm acc}$ depends on stellar luminosity and 
effective temperature, and how $\dot{M}_{\rm acc}$ depends on stellar mass and age. Comparison between our results with those obtained in star-forming regions 
of our Galaxy and in the Large Magellanic Clouds will be also discussed.

\subsection{Spatial distribution of accreting PMS stars}
\label{sec:spatial_Macc_Age}

From a first inspection of the spatial distribution of the most probable accreting objects in LH\,95 shown in Fig.\,\ref{fig:LH95_image}, younger low-mass PMS candidates 
(red diamonds) appear to be clustered around Be stars. In particular, they are nonuniformly distributed in the field of the LH\,95 association, but they are 
concentrated in small clusters around bright massive stars, with a clumpy spatial distribution on the scale of $\sim$5\,pc. Older PMS stars do not seem to form any 
significant concentration and are uniformly distributed within the region. \cite{gouliermisetal2007} have suggested for LH\,95 the existence of significant substructures 
of early-type stars containing candidate Herbig Ae/Be stars. Here, we support this scenario, but we also suggest that the sub-groups of early-type stars include also 
low-mass young PMS objects, similar to Galactic OB associations, like Orion (\citealt{bricenoetal2005, bricenoetal2019}).

\begin{figure}[h] 
\begin{center}
\includegraphics[width=9cm]{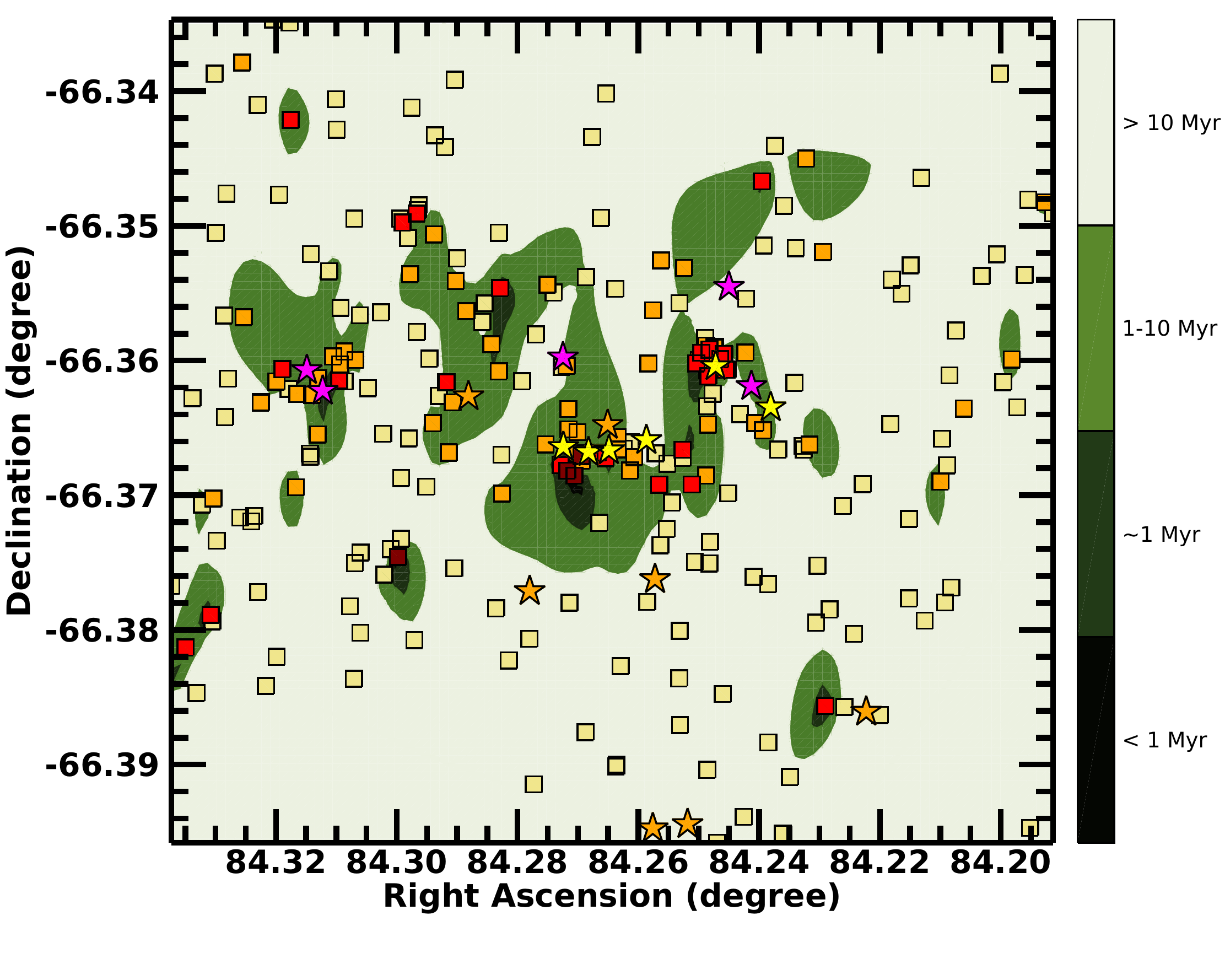}
\caption{Spatial distribution of PMS candidates in the field of LH\,95. Squares are color coded by mass accretion rate, where more highly accreting PMS stars 
($\dot{M}_{\rm acc} \gtsim 10^{-7}$\,$M_\odot$yr$^{-1}$) are marked with darker red. Filled contour regions colored in green show the position and density 
distribution of the PMS stars of different ages, with darker green regions corresponding to the density distribution of the youngest targets ($< 10$\,Myr), as 
indicated in the bar on the right. The lighter green background corresponds to the distribution of the older PMS stars with ages of some tens of 10\,Myr. As in 
Fig.\,\ref{fig:LH95_image}, star symbols mark the position of Be stars identified by \cite{gouliermisetal2002} (yellow), OB stars studied by \cite{darioetal2012} 
(purple), probable massive young targets selected from the $2MASS$ catalogue (\citealt{cutrietal2003}) with $J-H<0.8$ and $J<15$ mag (orange). North is up and 
East to the left. }
\label{fig:spatial_Macc_Age} 
\end{center} 
\end{figure}

Besides very different ages, as discussed in Section\,\ref{sec:mass_age}, the two populations of younger and older PMS stars also have considerably different spatial 
density distributions. We compare these distributions in Fig.\,\ref{fig:spatial_Macc_Age} by means of filled contours. In this figure, we considered the total 
population of 245 low-mass PMS candidates with well-defined masses and ages. The remarkable feature is the difference in the spatial distribution of younger 
and older PMS stars, with older objects much more widely distributed and not overlapping with the younger generation. 

The spatial distribution of the younger and strongly accreting PMS stars in Fig.\,\ref{fig:spatial_Macc_Age} suggests that a recent star formation episode occurred 
a few Myr ago in regions including also many of the early-type Be stars identified in the LH\,95 field. The older and less accreting PMS stars are instead uniformly 
distributed without any specific clumping within the field. They might have formed several tens of Myr ago in a more central configuration but later have had time 
to dissipate in a widespread configuration. Their mean age of $\sim 50$\,Myr and their spatial distribution within $\sim 40-50$\,pc at the distance of LH\,95 
(see also Fig.\,\ref{fig:LH95_image}) is indeed compatible with a velocity dispersion of a few km\,s$^{-1}$, which is typical for young star-forming regions. 

\subsection{Accretion luminosity versus stellar parameters} 
\label{sec:accr_lum}

\begin{figure}  
\hspace{-1cm}\includegraphics[width=10cm]{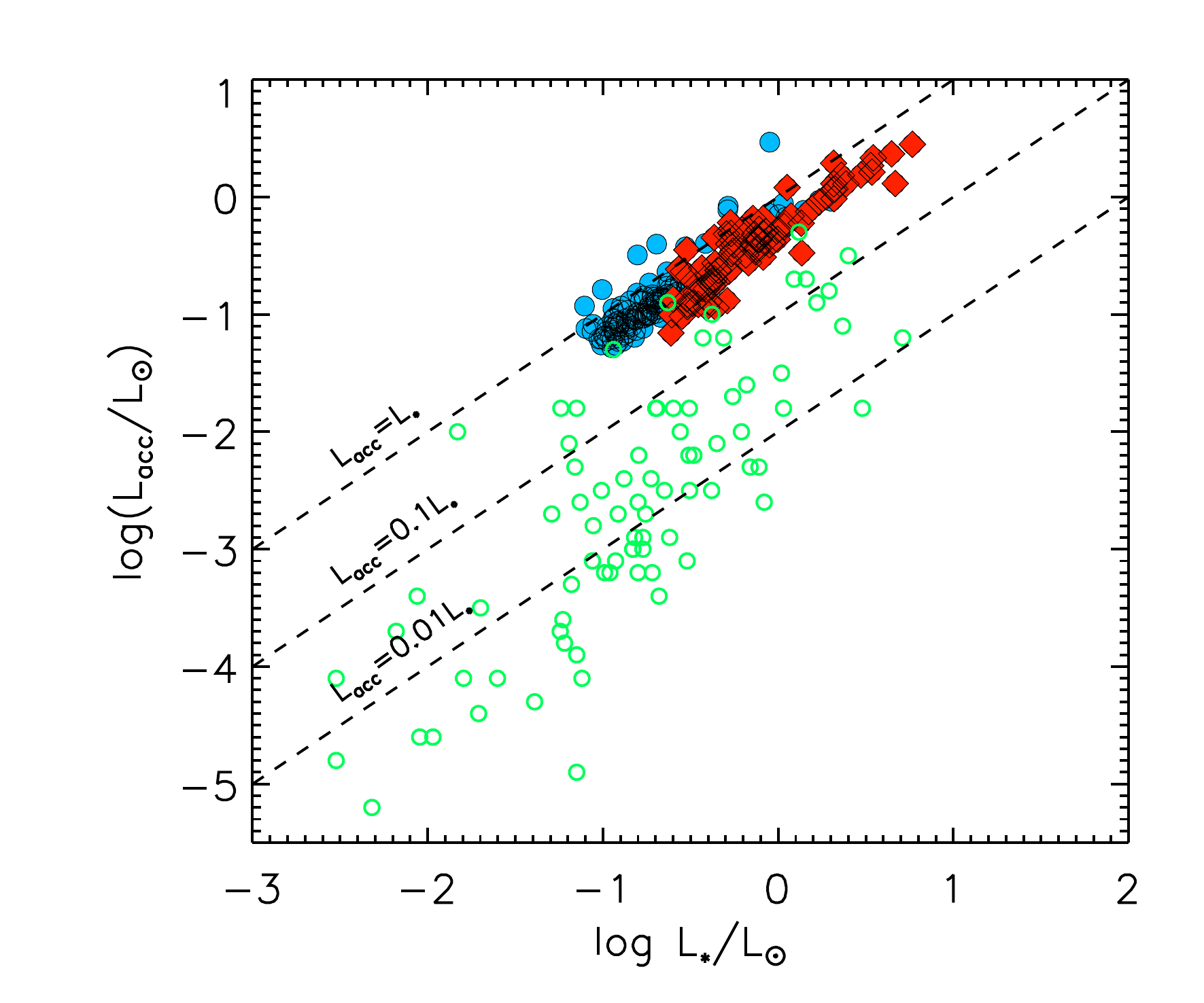}
\caption{Accretion luminosity versus stellar luminosity. Circle and diamond symbols are as in Fig.\,\ref{fig:HRdiagram}. Dashed lines represent the loci of the 
three $L_{\rm acc}-L_\star$ relations, as labeled. Open circles represent the \cite{alcalaetal2017} sample of low-mass stars in the Galactic Lupus star-forming 
region observed with the X-shooter spectrograph.} 
\label{fig:Lacc_Lum} 
\end{figure}

Figure\,\ref{fig:Lacc_Lum} shows the accretion luminosity as a function of the stellar luminosity for both younger and older PMS stars. As already observed in SFRs 
close to the Sun, $L_{\rm acc}$ increases with the stellar luminosity, with a dispersion appearing to be even smaller for our targets (the recent case of the Lupus SFR 
by \citealt{alcalaetal2017} is shown as an example). The accretion luminosity of our PMS candidates mainly falls in the range between 0.2\,$L_\star$ and $\sim L_\star$, 
with a peak of the accretion luminosity distribution around $\sim 0.5-0.6\,L_\star$, while those of regions in the solar neighbourhood, like the Lupus SFR by 
\cite{alcalaetal2017} and overimposed in the figure, is typically $\ltsim 0.1$\,$L_\star$ (see also, e.g., 
\citealt{muzerolleetal1998, whitehillenbrand2004, antoniuccietal2011, carattiogarattietal2012, biazzoetal2014}). In this context, we cannot make a real 
quantitative comparison between our results and the findings by \cite{alcalaetal2017}, but it is possible that the differences between the samples could be mainly 
due to the following reasons: $i)$ different selection criteria of accreting PMS candidates; $ii)$ different methods to derive stellar parameters; $iii)$ different 
$L_{\rm acc}$-$L_{\rm H\alpha}$ relationship, which can lead to differences up to $\sim 0.2-0.3$\,dex in $\log L_{\rm acc}/L_{\rm H\alpha}$ (for instance, the 
$L_{\rm acc}/L_{\rm H\alpha}$ ratio in the case of the \citealt{alcalaetal2017} empirical relationship is not exactly linear as in our case); $iv)$ different mass 
and metallicity ranges, our targets having $M_\star=0.1-1.8\,M_\odot$ (with a median of $\sim 0.7\,M_\odot$) and $Z=0.4\,Z_\odot$, compared to $0.02-2.0\,M_\odot$ 
(with a median of $\sim 0.2\,M_\odot$) and $Z \sim Z_\odot$ for the \cite{alcalaetal2017} sample; $v)$ other environmental conditions, such as the gas density and 
contamination.

\begin{figure}  
\hspace{-1cm}\includegraphics[width=10cm]{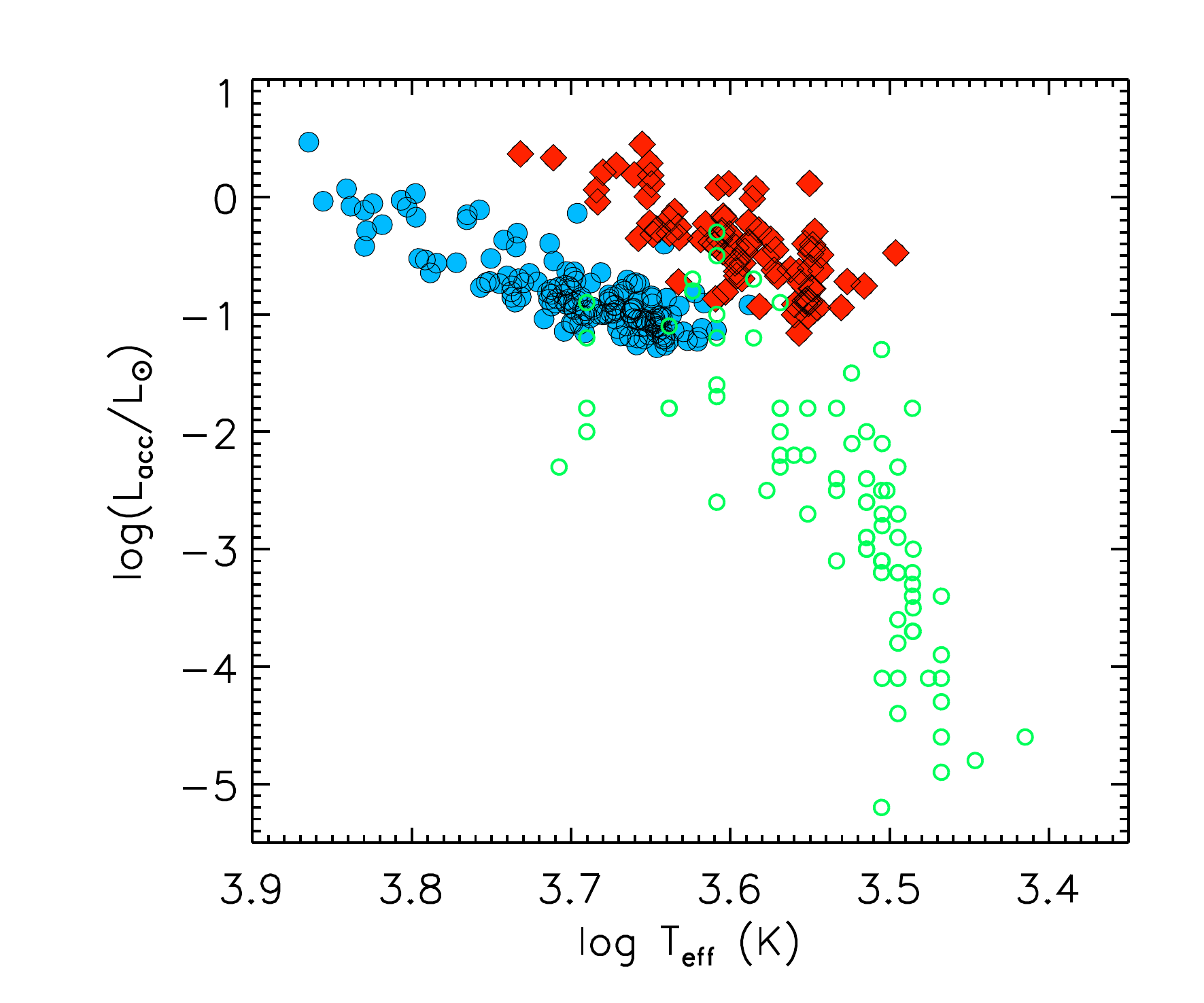} 
\caption{Accretion luminosity versus effective temperature. Filled circles and diamonds are as in Fig.\,\ref{fig:HRdiagram}. The Galactic Lupus objects by \cite{alcalaetal2017} 
are overlaid as open circles.}
\label{fig:Lacc_Teff} 
\end{figure}

In Fig.\,\ref{fig:Lacc_Teff}, the $L_{\rm acc}$ values are plotted in logarithmic scale as a function of the effective temperature 
of our PMS candidates, together with the sample of \cite{alcalaetal2017}. The $L_{\rm acc}$-$T_{\rm eff}$ plot appears to be very similar to the HR diagram shown in 
Fig.\,\ref{fig:HRdiagram}, with the younger PMS and older PMS candidates well separated in $T_{\rm eff}$. In the $T_{\rm eff}$ range between $\sim 3.6$ and $\sim 3.7$ 
some Lupus targets seem to have similar $\log L_{\rm acc}/L_\odot$ values as our PMS stars. This could be an indication of similar accretion properties 
at given $T_{\rm eff}$ range, and therefore stellar mass. Unfortunately, we do not have many objects with very low effective temperatures (in particular with 
$\log T_{\rm eff} < 3.5$) to verify the decreasing trend observed in \cite{alcalaetal2017} at the very low-mass regime. 

\subsection{Mass accretion rate versus Age} 
\label{sec:Macc_Age}

\begin{figure}[h] 
\begin{center}
\includegraphics[width=9.5cm]{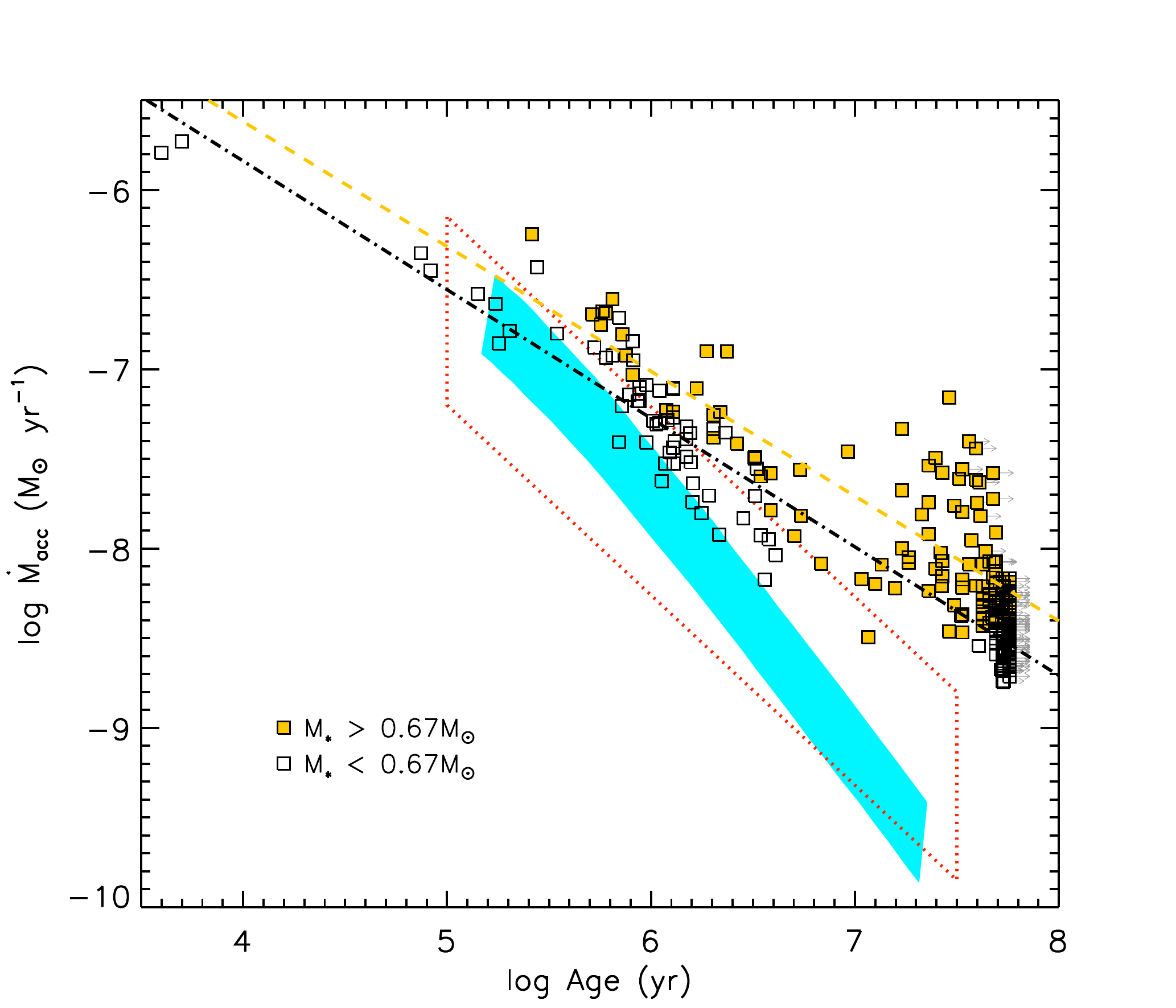} 
\caption{Mass accretion rate as a function of stellar age for our PMS candidates. The mean stellar mass is $\simeq 0.67\,M_\odot$. Filled and open squares
represent stars with $M_\star$ larger than or smaller than the mean. Regression fits for these two sub-sample are represented with dashed and dot-dashed lines, 
respectively. Arrows indicate lower limits in stellar ages. The shaded region represents a collection of viscous disk evolutionary models taken from 
\cite{hartmannetal1998} for solar-type stars with initial disk masses of $\sim 0.1-0.2\,M_\odot$, constant viscosity $\alpha=10^{-2}$, and viscosity exponent 
$\gamma=1$ (for details see \citealt{sicilia-aguilaretal2010}). The dotted region shows the best linear fit and $\pm 1\sigma$ scatter obtained considering 
$0.3-1.0 M_\odot$ stars in Galactic SFRs (see the recent review by \citealt{hartmannetal2016}).} 
\label{fig:Macc_Age}
\end{center} \end{figure}

In Fig.\,\ref{fig:Macc_Age}, the mass accretion rate is shown as a function of the stellar age. In this figure, we divided our sample in lower-mass and higher-mass 
targets, where $0.67\,M_\odot$ is the median mass of all PMS candidates. At a first glance, the slope of the $\dot{M}_{\rm acc}$-$\log t$ relationship appears 
to be similar for both lower-mass (i.e. $\simeq -0.72$) and higher-mass (i.e. $\simeq -0.70$) regimes and in good agreement with those measured in other MCs environments 
(see \citealt{demarchietal2011, demarchietal2013, demarchietal2017}). The shaded region in the same figure represents the prediction of viscous disk evolution by 
\cite{hartmannetal1998}. These models are able to reproduce the observed decreasing trend of $\dot{M}_{\rm acc}$ with age for low-mass T-Tauri stars in star-forming 
regions in the solar neighbourhood (see dotted region in the same figure and \citealt{hartmannetal2016} for a recent review on this issue\footnote{We note that 
the same authors have pointed out that this linear fit could be the consequence of correlated errors between age and accretion rate.}). The slope of this 
trend appears to be steeper than those obtained by us both for lower and higher-mass regimes ($\simeq -1.4$ against $\simeq -0.7$). This means that in the PMS candidates 
of LH\,95 $\dot M_{\rm acc}$ decreases more slowly with time than what is observed for low-mass T-Tauri stars in Galactic star-forming regions close to the Sun 
($\ltsim$ 1 kpc; \citealt{sicilia-aguilaretal2010}). This behaviour supports the recent suggestions by \cite{demarchietal2017} according to which when metallicity 
is higher, like in the local neighbourhood, there are more dust grains in the disk and therefore the radiation pressure is higher, limiting the accretion process in 
both its rate and duration, while the mass accretion process seems to last longer at low metallicity. Other authors (e.g., \citealt{yasuietal2009, yasuietal2010, yasuietal2016}) 
have concluded that in some low-metallicity environments of the outer Galaxy the disk lifetimes are shorter than in star forming regions in the solar vicinity. However, these 
works use the dust content of circumstellar disks as a proxy for the total mass of the disks and their lifetimes, while here we measure directly the infall of the probably 
more abundant gas onto the stars. Therefore, the results of the two studies are not directly comparable. And indeed, our findings are in agreement with optical spectroscopic 
observations of SFRs in the Galactic anticenter (\citealt{cusanoetal2011, kalarivink2015}), which indicate that a significant fraction of the young stellar objects have preserved 
their accretion disks, despite the low metallicity. These authors conclude that disk survival may depend not only on metallicity, but also on other environmental physical 
conditions or properties of the central objects. \cite{gallietal2015} have found the following empirical relationship between disk lifetime and stellar mass: 
$t_{\rm disk}=4\times10^{6}(M_\star/M_\odot)^{0.75}$. Such a kind of relationship was found by the authors for stars in the Taurus-Auriga association, 
i.e. with $Z \sim Z_\odot$, but similar results were previously obtained in other solar-metallicity environments (see the case of Lupus in \citealt{bertoutetal2007}). 
If applied to our PMS candidates, neglecting the effects due to different metallicity or binarity (to mention a few), this relationship would imply that disks around stars 
of $\sim 0.55$\,$M_\odot$ (median mass of our younger PMS stars) survive for $t_{\rm disk} \sim$2\,Myr, similar to the mean age of our younger population ($\sim 1$\,Myr). 
Disk dispersal time of the order of 2\,Myr is quite at odds with the ages of the older PMS candidates, whose disks have not totally been dissipated even at several tens 
Myr. We are therefore led to believe that, unless the older episodes of star formation were much more intense than the most recent one, it is very likely that circumstellar 
disks live longer in these metal-poor environments.

\subsection{Mass accretion rate versus Mass} 
\label{sec:Macc_Mass}

\begin{figure}[h] \begin{center}
\includegraphics[width=9.5cm]{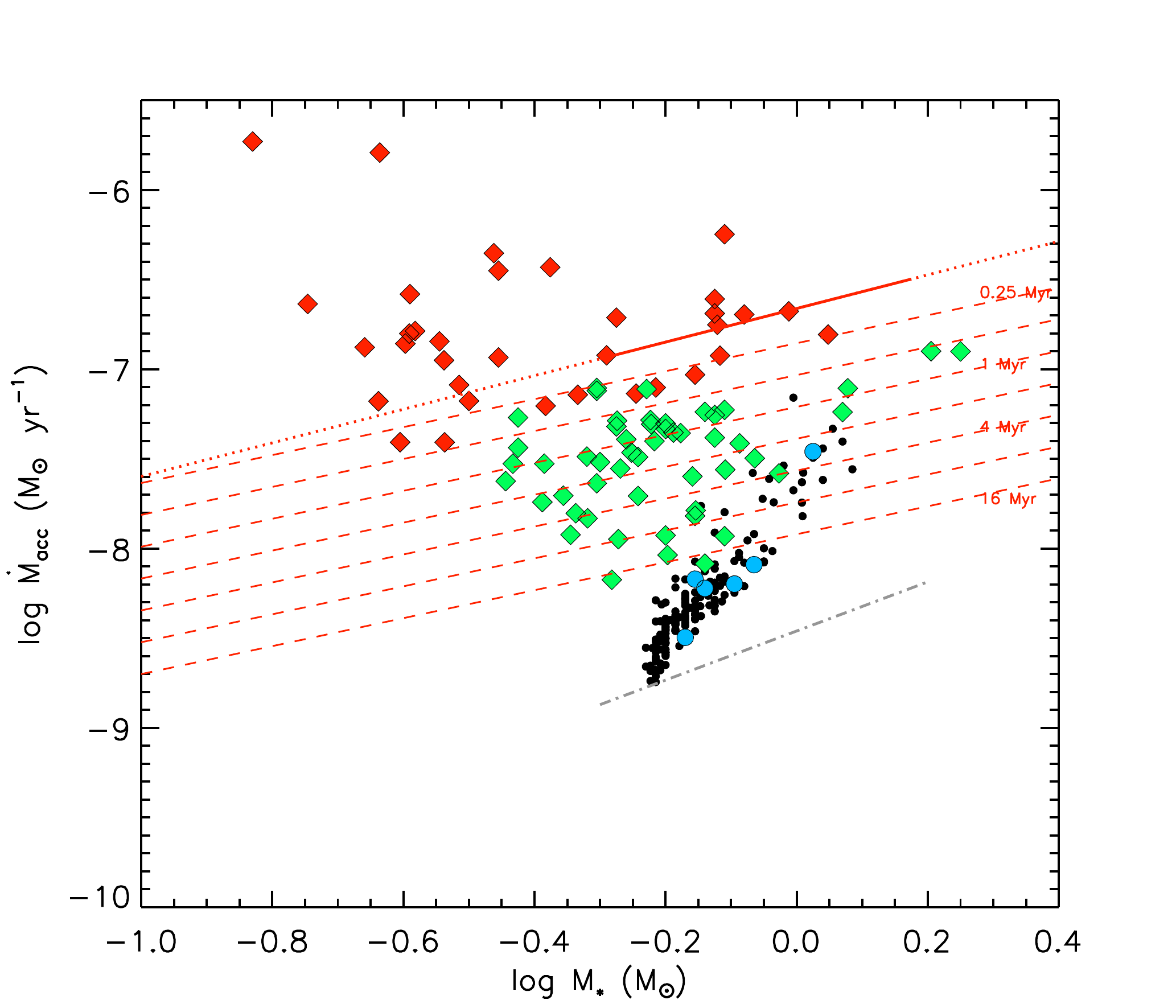}
\caption{Mass accretion rate versus stellar mass for the younger PMS candidates (filled diamonds) and older PMS candidates (filled circles). Colors refer 
to the PMS candidates with different ages (red: $\le 1$\,Myr, green: $1-8$\,Myr, blue: $8-16$\,Myr, black: $>$16\,Myr). Solid line represents the fit to the targets 
younger than 1\,Myr and with masses of $\sim 0.5-1.5$\,$M_\odot$ (see text), while dotted line is the extension of this fit to lower and higher masses. Dashed lines 
represent the $\log \dot{M}_{\rm acc}$-$\log M_\star$ relationship obtained by \cite{demarchietal2017} at given ages (0.25, 0.50, 1, 2, 4, 8, 16\,Myr) and considering in 
their Eq.\,3 the $a$ and $b$ coefficients obtained for the same mass range ($0.5-1.5$\,$M_\odot$) and ages ($<16$Myr) and $c$ coefficient obtained for 30\,Doradus, 
which we assume to be at the same metallicity as LH\,95. Dash-dotted line is the $\log \dot{M}_{\rm acc}$-$\log M_\star$ relationship obtained by \cite{alcalaetal2017} 
for stars with $M_\star \sim 0.5-1.5\,M_\odot$ in the Galactic Lupus star-forming region.} 
\label{fig:Macc_Mass} 
\end{center} 
\end{figure}

Figure\,\ref{fig:Macc_Mass} shows $\dot{M}_{\rm acc}$ versus $M_\star$ for all PMS candidates and contains several pieces of information in one graph. Younger PMS candidates 
are marked with diamonds (red for ages younger than 1\,Myr and green for $1-8$\,Myr), while older PMS candidates are represented by filled dots (the targets with ages 
of 8--16\,Myr are highlighted in blue, while those older than 16\,Myr are in black).

From a first glance, the stars in this plot appear to define a ``fan-shaped'' area. At a given stellar mass, we notice a wide spread in $\dot{M}_{\rm acc}$ for stars 
younger than 16\,Myr. In particular, this spread in $\log \dot{M}_{\rm acc}$ ranges from $\sim 1$\,dex at $\sim 0.25$\,$M_\odot$ up to $\sim 2$\,dex at 
$\sim 0.67$\,$M_\odot$, the mean mass of our targets. Splitting the sample of PMS stars in age bins, it is evident how the mass accretion rate is higher for younger 
stars, with mean values ranging from $\sim 2.6 \times 10^{-7}$\,$M_\odot$\,yr$^{-1}$ for stars younger than 1\,Myr (red diamonds), to 
$\sim 3.9 \times 10^{-8}$\,$M_\odot$\,yr$^{-1}$ at $\sim 1-8$\,Myr (green diamonds), to $\sim 1.1 \times 10^{-8}$\,$M_\odot$\,yr$^{-1}$ for the stars with ages 
of $\sim 8-16$\,Myr (blue circles). It is also clear that the slope of the $\dot{M}_{\rm acc}$-$M_\star$ relationship changes according to the stellar age, 
ranging from $\sim 0.0$ for ages younger than 1\,Myr, to $\sim 1.0$ between 1 and 8 Myr, up to $\sim 4$ for $8-16$\,Myr and older stars. Therefore, we conclude 
that attempting to define a relationship between $\dot{M}_{\rm acc}$ and $M_\star$ without taking the age of the star into account can give spurious results 
and should be avoided\footnote{We highlight here that part of the non detection of weakly accreting PMS candidates with high mass is due to our stringent 
selection (see Sect.\,\ref{sec:equiv_linelum}).}.

Another result we would like to point out concerns the slope of our targets younger than 16\,Myr. In Fig.\,\ref{fig:Macc_Mass}, we show with dashed lines the 
trends obtained, at given ages (0.25, 0.50, 1, 2, 4, 8, 16\,Myr), considering the relationship between $\log \dot{M}_{\rm acc}$ and ($\log M_\star$, $\log t$) 
by \cite{demarchietal2017} and fixing as coefficients related to age and mass those obtained for the mass range $0.5-1.5$\,$M_\odot$, i.e. $a=-0.59$ and $b=0.78$ 
(see their Eq.\,3), and as constant $c$, mainly related to the metallicity, that obtained by the same authors for 30\,Doradus in the LMC (i.e. $c=-3.67$; see 
their Table\,1), which we assume to be at the same metallicity as LH\,95. Considering our LH\,95 targets with ages younger than 1\,Myr and in the same mass 
range, the slope of the $\log \dot{M}_{\rm acc}$-$\log M_\star$ ($\sim 1$, solid line) is similar to that obtained by \cite{demarchietal2017} for 1\,Myr stars 
in 30\,Doradus. The slope of these targets is also similar to that found by \cite{alcalaetal2017} for $0.5-1.5$\,$M_\odot$ stars in the Galactic Lupus SFR 
at $\sim 1-3$\,Myr (dot-dashed line). This latter qualitative comparison seems to suggest that Galactic and extragalactic SFRs share similar slope of the 
$\log \dot{M}_{\rm acc}$-$\log M_\star$ relation. Moreover, this result also implies that the age is a parameter acting on the objects we can detect, as more massive 
targets reach the main sequence faster than lower mass objects and consequently have lower levels of H$\alpha$ emission, thus causing the fan-shape of 
Fig.\,\ref{fig:Macc_Mass}. This supports again the intercorrelation between mass accretion rate, mass, and age at given surrounding environments.

\subsection{Mass accretion rate versus stellar Age and Mass in the context of the Large Magellanic Cloud} 
\label{sec:Macc_LMC}

Assuming that all stars in our sample formed under similar conditions, we can study the simultaneous dependence of $\dot{M}_{\rm acc}$ on both $M_\star$ and $t$ 
through a multivariate least-squares fit of the type $\dot{M}_{\rm acc} \propto t^a \times M_\star^b$. Adopting this simple relationship in the mass range 
$0.5-1.5\,M_\odot$ and for stars younger than 16\,Myr, \cite{demarchietal2017} found $a \sim -0.6$ and $b \sim 1.3$ for $\sim 300$ stars in 
30\,Doradus in the LMC. If we consider the 54 objects with these characteristics in our LH\,95 sample, we find $a \sim -1.1$ 
and $b \sim 1.3$. We indeed are cautious about this result because of a relatively few number of targets. Therefore, we now compare the mass accretion properties 
of the stars in LH\,95 with those in several SFRs of our Galaxy and the Magellanic Clouds obtained with the same method. This provides us quantitative information on 
the effects of the environment during the final stages of the star formation. Following the prescriptions given by \cite{demarchietal2017}, it is convenient 
to use a power-law dependence on mass and age like 

\begin{equation}
\label{eq:macc-age-mass}
\log \dot{M}_{\rm acc} = a \times \log \frac{t}{{\rm Myr}} + b \times \log \frac{M_\star}{M_\odot} + c\,, 
\end{equation}

\noindent{where the $c$ term reflects environmental effects, such as the metallicity, on the mass accretion rate. These authors studied a homogeneous sample of 
1307 objects with $0.5-1.5\,M_\odot$ younger than 16\,Myr in six regions of the Milky Way, LMC, and SMC and analyzed them with the same method as our 
targets, finding $a=-0.59\pm0.02$ and $b=0.78\pm0.08$, respectively. Using the same values of $a$ and $b$ also for the PMS candidates in LH\,95 with the 
same restriction in mass and age, we derive $c=-3.54$. This latter value is consistent with the results for the two clusters in the LMC analyzed by the authors 
(namely, 30\,Doradus and the SN1987A field) and more generally with their relationship $c=(-3.69 \pm 0.02)-(0.30 \pm 0.04) \log Z/Z_\odot$ obtained for the six clusters 
in the LMC, SMC, and the MW (see their Table\,1). In particular, our value is between $-3.41$ obtained in NGC\,346 (a cluster in the SMC with $Z \sim 0.002$) 
and $-3.65$ found for the MW clusters Trumpler\,14 and NGC\,3603 with $Z \sim Z_\odot$. Even if we do not draw more 
quantitative conclusions here, our analysis confirms the importance of considering cluster metallicity, besides stellar mass and age, when mass accretion is 
studied in different environments. Clearly, other physical conditions (like mean gas density or local magnetic field) of the environment might have an effect on the 
extend and duration of the star formation process in general, and on the evolution of the mass accretion rate, as suggested by \cite{demarchietal2017}.}

\section{Summary and conclusions} 
\label{sec:summ_Concl} 
We have applied to the young association LH\,95 in the Large Magellanic Cloud a photometric detection method to reliably identify PMS candidates actively undergoing 
mass accretion in a resolved stellar population without requiring spectroscopic observations. The method combines HST wide-band $F555W$ and $F814W$ photometry 
with narrow-band $F658N$ imaging to $i)$ identify stars with H$\alpha$ excess using as a reference template of the photospheric level the mean $m_{555}-m_{658}$ 
color of normal stars with very small photometric uncertainties; $ii)$ convert the excess H$\alpha$ magnitude into luminosity and equivalent width; $iii)$ derive 
accretion luminosity and mass accretion rate with similar accuracy as allowed by spectral line analysis. The main results of our study are summarized in the 
following items.

\begin{enumerate} 

\item From the original photometric catalogue of 24515 sources, we extracted 1294 targets, taken as reference for our selection of PMS candidates, as they have errors 
in all three bands of less than 0.05\,mag. Then, we identified 245 low-mass PMS candidates as those having $m_{555}-m_{658}$ color exceding that of the reference 
stars by at least four times the photometric uncertainty at the same $m_{555}-m_{814}$ color. 

\item From the measured H$\alpha$ luminosity of these PMS candidates, we derived the accretion luminosity and, through other stellar physical parameters obtained 
thanks to the \cite{besselletal1998} stellar atmospheric models, \cite{pecautmamajek2013} calibrations, and evolutionary tracks by \cite{bressanetal2012}, their mass 
accretion rates. The PMS candidates have a median value of the mass accretion rate of $\sim 7.5 \times 10^{-9}$\,$M_\odot$\,yr$^{-1}$. 

\item Within the sample of PMS candidates we have identified two populations, which we call younger PMS candidates ($t \ltsim 8$\,Myr; median age of $\sim$1\,Myr) 
and older PMS candidates ($9 \ltsim t \ltsim 60$\,Myr; median age of $\sim$50\,Myr) with higher median values of the mass accretion rate for the former group compared to 
the latter ($\sim 5.4 \times 10^{-8}$\,$M_\odot$\,yr$^{-1}$ against $\sim 4.8 \times 10^{-9}$\,$M_\odot$\,yr$^{-1}$).

\item We have studied how the mass accretion rate changes with time as our PMS candidates approach the main sequence. We find that $\dot M_{\rm acc}$ decreases 
more slowly with time than what is predicted by models of viscous disk evolution (\citealt{hartmannetal1998}) and observed for low-mass T-Tauri stars in Galactic 
star-forming regions within 1\,kpc (e.g., \citealt{sicilia-aguilaretal2010, hartmannetal2016}). This is in line with previous findings in the Magellanic Clouds. 
Analyzing the $\dot{M}_{\rm acc}$-$M_\star$ relationship, a clear dependence on age is evident, with a slope increasing with age. 

\item We have studied the relationships between mass accretion rate, stellar mass and age, and we confirm previous findings obtained in the Magellanic Clouds, namely 
that attemps to derive correlations by fitting separately the observed dependence of $\dot M_{\rm acc}$ on $M_\star$ or $t$ may fail and introduce biases. Since 
these three stellar properties are intercorrelated, a proper multivariate fit is needed. Adopting for the PMS candidates in LH\,95 a simple regression fit of the type 
$\dot M_{\rm acc} \propto Age^a \times M_\star^b$ we find $a \sim -1.1$ and $b \sim 1.3$ for the mass range $0.5-1.5\,M_\odot$ and ages younger than 16\,Myr. 
Since the small number of targets, we are cautious about this result, and therefore we have also compared the mass accretion properties of the PMS in LH\,95 with 
those homogeneosly derived in several regions of the MW, LMC, and SMC (see next item). 

\item We have applied to LH\,95 the multivariate regression fit 
$\log \dot M_{\rm acc} = a \times \log \frac{t}{{\rm Myr}} + b \times \log \frac{M_\star}{M_\odot} + c$ 
of \cite{demarchietal2017} obtained for a uniform sample of 1307 PMS stars, with masses of $0.5-1.5\,M_\odot$ and younger than 16\,Myr contained in six different 
SFRs in the Milky Way, LMC, and SMC. The $c$ value we find for LH\,95 results to be close to that of the two regions in the LMC at the same metallicity (namely, 
30\,Doradus and the SN\,1987A field). Moreover, it is lower than that achieved at lower-$Z$ environments and higher than that found in solar-metallicity regions, thus 
confirming that metallicity is an important parameter to be taken into account when studying accretion properties and evolution.

\item We find that the younger PMS stars are clustered in sub-groups around early-type stars (mainly B type stars), while the older PMS stars are more uniformly distributed 
over the whole field of LH\,95. We note that the presence of this sub-clustering suggests it may have its origin in short-lived parental molecular clouds within 
a giant molecular cloud complex, as in the case of Galactic OB associations (see, e.g., the Orion association; \citealt{bricenoetal2007}). 

\item From a morphological study of age, spatial distribution, and accretion diagnostics, we find multiple generations of stars due to at least two star formation 
bursts, with the most recent one occurring some Myr ago and the previous one some tens Myr ago. The high values of $\dot M_{\rm acc}$ of the younger PMS stars 
and their vicinity to the early-type stars suggest that their circumstellar disks have not still considerably dispersed.

\end{enumerate}

Since no infrared observations are available for this region, we can not drawn any conclusion about the relationship between accretion properties and inner disk tracers. 
The advent of the {\it James Webb Space Telescope} will allow us to link mass accretion rate and grain properties. This will be also important to have information 
about the disk geometry and to explain which mechanisms allow circumstellar disks to feed their central PMS stars for tens of Myr in a low-metallicity, low-density 
environment such as the field of LH\,95 and give rise to a certain level of measurable mass accretion rate. Moreover, future spectroscopic observations of the region 
to derive accurate metallicity from the measurement, e.g., of the [O/H] ratio are very much needed if we want to understand the different contributions of metallicity 
and other effects, such as the environmental gas density, on the accretion process. 

\acknowledgments

We thank our referee, Prof. Gregory Herczeg, whose extensive and insightful comments have helped us to improve the presentation of this work. 
KB is grateful to ESA for the support, via its Science Visitor programme, during the data analysis useful for the preparation of this paper. 
KB also thanks the {\it Osservatorio Astronomico di Roma} for the hospitality during the preparation of the paper. This work was based on observations made 
with the NASA/ESA Hubble Space Telescope, and obtained from the Hubble Legacy Archive, which is a collaboration between the Space Telescope Science Institute (STScI/NASA), 
the Spacte Telescope European Coordinating Facility (ST-ECF/ESA) and the Canadian Astronomy Data Centre (CADC/NRC/CSA). Some of the data presented in this
paper were obtained from the Mikulski Archive for Space Telescopes (MAST). STScI is operated by the Association of Universities for Research in 
Astronomy, Inc., under NASA contract NAS5-26555. This research made use of the SIMBAD database, operated at the CDS (Strasbourg, France) and data 
products from the Two Micron All Sky Survey, which is a joint project of the University of Massachusetts and the Infrared Processing and Analysis 
Center/California Institute of Technology, funded by the National Aeronautics and Space Administration and the National Science Foundation. This 
research has made also use of the SVO Filter Profile Service supported from the Spanish MINECO through grant AyA2014-55216.

\appendix

\section{Mass, age, and mass accretion rate as determined from different 
evolutionary tracks and isochrones}
\label{sec:appendixA}

Masses and ages computed from different evolutionary tracks allow us to estimate the model-dependent uncertainty on the relationship between $\dot{M}_{\rm acc}$ and 
the stellar mass $M_\star$ and age $t$.

\begin{figure}  
\begin{center}
\includegraphics[width=6.8cm,height=7.5cm]{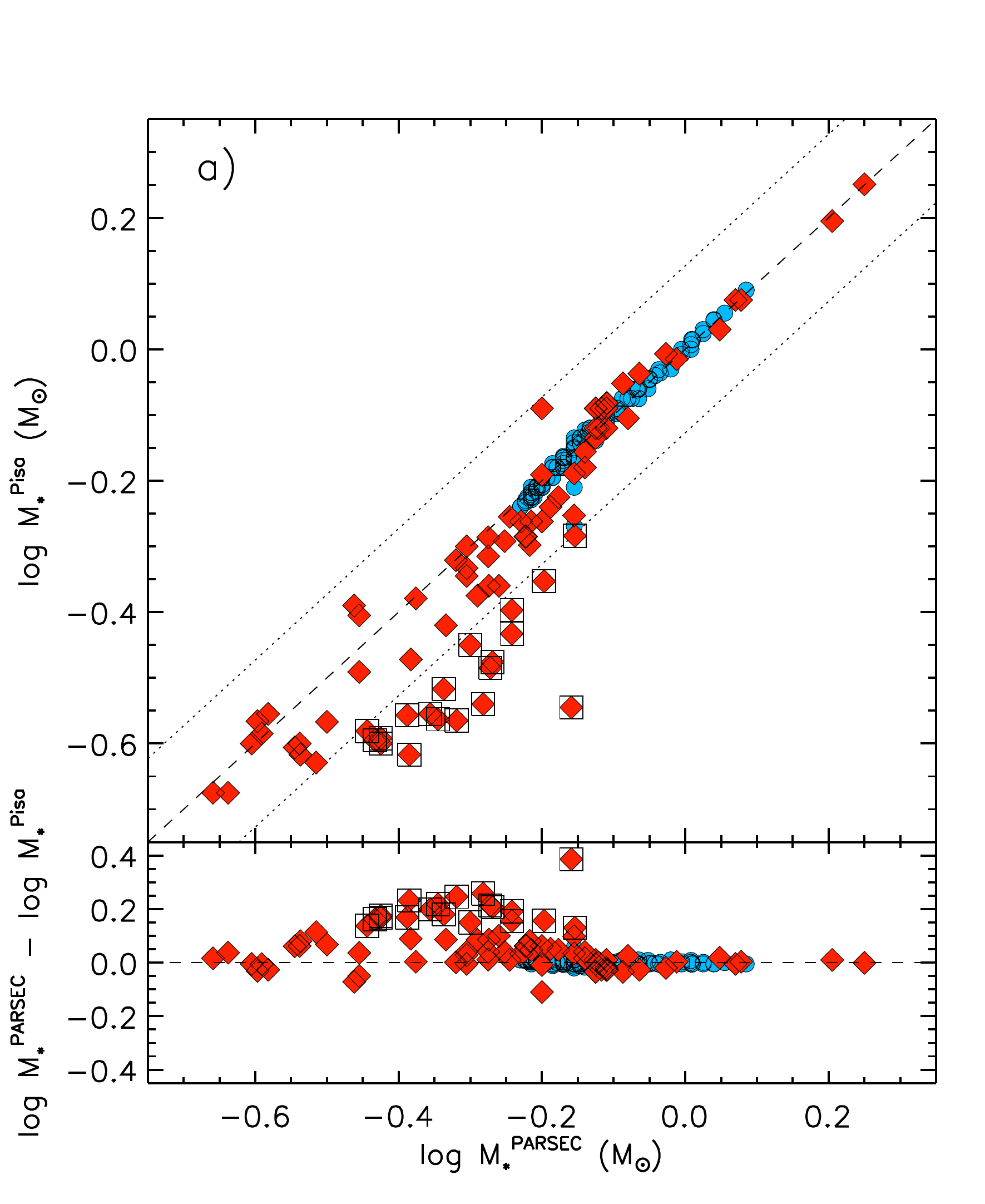}
\includegraphics[width=6.8cm,height=7.5cm]{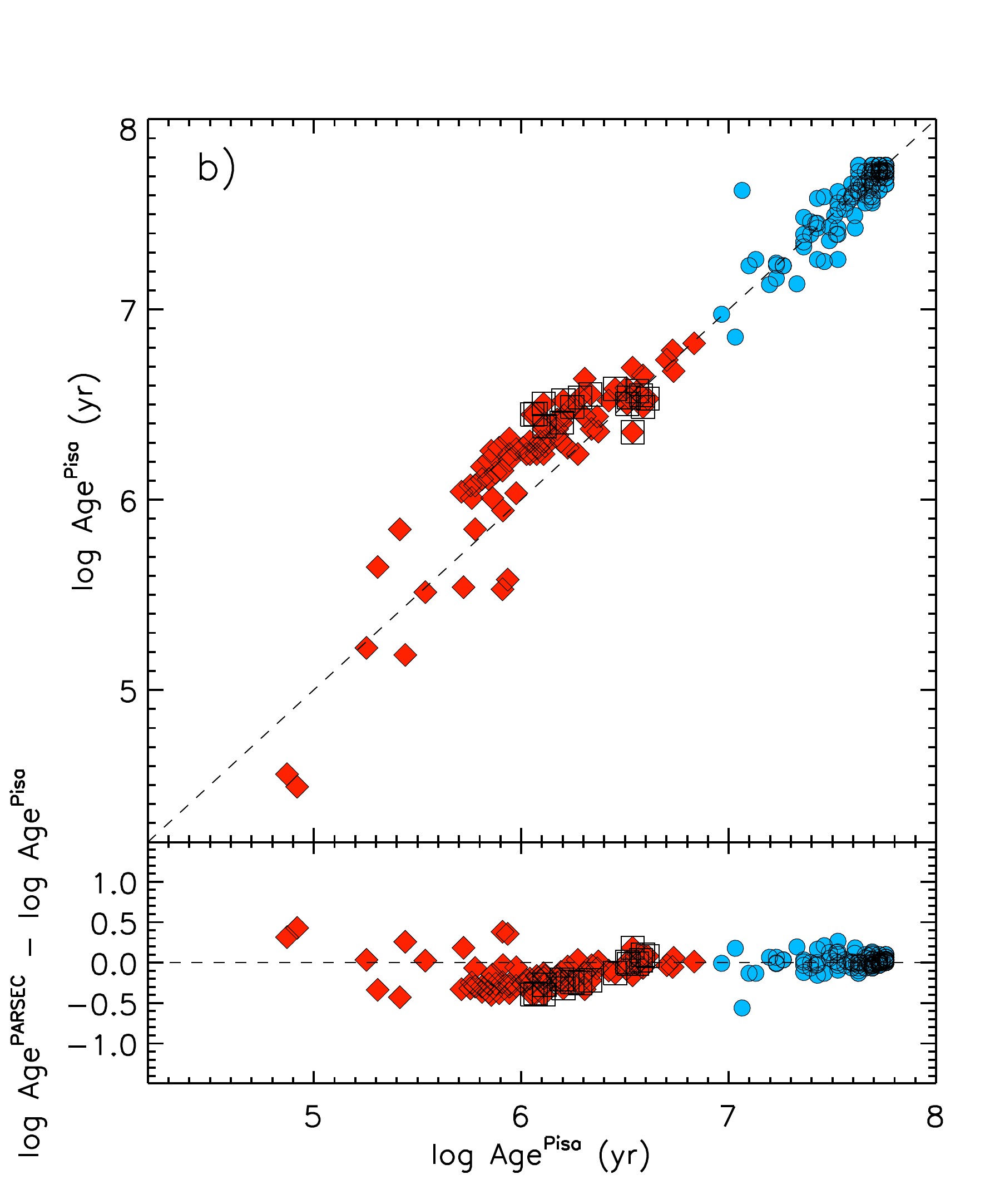}
\includegraphics[width=6.8cm,height=7.5cm]{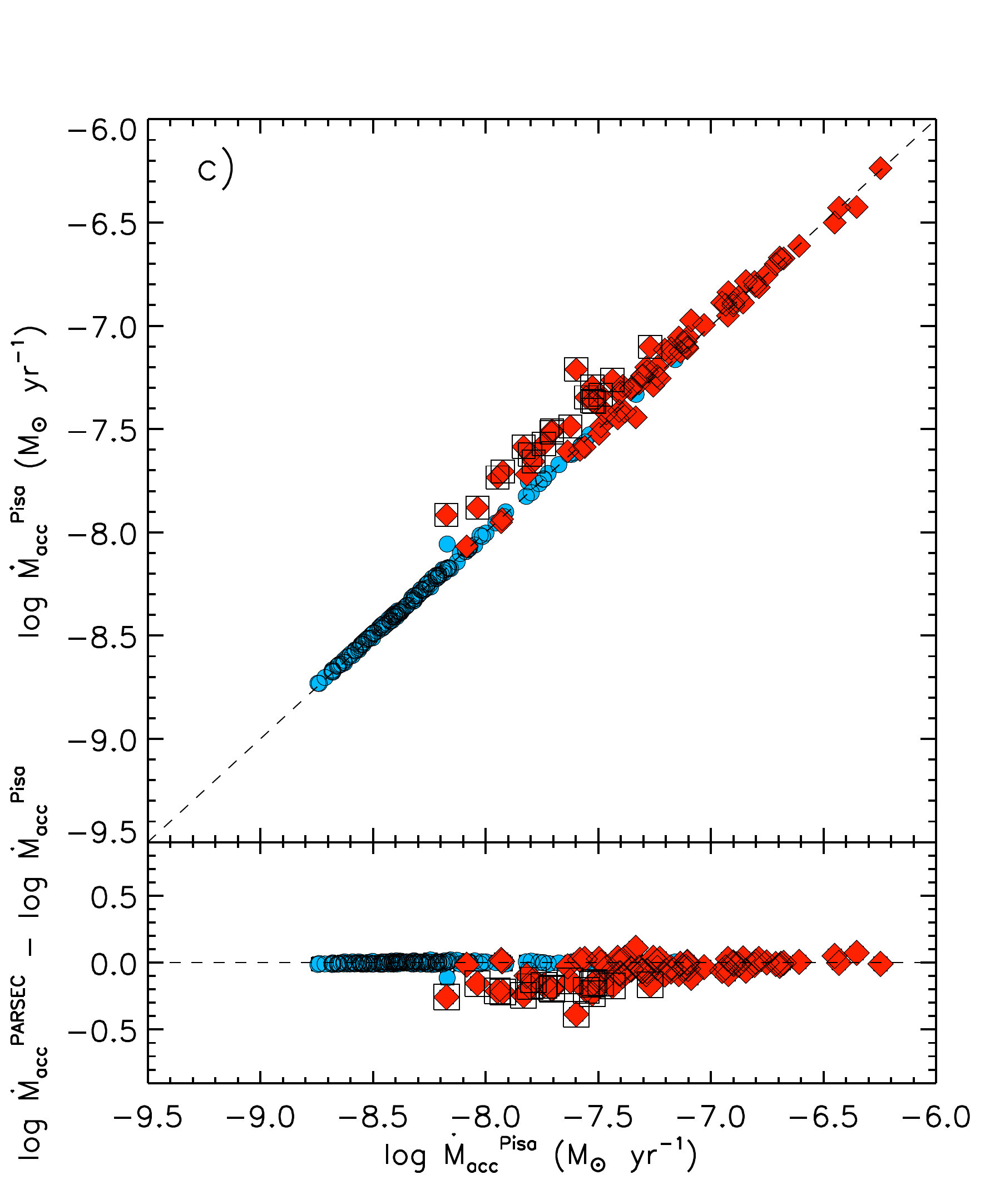}
\caption{Comparison between masses (panel $a$), ages (panel $b$), and mass accretion rates (panel $c$) derived from the PARSEC and Pisa PMS models. Diamonds 
and circles refer to younger and older PMS stars. The dashed lines represent the 1:1 relation. The dotted lines in the panel $a)$ are shifted by twice the rms 
difference between the $\log M_\star$ values. Open squares mark the position of the targets outside the dotted lines in panel $a)$.}
\label{fig:tracks_comp} 
\end{center}
\end{figure}

In Fig.\,\ref{fig:tracks_comp}, we show the comparison between the masses, ages, and mass accretion rates as derived from two sets of PMS tracks for the same metallicity, 
namely the PARSEC stellar evolution model (\citealt{bressanetal2012}) and the Pisa stellar models (\citealt{tognellietal2011}). Filled diamonds and circles refer to 
younger and older stars. The largest residuals between the two sets of tracks are seen for the younger low-mass stars, while for the older stars with higher masses 
the agreement is good (see panels $a$ and $b$). In particular, the two sets of models differ significantly for $\log T_{\rm eff}$ in the range $\sim 3.5-3.6$ 
and $\log L_\star$ in the range between $-0.8\,L_\odot$ and $-0.1\,L_\odot$, which translates into the spread in $M_\star$ and $\dot M_{\rm acc}$ of young low-mass targets 
observed in panel $a$ and $c$ of Fig.\,\ref{fig:tracks_comp}. Squares in all three panels represent the 19 PMS younger stars departing from the 1:1 relation by 
twice the rms difference. The $T_{\rm eff}$ and $L_\star$ values of these targets place them in an area of the HR diagram where the two sets of tracks are more 
discrepant most probably because of different treatment of the mixing length and opacity (P. Marigo, priv. comm.). In any case, the difference in mass of these 19 
targets, representing 8\% of the total sample with well determined masses and ages with both sets of tracks, affects the determination of the mass accretion rate 
slightly, with mean differences of about 0.2\,dex in $\log M_\star$ between the PARSEC and Pisa models producing at most differences of about 0.2\,dex in 
$\log \dot M_{\rm acc}$. For the rest of the sample ($\sim 92$\%) the agreement in $\dot M_{\rm acc}$ is very good (panel $c$). Similar findings were also reported 
by \cite{biazzoetal2014} in the case of L1616/L1615, a Galactic cometary cloud in the solar vicinity. 

\section{Stellar parameters}
\label{sec:appendixB}

\startlongtable
\setlength{\tabcolsep}{3pt}
\begin{deluxetable*}{ccccccccccrrr}
\tabletypesize{\scriptsize}
\tablecaption{Stellar parameters of our sample of PMS accreting candidates. Columns list: object ID within our catalogue, right ascension, declination, 
dereddened magnitude in the $F555W$, $F814W$, and $F658N$ bands, H$\alpha$ equivalent width, effective temperature, luminosity, mass, age, accretion luminosity, 
and mass accretion rate. Typical uncertainties in mass, age, $\log L_{\rm acc}$, and $\log \dot{M}_{\rm acc}$ are discussed in the text. \label{tab:param}} 
\tablehead{
\colhead{Number}&\colhead{RA}&\colhead{DEC}&\colhead{$(m_{555})_0$}&\colhead{$(m_{814})_0$}&
\colhead{$(m_{658})_0$}&\colhead{$EW_{\rm H\alpha}$}&\colhead{$T_{\rm eff}$}&\colhead{$L_\star$}&
\colhead{$M_\star$}&\colhead{$\log t$}&\colhead{$\log L_{\rm acc}$}&\colhead{$\log \dot{M}_{\rm acc}$}\\ 
\colhead{(ID)} & \colhead{(deg)}& \colhead{(deg)}&\colhead{(mag)}& \colhead{(mag)}& \colhead{(mag)}&
\colhead{(\AA)}&\colhead{(K)}&\colhead{($L_\odot$)}&\colhead{($M_\odot$)}&\colhead{(yr)}&
\colhead{($L_\odot$)}&\colhead{($M_\odot$/yr)}
}
\startdata
\multicolumn{13}{c}{{\it Younger PMS candidates}}\\
\hline  
  100583 & 84.31741 & $-$66.34244 & 21.873$\pm$0.010 & 21.014$\pm$0.010 & 20.977$\pm$0.089 & 14.9$\pm$1.5 & 5393$\pm$29 & 4.44$\pm$0.17 & 1.8 & 6.4 &    0.4 & $-$6.9\\
  100659 & 84.27160 & $-$66.36848 & 21.991$\pm$0.013 & 20.738$\pm$0.011 & 20.773$\pm$0.061 & 18.7$\pm$1.0 & 4523$\pm$20 & 5.83$\pm$0.20 & 0.8 & 5.4 &    0.4 & $-$6.2\\
  101386 & 84.28833 & $-$66.35663 & 23.130$\pm$0.016 & 22.046$\pm$0.011 & 21.999$\pm$0.064 & 19.9$\pm$1.0 & 4823$\pm$28 & 1.71$\pm$0.06 & 1.2 & 6.3 & $-$0.0 & $-$7.2\\
  101473 & 84.29159 & $-$66.36192 & 23.237$\pm$0.018 & 21.963$\pm$0.012 & 21.887$\pm$0.079 & 24.6$\pm$1.1 & 4490$\pm$20 & 1.88$\pm$0.06 & 0.8 & 5.9 &    0.0 & $-$6.9\\
  101511 & 84.27044 & $-$66.36885 & 23.275$\pm$0.020 & 21.103$\pm$0.011 & 21.607$\pm$0.089 & 19.2$\pm$1.4 & 3551$\pm$10 & 4.65$\pm$0.13 & 0.2 & 3.6 &    0.1 & $-$5.8\\
  101783 & 84.29366 & $-$66.35094 & 23.607$\pm$0.018 & 22.321$\pm$0.013 & 22.453$\pm$0.085 & 14.1$\pm$1.5 & 4474$\pm$20 & 1.35$\pm$0.04 & 0.8 & 6.1 & $-$0.2 & $-$7.2\\
  101786 & 84.31878 & $-$66.36094 & 23.610$\pm$0.022 & 22.205$\pm$0.011 & 22.206$\pm$0.103 & 23.1$\pm$1.5 & 4315$\pm$23 & 1.49$\pm$0.04 & 0.7 & 5.9 & $-$0.1 & $-$7.0\\
  101906 & 84.26928 & $-$66.36739 & 23.750$\pm$0.027 & 21.972$\pm$0.016 & 21.723$\pm$0.088 & 39.5$\pm$0.9 & 3836$\pm$28 & 2.11$\pm$0.05 & 0.4 & 4.9 &    0.1 & $-$6.4\\
  102386 & 84.30670 & $-$66.36026 & 24.203$\pm$0.027 & 22.965$\pm$0.016 & 22.775$\pm$0.089 & 29.0$\pm$1.2 & 4548$\pm$37 & 0.75$\pm$0.02 & 0.9 & 6.6 & $-$0.4 & $-$7.6\\
  102422 & 84.27140 & $-$66.36541 & 24.236$\pm$0.030 & 22.855$\pm$0.016 & 22.668$\pm$0.178 & 31.0$\pm$2.2 & 4347$\pm$31 & 0.82$\pm$0.02 & 0.8 & 6.3 & $-$0.3 & $-$7.4\\
  102483 & 84.30853 & $-$66.35963 & 24.285$\pm$0.036 & 22.982$\pm$0.017 & 22.693$\pm$0.108 & 33.9$\pm$1.3 & 4451$\pm$37 & 0.73$\pm$0.01 & 0.9 & 6.5 & $-$0.3 & $-$7.5\\
  102676 & 84.31286 & $-$66.36160 & 24.464$\pm$0.034 & 22.936$\pm$0.020 & 22.798$\pm$0.118 & 31.2$\pm$1.5 & 4156$\pm$35 & 0.77$\pm$0.01 & 0.7 & 6.2 & $-$0.4 & $-$7.4\\
  102769 & 84.29126 & $-$66.36712 & 24.552$\pm$0.035 & 22.937$\pm$0.017 & 22.799$\pm$0.135 & 32.7$\pm$1.7 & 4046$\pm$35 & 0.78$\pm$0.01 & 0.6 & 6.1 & $-$0.4 & $-$7.3\\
  102891 & 84.31294 & $-$66.36579 & 24.640$\pm$0.032 & 22.892$\pm$0.015 & 22.856$\pm$0.175 & 31.2$\pm$2.2 & 3875$\pm$32 & 0.84$\pm$0.01 & 0.5 & 5.9 & $-$0.4 & $-$7.1\\
  102986 & 84.27272 & $-$66.36806 & 24.712$\pm$0.041 & 22.872$\pm$0.017 & 22.795$\pm$0.068 & 34.8$\pm$0.9 & 3756$\pm$34 & 0.99$\pm$0.01 & 0.4 & 5.8 & $-$0.4 & $-$6.9\\
  103104 & 84.27522 & $-$66.36653 & 24.800$\pm$0.040 & 23.114$\pm$0.019 & 23.295$\pm$0.096 & 20.0$\pm$1.6 & 3955$\pm$40 & 0.68$\pm$0.01 & 0.6 & 6.1 & $-$0.6 & $-$7.5\\
  103237 & 84.28419 & $-$66.35909 & 24.900$\pm$0.043 & 23.079$\pm$0.019 & 23.175$\pm$0.087 & 27.3$\pm$1.3 & 3781$\pm$40 & 0.82$\pm$0.01 & 0.4 & 5.9 & $-$0.5 & $-$7.2\\
  103338 & 84.27172 & $-$66.36070 & 24.969$\pm$0.051 & 23.345$\pm$0.032 & 22.669$\pm$0.260 & 49.3$\pm$1.9 & 4034$\pm$53 & 0.54$\pm$0.01 & 0.6 & 6.3 & $-$0.3 & $-$7.3\\
  103348 & 84.31655 & $-$66.36971 & 24.977$\pm$0.036 & 23.550$\pm$0.022 & 22.757$\pm$0.157 & 50.2$\pm$1.1 & 4286$\pm$40 & 0.43$\pm$0.01 & 0.8 & 6.7 & $-$0.3 & $-$7.6\\
  103489 & 84.29056 & $-$66.36341 & 25.079$\pm$0.045 & 23.417$\pm$0.022 & 22.682$\pm$0.164 & 51.1$\pm$1.2 & 3986$\pm$45 & 0.51$\pm$0.01 & 0.7 & 6.4 & $-$0.3 & $-$7.4\\
  103560 & 84.27018 & $-$66.36720 & 25.124$\pm$0.061 & 23.334$\pm$0.021 & 22.596$\pm$0.151 & 52.5$\pm$1.0 & 3821$\pm$58 & 0.62$\pm$0.01 & 0.5 & 6.0 & $-$0.3 & $-$7.1\\
  103566 & 84.32519 & $-$66.35709 & 25.128$\pm$0.046 & 23.383$\pm$0.020 & 22.440$\pm$0.204 & 56.0$\pm$1.1 & 3879$\pm$45 & 0.53$\pm$0.01 & 0.5 & 6.1 & $-$0.2 & $-$7.1\\
  103652 & 84.33067 & $-$66.37919 & 25.182$\pm$0.042 & 22.940$\pm$0.026 & 22.629$\pm$0.066 & 50.0$\pm$0.6 & 3524$\pm$13 & 0.85$\pm$0.01 & 0.3 & 5.2 & $-$0.3 & $-$6.6\\
  103776 & 84.30940 & $-$66.36179 & 25.283$\pm$0.052 & 23.271$\pm$0.034 & 22.891$\pm$0.092 & 47.5$\pm$0.8 & 3574$\pm$46 & 0.72$\pm$0.01 & 0.3 & 5.3 & $-$0.4 & $-$6.9\\
  103819 & 84.27144 & $-$66.36389 & 25.320$\pm$0.053 & 23.600$\pm$0.021 & 23.308$\pm$0.105 & 39.9$\pm$1.2 & 3911$\pm$51 & 0.44$\pm$0.01 & 0.5 & 6.2 & $-$0.6 & $-$7.5\\
  103825 & 84.29388 & $-$66.36495 & 25.327$\pm$0.052 & 23.611$\pm$0.028 & 23.633$\pm$0.193 & 28.1$\pm$2.7 & 3916$\pm$54 & 0.43$\pm$0.01 & 0.5 & 6.2 & $-$0.7 & $-$7.6\\
  103914 & 84.26993 & $-$66.36562 & 25.389$\pm$0.058 & 23.695$\pm$0.024 & 23.574$\pm$0.120 & 33.5$\pm$1.6 & 3945$\pm$57 & 0.40$\pm$0.01 & 0.7 & 6.6 & $-$0.7 & $-$7.8\\
  104186 & 84.26853 & $-$66.35411 & 25.570$\pm$0.066 & 23.973$\pm$0.031 & 24.063$\pm$0.114 & 22.5$\pm$2.0 & 4068$\pm$65 & 0.30$\pm$0.01 & 0.7 & 6.8 & $-$0.9 & $-$8.1\\
  104237 & 84.28295 & $-$66.36111 & 25.605$\pm$0.058 & 23.479$\pm$0.034 & 23.479$\pm$0.171 & 38.8$\pm$1.8 & 3605$\pm$56 & 0.52$\pm$0.01 & 0.3 & 5.9 & $-$0.6 & $-$7.2\\
  104260 & 84.29657 & $-$66.34939 & 25.623$\pm$0.062 & 23.441$\pm$0.019 & 23.101$\pm$0.142 & 49.5$\pm$1.1 & 3546$\pm$17 & 0.54$\pm$0.01 & 0.3 & 5.9 & $-$0.5 & $-$7.0\\
  104457 & 84.30190 & $-$66.37620 & 25.741$\pm$0.072 & 24.047$\pm$0.027 & 23.478$\pm$0.048 & 47.5$\pm$0.7 & 3945$\pm$70 & 0.29$\pm$0.01 & 0.7 & 6.7 & $-$0.6 & $-$7.8\\
  104609 & 84.29965 & $-$66.37489 & 25.848$\pm$0.070 & 22.860$\pm$0.016 & 23.084$\pm$0.119 & 52.7$\pm$0.9 & 3135$\pm$21 & 1.36$\pm$0.02 & 0.2 & 3.7 & $-$0.5 & $-$5.7\\
  104680 & 84.29765 & $-$66.35389 & 25.884$\pm$0.063 & 23.896$\pm$0.038 & 23.710$\pm$0.047 & 41.6$\pm$0.7 & 3599$\pm$55 & 0.41$\pm$0.01 & 0.4 & 6.1 & $-$0.7 & $-$7.5\\
  104742 & 84.33483 & $-$66.38161 & 25.918$\pm$0.068 & 23.604$\pm$0.021 & 23.460$\pm$0.097 & 47.3$\pm$0.9 & 3493$\pm$39 & 0.46$\pm$0.01 & 0.3 & 6.0 & $-$0.6 & $-$7.1\\
  104743 & 84.27491 & $-$66.35468 & 25.918$\pm$0.071 & 23.641$\pm$0.021 & 24.254$\pm$0.098 & 17.9$\pm$2.0 & 3510$\pm$29 & 0.44$\pm$0.01 & 0.3 & 6.0 & $-$0.9 & $-$7.4\\
  104889 & 84.30924 & $-$66.36060 & 25.999$\pm$0.076 & 23.950$\pm$0.028 & 23.848$\pm$0.291 & 40.3$\pm$2.9 & 3535$\pm$60 & 0.39$\pm$0.01 & 0.4 & 6.1 & $-$0.8 & $-$7.5\\
  104905 & 84.28243 & $-$66.37018 & 26.006$\pm$0.080 & 24.131$\pm$0.031 & 23.581$\pm$0.156 & 49.5$\pm$1.3 & 3718$\pm$63 & 0.31$\pm$0.01 & 0.6 & 6.5 & $-$0.7 & $-$7.7\\
  104961 & 84.31383 & $-$66.36284 & 26.044$\pm$0.079 & 23.861$\pm$0.025 & 23.408$\pm$0.147 & 52.1$\pm$1.1 & 3546$\pm$22 & 0.37$\pm$0.01 & 0.4 & 6.1 & $-$0.6 & $-$7.3\\
  104962 & 84.31038 & $-$66.36000 & 26.044$\pm$0.073 & 23.852$\pm$0.025 & 23.843$\pm$0.057 & 40.7$\pm$0.9 & 3542$\pm$20 & 0.37$\pm$0.01 & 0.4 & 6.1 & $-$0.8 & $-$7.4\\
  105165 & 84.29009 & $-$66.35439 & 26.159$\pm$0.092 & 24.131$\pm$0.032 & 24.158$\pm$0.098 & 35.4$\pm$1.5 & 3557$\pm$72 & 0.32$\pm$0.01 & 0.4 & 6.2 & $-$0.9 & $-$7.7\\
  105410 & 84.29890 & $-$66.35005 & 26.283$\pm$0.090 & 23.788$\pm$0.022 & 23.695$\pm$0.136 & 49.8$\pm$1.2 & 3363$\pm$40 & 0.42$\pm$0.02 & 0.2 & 5.7 & $-$0.7 & $-$6.9\\
  105418 & 84.31978 & $-$66.36187 & 26.285$\pm$0.095 & 24.197$\pm$0.029 & 23.432$\pm$0.047 & 56.6$\pm$0.5 & 3650$\pm$82 & 0.27$\pm$0.01 & 0.5 & 6.5 & $-$0.6 & $-$7.6\\
  105464 & 84.31105 & $-$66.35368 & 26.305$\pm$0.083 & 24.279$\pm$0.032 & 24.432$\pm$0.224 & 30.5$\pm$3.0 & 3559$\pm$66 & 0.28$\pm$0.01 & 0.5 & 6.3 & $-$1.0 & $-$7.9\\
  105643 & 84.33208 & $-$66.37099 & 26.400$\pm$0.099 & 24.294$\pm$0.030 & 24.401$\pm$0.199 & 34.5$\pm$2.5 & 3629$\pm$82 & 0.25$\pm$0.01 & 0.5 & 6.6 & $-$1.0 & $-$7.9\\
  105709 & 84.28530 & $-$66.35606 & 26.432$\pm$0.098 & 24.449$\pm$0.036 & 24.792$\pm$0.022 & 20.5$\pm$1.5 & 3605$\pm$77 & 0.24$\pm$0.01 & 0.5 & 6.6 & $-$1.2 & $-$8.2\\
  105710 & 84.30597 & $-$66.35695 & 26.433$\pm$0.090 & 24.409$\pm$0.034 & 24.071$\pm$0.265 & 46.6$\pm$2.2 & 3561$\pm$71 & 0.25$\pm$0.01 & 0.5 & 6.5 & $-$0.9 & $-$7.8\\
  105729 & 84.28271 & $-$66.35492 & 26.445$\pm$0.095 & 23.816$\pm$0.025 & 23.788$\pm$0.145 & 51.1$\pm$1.2 & 3281$\pm$37 & 0.42$\pm$0.02 & 0.2 & 5.2 & $-$0.8 & $-$6.6\\
  200884 & 84.25101 & $-$66.36951 & 22.206$\pm$0.014 & 21.259$\pm$0.010 & 21.059$\pm$0.178 & 24.8$\pm$2.5 & 5142$\pm$31 & 3.48$\pm$0.12 & 1.6 & 6.3 &    0.3 & $-$6.9\\
  201022 & 84.24808 & $-$66.35949 & 22.400$\pm$0.014 & 21.299$\pm$0.011 & 21.365$\pm$0.090 & 14.1$\pm$1.6 & 4787$\pm$27 & 3.43$\pm$0.12 & 1.1 & 5.9 &    0.2 & $-$6.8\\
  201076 & 84.25639 & $-$66.36952 & 22.478$\pm$0.012 & 21.327$\pm$0.011 & 21.225$\pm$0.107 & 23.7$\pm$1.6 & 4695$\pm$19 & 3.37$\pm$0.12 & 1.0 & 5.8 &    0.3 & $-$6.7\\
  201235 & 84.24942 & $-$66.35973 & 22.694$\pm$0.014 & 21.472$\pm$0.013 & 21.424$\pm$0.138 & 22.3$\pm$2.1 & 4575$\pm$22 & 2.96$\pm$0.10 & 0.8 & 5.7 &    0.2 & $-$6.7\\
  201461 & 84.26330 & $-$66.36691 & 22.928$\pm$0.015 & 21.848$\pm$0.013 & 21.741$\pm$0.093 & 22.7$\pm$1.4 & 4832$\pm$30 & 2.05$\pm$0.07 & 1.2 & 6.2 &    0.1 & $-$7.1\\
  201512 & 84.26661 & $-$66.36714 & 22.976$\pm$0.025 & 21.681$\pm$0.017 & 21.620$\pm$0.162 & 24.2$\pm$2.4 & 4462$\pm$28 & 2.44$\pm$0.06 & 0.8 & 5.8 &    0.1 & $-$6.8\\
  201570 & 84.24540 & $-$66.36099 & 23.038$\pm$0.017 & 21.747$\pm$0.012 & 21.436$\pm$0.223 & 34.6$\pm$2.6 & 4467$\pm$19 & 2.29$\pm$0.07 & 0.8 & 5.8 &    0.2 & $-$6.7\\
  201673 & 84.23938 & $-$66.34700 & 23.146$\pm$0.017 & 21.859$\pm$0.014 & 21.176$\pm$0.076 & 46.3$\pm$0.6 & 4473$\pm$20 & 2.07$\pm$0.06 & 0.8 & 5.8 &    0.3 & $-$6.6\\
  202098 & 84.24630 & $-$66.36020 & 23.544$\pm$0.025 & 21.885$\pm$0.014 & 21.610$\pm$0.155 & 38.4$\pm$1.6 & 3990$\pm$25 & 2.07$\pm$0.05 & 0.4 & 5.4 &    0.1 & $-$6.4\\
  202261 & 84.22890 & $-$66.38596 & 23.699$\pm$0.018 & 21.937$\pm$0.012 & 21.931$\pm$0.061 & 30.3$\pm$0.8 & 3857$\pm$19 & 2.08$\pm$0.06 & 0.4 & 4.9 & $-$0.0 & $-$6.5\\
  202575 & 84.24827 & $-$66.35983 & 23.974$\pm$0.025 & 22.551$\pm$0.020 & 22.526$\pm$0.114 & 24.7$\pm$1.7 & 4291$\pm$29 & 1.08$\pm$0.03 & 0.7 & 6.1 & $-$0.3 & $-$7.2\\
  202742 & 84.26123 & $-$66.36847 & 24.107$\pm$0.031 & 22.556$\pm$0.015 & 22.465$\pm$0.057 & 29.7$\pm$0.8 & 4127$\pm$30 & 1.10$\pm$0.02 & 0.6 & 5.9 & $-$0.2 & $-$7.1\\
  202761 & 84.25253 & $-$66.36693 & 24.122$\pm$0.026 & 22.489$\pm$0.016 & 22.301$\pm$0.073 & 34.9$\pm$0.9 & 4023$\pm$27 & 1.18$\pm$0.03 & 0.5 & 5.8 & $-$0.2 & $-$6.9\\
  202795 & 84.26528 & $-$66.36755 & 24.152$\pm$0.040 & 22.543$\pm$0.018 & 21.693$\pm$0.055 & 53.0$\pm$0.4 & 4053$\pm$39 & 1.12$\pm$0.01 & 0.5 & 5.8 &    0.1 & $-$6.7\\
  202903 & 84.19248 & $-$66.34855 & 24.232$\pm$0.026 & 22.852$\pm$0.015 & 22.354$\pm$0.089 & 42.0$\pm$0.9 & 4348$\pm$28 & 0.82$\pm$0.02 & 0.8 & 6.3 & $-$0.2 & $-$7.3\\
  202916 & 84.24573 & $-$66.36068 & 24.240$\pm$0.033 & 22.917$\pm$0.019 & 22.586$\pm$0.297 & 35.7$\pm$3.3 & 4425$\pm$35 & 0.78$\pm$0.01 & 0.8 & 6.4 & $-$0.3 & $-$7.4\\
  203010 & 84.24858 & $-$66.35919 & 24.319$\pm$0.030 & 22.692$\pm$0.016 & 22.616$\pm$0.176 & 30.5$\pm$2.2 & 4030$\pm$30 & 0.98$\pm$0.02 & 0.6 & 5.9 & $-$0.3 & $-$7.1\\
  203019 & 84.24731 & $-$66.35934 & 24.323$\pm$0.034 & 22.745$\pm$0.017 & 22.856$\pm$0.131 & 21.1$\pm$2.1 & 4092$\pm$34 & 0.93$\pm$0.01 & 0.6 & 6.0 & $-$0.4 & $-$7.3\\
  203097 & 84.24862 & $-$66.36884 & 24.388$\pm$0.028 & 22.793$\pm$0.015 & 22.911$\pm$0.099 & 21.0$\pm$1.6 & 4071$\pm$28 & 0.89$\pm$0.02 & 0.6 & 6.0 & $-$0.4 & $-$7.3\\
  203406 & 84.24845 & $-$66.36102 & 24.623$\pm$0.035 & 22.992$\pm$0.020 & 23.062$\pm$0.130 & 24.1$\pm$1.9 & 4025$\pm$36 & 0.74$\pm$0.01 & 0.6 & 6.1 & $-$0.5 & $-$7.4\\
  203473 & 84.23920 & $-$66.36548 & 24.672$\pm$0.036 & 23.033$\pm$0.018 & 22.351$\pm$0.218 & 49.6$\pm$1.6 & 4015$\pm$36 & 0.72$\pm$0.01 & 0.6 & 6.1 & $-$0.2 & $-$7.1\\
  203480 & 84.24783 & $-$66.35997 & 24.678$\pm$0.041 & 22.988$\pm$0.027 & 22.973$\pm$0.162 & 29.1$\pm$2.2 & 3950$\pm$44 & 0.76$\pm$0.01 & 0.5 & 6.0 & $-$0.4 & $-$7.3\\
  203840 & 84.24845 & $-$66.36369 & 24.935$\pm$0.042 & 23.512$\pm$0.023 & 23.704$\pm$0.075 & 13.5$\pm$1.5 & 4291$\pm$44 & 0.45$\pm$0.01 & 0.8 & 6.7 & $-$0.7 & $-$7.9\\
  204035 & 84.24983 & $-$66.36026 & 25.077$\pm$0.052 & 23.347$\pm$0.028 & 23.051$\pm$0.260 & 40.2$\pm$2.6 & 3898$\pm$53 & 0.55$\pm$0.01 & 0.6 & 6.2 & $-$0.5 & $-$7.4\\
  204063 & 84.26327 & $-$66.36600 & 25.100$\pm$0.050 & 23.405$\pm$0.024 & 23.259$\pm$0.198 & 34.4$\pm$2.3 & 3943$\pm$50 & 0.52$\pm$0.01 & 0.7 & 6.5 & $-$0.5 & $-$7.6\\
  204149 & 84.23152 & $-$66.36654 & 25.158$\pm$0.040 & 23.408$\pm$0.020 & 22.894$\pm$0.058 & 46.8$\pm$0.6 & 3872$\pm$40 & 0.52$\pm$0.01 & 0.5 & 6.2 & $-$0.4 & $-$7.3\\
  204358 & 84.24573 & $-$66.35980 & 25.287$\pm$0.058 & 23.082$\pm$0.021 & 23.009$\pm$0.244 & 43.0$\pm$2.3 & 3537$\pm$16 & 0.75$\pm$0.01 & 0.3 & 5.3 & $-$0.4 & $-$6.8\\
  204645 & 84.25251 & $-$66.36754 & 25.486$\pm$0.048 & 23.843$\pm$0.026 & 23.918$\pm$0.074 & 24.2$\pm$1.3 & 4010$\pm$49 & 0.34$\pm$0.01 & 0.6 & 6.5 & $-$0.8 & $-$7.9\\
  204708 & 84.25025 & $-$66.36051 & 25.530$\pm$0.075 & 23.217$\pm$0.024 & 23.126$\pm$0.199 & 45.9$\pm$1.7 & 3494$\pm$43 & 0.65$\pm$0.01 & 0.3 & 5.5 & $-$0.5 & $-$6.8\\
  204775 & 84.20983 & $-$66.36929 & 25.572$\pm$0.054 & 23.715$\pm$0.023 & 23.454$\pm$0.212 & 41.3$\pm$2.1 & 3737$\pm$43 & 0.45$\pm$0.01 & 0.5 & 6.2 & $-$0.6 & $-$7.5\\
  204836 & 84.24826 & $-$66.36148 & 25.609$\pm$0.072 & 23.395$\pm$0.024 & 22.914$\pm$0.134 & 53.1$\pm$0.9 & 3534$\pm$20 & 0.56$\pm$0.01 & 0.3 & 5.9 & $-$0.4 & $-$6.8\\
  204896 & 84.19803 & $-$66.36023 & 25.654$\pm$0.053 & 23.640$\pm$0.021 & 24.103$\pm$0.077 & 15.3$\pm$1.6 & 3572$\pm$42 & 0.51$\pm$0.01 & 0.3 & 5.8 & $-$0.9 & $-$7.4\\
  205280 & 84.26629 & $-$66.37236 & 25.917$\pm$0.073 & 24.124$\pm$0.034 & 24.228$\pm$0.162 & 26.3$\pm$2.4 & 3817$\pm$72 & 0.30$\pm$0.01 & 0.6 & 6.6 & $-$0.9 & $-$8.0\\
  205303 & 84.24829 & $-$66.36506 & 25.934$\pm$0.074 & 23.900$\pm$0.030 & 24.124$\pm$0.011 & 27.7$\pm$1.0 & 3551$\pm$59 & 0.40$\pm$0.01 & 0.4 & 6.1 & $-$0.9 & $-$7.6\\
  205453 & 84.25227 & $-$66.35344 & 26.021$\pm$0.082 & 24.156$\pm$0.030 & 23.020$\pm$0.141 & 60.0$\pm$0.7 & 3729$\pm$65 & 0.30$\pm$0.01 & 0.6 & 6.5 & $-$0.5 & $-$7.5\\
  205455 & 84.25505 & $-$66.36796 & 26.022$\pm$0.074 & 24.061$\pm$0.027 & 24.184$\pm$0.166 & 29.9$\pm$2.3 & 3628$\pm$58 & 0.35$\pm$0.01 & 0.5 & 6.2 & $-$0.9 & $-$7.8\\
  205790 & 84.24213 & $-$66.35971 & 26.220$\pm$0.082 & 24.069$\pm$0.030 & 24.180$\pm$0.238 & 35.6$\pm$2.8 & 3576$\pm$41 & 0.30$\pm$0.01 & 0.4 & 6.3 & $-$0.9 & $-$7.7\\
  205990 & 84.23206 & $-$66.34531 & 26.322$\pm$0.088 & 23.874$\pm$0.027 & 24.248$\pm$0.278 & 34.8$\pm$3.3 & 3392$\pm$40 & 0.39$\pm$0.01 & 0.2 & 5.9 & $-$0.9 & $-$7.2\\
\hline
\multicolumn{13}{c}{{\it Older PMS candidates}}\\
\hline
  100952 & 84.32237 & $-$66.36344 & 22.520$\pm$0.012 & 21.902$\pm$0.013 & 21.820$\pm$0.081 & 12.7$\pm$1.5 & 6277$\pm$56 & 2.15$\pm$0.08 & 1.1 &     7.2 &    0.0 & $-$7.3\\
  100959 & 84.26922 & $-$66.36774 & 22.528$\pm$0.021 & 22.064$\pm$0.022 & 21.720$\pm$0.151 & 24.0$\pm$2.2 & 6934$\pm$92 & 2.05$\pm$0.06 & 1.2 &  $>$7.6 &    0.1 & $-$7.4\\
  100973 & 84.29169 & $-$66.36190 & 22.545$\pm$0.012 & 22.135$\pm$0.014 & 21.985$\pm$0.064 & 13.2$\pm$1.1 & 7172$\pm$57 & 2.01$\pm$0.07 & 1.2 &  $>$7.5 & $-$0.0 & $-$7.6\\
  101478 & 84.32236 & $-$66.36340 & 23.241$\pm$0.017 & 22.622$\pm$0.016 & 22.325$\pm$0.161 & 23.9$\pm$2.3 & 6272$\pm$71 & 1.11$\pm$0.03 & 1.0 &  $>$7.6 & $-$0.2 & $-$7.6\\
  101503 & 84.32544 & $-$66.33817 & 23.267$\pm$0.025 & 22.716$\pm$0.021 & 22.483$\pm$0.115 & 19.8$\pm$1.8 & 6584$\pm$84 & 1.06$\pm$0.03 & 1.0 &  $>$7.6 & $-$0.2 & $-$7.7\\
  101505 & 84.29621 & $-$66.34878 & 23.273$\pm$0.017 & 22.763$\pm$0.015 & 22.611$\pm$0.053 & 14.8$\pm$0.9 & 6736$\pm$60 & 1.05$\pm$0.03 & 1.0 &  $>$7.6 & $-$0.3 & $-$7.8\\
  101794 & 84.29124 & $-$66.36714 & 23.622$\pm$0.021 & 22.771$\pm$0.016 & 22.663$\pm$0.051 & 18.6$\pm$0.9 & 5417$\pm$57 & 0.88$\pm$0.02 & 1.0 &     7.2 & $-$0.3 & $-$7.7\\
  101828 & 84.31632 & $-$66.36279 & 23.660$\pm$0.019 & 22.929$\pm$0.015 & 22.377$\pm$0.141 & 36.0$\pm$1.6 & 5828$\pm$65 & 0.79$\pm$0.02 & 0.9 &     7.5 & $-$0.2 & $-$7.6\\
  101864 & 84.33022 & $-$66.37055 & 23.698$\pm$0.018 & 22.881$\pm$0.017 & 22.811$\pm$0.086 & 15.9$\pm$1.5 & 5525$\pm$55 & 0.80$\pm$0.02 & 0.9 &     7.4 & $-$0.4 & $-$7.7\\
  101947 & 84.27251 & $-$66.36078 & 23.792$\pm$0.028 & 23.287$\pm$0.025 & 22.945$\pm$0.161 & 24.4$\pm$2.3 & 6755$\pm$100& 0.65$\pm$0.01 & 0.9 &  $>$7.6 & $-$0.4 & $-$8.0\\
  102266 & 84.32281 & $-$66.37748 & 24.089$\pm$0.023 & 23.425$\pm$0.021 & 23.303$\pm$0.094 & 15.8$\pm$1.6 & 6086$\pm$88 & 0.52$\pm$0.01 & 0.9 &     7.6 & $-$0.6 & $-$8.1\\
  102449 & 84.29354 & $-$66.34358 & 24.260$\pm$0.025 & 23.312$\pm$0.018 & 23.258$\pm$0.075 & 17.6$\pm$1.3 & 5139$\pm$57 & 0.53$\pm$0.01 & 0.9 &     7.4 & $-$0.5 & $-$7.9\\
  102510 & 84.27681 & $-$66.35836 & 24.317$\pm$0.030 & 23.667$\pm$0.024 & 23.510$\pm$0.069 & 17.4$\pm$1.2 & 6143$\pm$110& 0.42$\pm$0.01 & 0.8 &     7.6 & $-$0.6 & $-$8.2\\
  102776 & 84.26749 & $-$66.34370 & 24.559$\pm$0.032 & 23.575$\pm$0.021 & 23.481$\pm$0.062 & 20.4$\pm$1.1 & 5045$\pm$65 & 0.41$\pm$0.01 & 0.8 &     7.4 & $-$0.6 & $-$8.0\\
  102784 & 84.31930 & $-$66.34799 & 24.566$\pm$0.030 & 23.790$\pm$0.022 & 23.718$\pm$0.097 & 15.2$\pm$1.7 & 5666$\pm$90 & 0.35$\pm$0.01 & 0.8 &  $>$7.7 & $-$0.7 & $-$8.2\\
  102912 & 84.32805 & $-$66.34791 & 24.657$\pm$0.030 & 23.895$\pm$0.023 & 23.819$\pm$0.065 & 15.1$\pm$1.2 & 5714$\pm$91 & 0.32$\pm$0.01 & 0.8 &  $>$7.7 & $-$0.8 & $-$8.3\\
  102964 & 84.29495 & $-$66.36967 & 24.695$\pm$0.033 & 23.870$\pm$0.027 & 23.574$\pm$0.133 & 27.1$\pm$1.9 & 5497$\pm$91 & 0.32$\pm$0.01 & 0.8 &     7.7 & $-$0.7 & $-$8.2\\
  103085 & 84.31777 & $-$66.36240 & 24.784$\pm$0.037 & 23.882$\pm$0.025 & 23.702$\pm$0.085 & 23.1$\pm$1.4 & 5261$\pm$83 & 0.31$\pm$0.01 & 0.8 &  $>$7.8 & $-$0.7 & $-$8.2\\
  103112 & 84.26605 & $-$66.34971 & 24.805$\pm$0.041 & 23.957$\pm$0.030 & 22.957$\pm$0.246 & 50.2$\pm$1.7 & 5427$\pm$111& 0.30$\pm$0.01 & 0.8 &     7.7 & $-$0.4 & $-$7.9\\
  103234 & 84.26518 & $-$66.34048 & 24.899$\pm$0.043 & 23.903$\pm$0.025 & 23.840$\pm$0.082 & 19.0$\pm$1.5 & 5013$\pm$78 & 0.30$\pm$0.01 & 0.8 &     7.7 & $-$0.8 & $-$8.2\\
  103235 & 84.32830 & $-$66.36450 & 24.899$\pm$0.039 & 24.063$\pm$0.029 & 23.910$\pm$0.083 & 20.6$\pm$1.4 & 5463$\pm$105& 0.27$\pm$0.01 & 0.8 &  $>$7.8 & $-$0.8 & $-$8.3\\
  103256 & 84.29654 & $-$66.35818 & 24.917$\pm$0.041 & 24.110$\pm$0.028 & 23.738$\pm$0.188 & 30.1$\pm$2.4 & 5559$\pm$113& 0.26$\pm$0.01 & 0.7 &  $>$7.8 & $-$0.7 & $-$8.3\\
  103399 & 84.30085 & $-$66.37431 & 25.015$\pm$0.043 & 23.772$\pm$0.027 & 23.753$\pm$0.078 & 21.3$\pm$1.4 & 4540$\pm$56 & 0.36$\pm$0.01 & 0.8 &     7.3 & $-$0.7 & $-$8.1\\
  103452 & 84.27387 & $-$66.35526 & 25.055$\pm$0.042 & 24.094$\pm$0.030 & 23.816$\pm$0.178 & 28.5$\pm$2.4 & 5105$\pm$94 & 0.25$\pm$0.01 & 0.7 &  $>$7.8 & $-$0.8 & $-$8.2\\
  103515 & 84.29445 & $-$66.36017 & 25.094$\pm$0.046 & 23.910$\pm$0.029 & 23.902$\pm$0.083 & 19.7$\pm$1.5 & 4639$\pm$64 & 0.31$\pm$0.01 & 0.8 &     7.4 & $-$0.8 & $-$8.2\\
  103517 & 84.29024 & $-$66.33946 & 25.095$\pm$0.043 & 24.092$\pm$0.029 & 24.138$\pm$0.050 & 13.3$\pm$1.1 & 4996$\pm$78 & 0.25$\pm$0.01 & 0.7 &  $>$7.7 & $-$0.9 & $-$8.3\\
  103545 & 84.30585 & $-$66.37456 & 25.109$\pm$0.046 & 24.168$\pm$0.033 & 23.862$\pm$0.064 & 29.4$\pm$1.0 & 5157$\pm$104& 0.24$\pm$0.01 & 0.7 &  $>$7.8 & $-$0.8 & $-$8.2\\
  103583 & 84.32980 & $-$66.35082 & 25.138$\pm$0.058 & 24.197$\pm$0.040 & 23.963$\pm$0.156 & 26.3$\pm$2.3 & 5157$\pm$131& 0.23$\pm$0.01 & 0.7 &     7.7 & $-$0.8 & $-$8.3\\
  103644 & 84.27910 & $-$66.36184 & 25.177$\pm$0.046 & 24.187$\pm$0.032 & 23.796$\pm$0.137 & 33.5$\pm$1.7 & 5029$\pm$89 & 0.23$\pm$0.01 & 0.7 &     7.5 & $-$0.8 & $-$8.2\\
  103715 & 84.31757 & $-$66.33519 & 25.230$\pm$0.043 & 24.393$\pm$0.034 & 24.029$\pm$0.107 & 30.2$\pm$1.5 & 5460$\pm$117& 0.20$\pm$0.01 & 0.7 &     7.7 & $-$0.9 & $-$8.4\\
  103717 & 84.31974 & $-$66.38231 & 25.230$\pm$0.045 & 23.934$\pm$0.026 & 24.042$\pm$0.069 & 15.6$\pm$1.4 & 4461$\pm$48 & 0.31$\pm$0.01 & 0.8 &     7.1 & $-$0.9 & $-$8.2\\
  103730 & 84.29031 & $-$66.37573 & 25.238$\pm$0.047 & 24.249$\pm$0.030 & 24.074$\pm$0.129 & 24.4$\pm$2.0 & 5032$\pm$88 & 0.22$\pm$0.01 & 0.7 &     7.6 & $-$0.9 & $-$8.3\\
  103786 & 84.32967 & $-$66.37369 & 25.294$\pm$0.040 & 24.361$\pm$0.036 & 24.076$\pm$0.079 & 28.4$\pm$1.2 & 5178$\pm$98 & 0.20$\pm$0.01 & 0.7 &  $>$7.7 & $-$0.9 & $-$8.4\\
  103815 & 84.30586 & $-$66.38051 & 25.318$\pm$0.051 & 24.288$\pm$0.029 & 24.031$\pm$0.070 & 28.7$\pm$1.1 & 4938$\pm$88 & 0.21$\pm$0.01 & 0.7 &     7.7 & $-$0.9 & $-$8.3\\
  103898 & 84.32343 & $-$66.37184 & 25.383$\pm$0.049 & 24.221$\pm$0.030 & 23.968$\pm$0.176 & 30.6$\pm$2.3 & 4677$\pm$69 & 0.23$\pm$0.01 & 0.8 &     7.4 & $-$0.8 & $-$8.2\\
  103926 & 84.32151 & $-$66.38447 & 25.398$\pm$0.056 & 24.329$\pm$0.034 & 24.287$\pm$0.080 & 19.3$\pm$1.5 & 4855$\pm$98 & 0.21$\pm$0.01 & 0.7 &     7.7 & $-$1.0 & $-$8.4\\
  103990 & 84.30760 & $-$66.37855 & 25.446$\pm$0.057 & 24.403$\pm$0.033 & 24.312$\pm$0.133 & 21.3$\pm$2.2 & 4911$\pm$98 & 0.19$\pm$0.01 & 0.7 &     7.7 & $-$1.0 & $-$8.4\\
  103998 & 84.32394 & $-$66.37225 & 25.454$\pm$0.054 & 24.448$\pm$0.038 & 24.292$\pm$0.158 & 23.8$\pm$2.4 & 4989$\pm$99 & 0.18$\pm$0.01 & 0.7 &     7.6 & $-$1.0 & $-$8.4\\
  104055 & 84.29799 & $-$66.35120 & 25.492$\pm$0.048 & 24.373$\pm$0.036 & 24.132$\pm$0.209 & 29.4$\pm$2.8 & 4749$\pm$71 & 0.20$\pm$0.01 & 0.7 &     7.7 & $-$0.9 & $-$8.3\\
  104080 & 84.29913 & $-$66.36902 & 25.507$\pm$0.055 & 24.477$\pm$0.042 & 24.414$\pm$0.056 & 19.7$\pm$1.2 & 4938$\pm$103& 0.18$\pm$0.01 & 0.7 &  $>$7.8 & $-$1.0 & $-$8.5\\
  104086 & 84.29786 & $-$66.36609 & 25.511$\pm$0.055 & 24.382$\pm$0.031 & 24.367$\pm$0.066 & 19.0$\pm$1.4 & 4732$\pm$75 & 0.20$\pm$0.01 & 0.7 &     7.7 & $-$1.0 & $-$8.4\\
  104094 & 84.28131 & $-$66.38257 & 25.516$\pm$0.060 & 24.371$\pm$0.033 & 24.195$\pm$0.126 & 27.0$\pm$1.9 & 4705$\pm$82 & 0.20$\pm$0.01 & 0.7 &     7.7 & $-$0.9 & $-$8.3\\
  104107 & 84.32576 & $-$66.37196 & 25.526$\pm$0.062 & 24.485$\pm$0.033 & 24.418$\pm$0.092 & 20.1$\pm$1.7 & 4915$\pm$105& 0.18$\pm$0.01 & 0.7 &  $>$7.8 & $-$1.0 & $-$8.5\\
  104266 & 84.31422 & $-$66.36722 & 25.628$\pm$0.061 & 24.261$\pm$0.029 & 24.048$\pm$0.007 & 31.8$\pm$0.8 & 4366$\pm$64 & 0.22$\pm$0.01 & 0.7 &     7.4 & $-$0.9 & $-$8.2\\
  104298 & 84.29190 & $-$66.34444 & 25.652$\pm$0.064 & 24.540$\pm$0.037 & 24.381$\pm$0.033 & 25.7$\pm$1.0 & 4764$\pm$91 & 0.17$\pm$0.01 & 0.7 &  $>$7.8 & $-$1.0 & $-$8.4\\
  104316 & 84.26365 & $-$66.35498 & 25.661$\pm$0.068 & 24.314$\pm$0.032 & 24.457$\pm$0.003 & 14.7$\pm$1.2 & 4393$\pm$71 & 0.21$\pm$0.01 & 0.7 &     7.5 & $-$1.0 & $-$8.4\\
  104426 & 84.33368 & $-$66.36310 & 25.727$\pm$0.056 & 24.586$\pm$0.038 & 24.368$\pm$0.049 & 28.8$\pm$1.0 & 4712$\pm$80 & 0.17$\pm$0.01 & 0.7 &     7.7 & $-$1.0 & $-$8.4\\
  104566 & 84.30245 & $-$66.35672 & 25.813$\pm$0.066 & 24.772$\pm$0.043 & 24.791$\pm$0.047 & 15.6$\pm$1.4 & 4915$\pm$118& 0.14$\pm$0.01 & 0.6 &  $>$7.8 & $-$1.2 & $-$8.6\\
  104619 & 84.28295 & $-$66.35082 & 25.855$\pm$0.072 & 24.858$\pm$0.040 & 24.573$\pm$0.095 & 29.4$\pm$1.5 & 5011$\pm$125& 0.13$\pm$0.01 & 0.6 &     7.7 & $-$1.1 & $-$8.6\\
  104642 & 84.29637 & $-$66.34912 & 25.866$\pm$0.073 & 24.537$\pm$0.037 & 24.433$\pm$0.054 & 26.7$\pm$1.2 & 4417$\pm$77 & 0.17$\pm$0.01 & 0.7 &     7.7 & $-$1.0 & $-$8.4\\
  104643 & 84.27788 & $-$66.38097 & 25.866$\pm$0.087 & 24.660$\pm$0.040 & 24.360$\pm$0.126 & 33.0$\pm$1.8 & 4602$\pm$111& 0.16$\pm$0.01 & 0.6 &  $>$7.7 & $-$1.0 & $-$8.4\\
  104738 & 84.32780 & $-$66.36165 & 25.917$\pm$0.078 & 24.682$\pm$0.035 & 24.500$\pm$0.108 & 28.6$\pm$1.7 & 4553$\pm$92 & 0.15$\pm$0.01 & 0.6 &  $>$7.8 & $-$1.0 & $-$8.4\\
  104757 & 84.30210 & $-$66.36574 & 25.927$\pm$0.077 & 24.694$\pm$0.039 & 24.513$\pm$0.099 & 28.6$\pm$1.6 & 4557$\pm$93 & 0.15$\pm$0.01 & 0.6 &  $>$7.8 & $-$1.0 & $-$8.5\\
  104762 & 84.29932 & $-$66.34977 & 25.928$\pm$0.064 & 24.592$\pm$0.037 & 24.477$\pm$0.098 & 27.3$\pm$1.6 & 4407$\pm$70 & 0.17$\pm$0.01 & 0.7 &     7.7 & $-$1.0 & $-$8.4\\
  104781 & 84.32844 & $-$66.35696 & 25.942$\pm$0.063 & 24.398$\pm$0.029 & 24.152$\pm$0.070 & 35.7$\pm$1.0 & 4136$\pm$62 & 0.20$\pm$0.01 & 0.7 &     7.2 & $-$0.9 & $-$8.2\\
  104797 & 84.30850 & $-$66.36185 & 25.952$\pm$0.068 & 24.689$\pm$0.036 & 24.502$\pm$0.211 & 29.3$\pm$2.9 & 4506$\pm$74 & 0.15$\pm$0.01 & 0.6 &  $>$7.8 & $-$1.0 & $-$8.4\\
  104931 & 84.28252 & $-$66.36729 & 26.027$\pm$0.093 & 24.715$\pm$0.043 & 24.490$\pm$0.081 & 31.5$\pm$1.5 & 4439$\pm$96 & 0.15$\pm$0.01 & 0.6 &  $>$7.8 & $-$1.0 & $-$8.4\\
  105002 & 84.30695 & $-$66.38393 & 26.066$\pm$0.075 & 24.725$\pm$0.041 & 24.421$\pm$0.045 & 35.0$\pm$1.0 & 4401$\pm$80 & 0.15$\pm$0.01 & 0.6 &  $>$7.8 & $-$1.0 & $-$8.4\\
  105009 & 84.28341 & $-$66.37871 & 26.070$\pm$0.082 & 24.819$\pm$0.047 & 24.484$\pm$0.075 & 35.0$\pm$1.3 & 4526$\pm$94 & 0.14$\pm$0.01 & 0.6 &     7.7 & $-$1.0 & $-$8.5\\
  105084 & 84.33037 & $-$66.37966 & 26.114$\pm$0.072 & 24.580$\pm$0.040 & 24.690$\pm$0.081 & 20.2$\pm$1.7 & 4148$\pm$73 & 0.17$\pm$0.01 & 0.7 &     7.5 & $-$1.1 & $-$8.5\\
  105088 & 84.29739 & $-$66.34153 & 26.116$\pm$0.062 & 24.822$\pm$0.041 & 24.549$\pm$0.222 & 33.2$\pm$2.7 & 4463$\pm$70 & 0.14$\pm$0.01 & 0.6 &     7.7 & $-$1.1 & $-$8.5\\
  105103 & 84.29914 & $-$66.37354 & 26.121$\pm$0.083 & 24.907$\pm$0.043 & 24.297$\pm$0.221 & 43.5$\pm$2.1 & 4589$\pm$106& 0.13$\pm$0.01 & 0.6 &  $>$7.7 & $-$1.0 & $-$8.4\\
  105147 & 84.30678 & $-$66.37535 & 26.153$\pm$0.083 & 25.006$\pm$0.047 & 24.493$\pm$0.074 & 39.8$\pm$1.1 & 4702$\pm$113& 0.11$\pm$0.01 & 0.6 &  $>$7.8 & $-$1.0 & $-$8.5\\
  105168 & 84.28984 & $-$66.35272 & 26.159$\pm$0.086 & 24.813$\pm$0.041 & 24.711$\pm$0.059 & 26.9$\pm$1.4 & 4394$\pm$90 & 0.14$\pm$0.01 & 0.6 &     7.7 & $-$1.1 & $-$8.5\\
  105223 & 84.31408 & $-$66.35240 & 26.182$\pm$0.084 & 24.837$\pm$0.037 & 24.836$\pm$0.082 & 22.2$\pm$1.8 & 4395$\pm$86 & 0.13$\pm$0.01 & 0.6 &     7.7 & $-$1.2 & $-$8.6\\
  105280 & 84.32034 & $-$66.33499 & 26.214$\pm$0.074 & 24.697$\pm$0.038 & 24.882$\pm$0.073 & 15.9$\pm$1.7 & 4170$\pm$74 & 0.15$\pm$0.01 & 0.7 &     7.6 & $-$1.2 & $-$8.5\\
  105299 & 84.30460 & $-$66.36235 & 26.223$\pm$0.084 & 25.008$\pm$0.047 & 24.269$\pm$0.232 & 47.1$\pm$2.0 & 4587$\pm$109& 0.11$\pm$0.01 & 0.6 &  $>$7.8 & $-$1.0 & $-$8.4\\
  105485 & 84.30917 & $-$66.35639 & 26.314$\pm$0.084 & 25.002$\pm$0.050 & 24.681$\pm$0.027 & 35.2$\pm$1.0 & 4439$\pm$92 & 0.11$\pm$0.01 & 0.6 &  $>$7.8 & $-$1.1 & $-$8.5\\
  105499 & 84.30689 & $-$66.34977 & 26.322$\pm$0.086 & 25.380$\pm$0.055 & 24.216$\pm$0.124 & 54.3$\pm$0.9 & 5155$\pm$183& 0.08$\pm$0.01 & 0.6 &  $>$7.8 & $-$0.9 & $-$8.6\\
  105518 & 84.31413 & $-$66.36744 & 26.335$\pm$0.089 & 25.160$\pm$0.054 & 23.865$\pm$0.094 & 58.0$\pm$0.6 & 4655$\pm$124& 0.10$\pm$0.01 & 0.6 &     7.7 & $-$0.8 & $-$8.3\\
  105523 & 84.32288 & $-$66.34132 & 26.338$\pm$0.091 & 25.017$\pm$0.042 & 25.103$\pm$0.009 & 17.3$\pm$1.5 & 4427$\pm$95 & 0.11$\pm$0.01 & 0.6 &  $>$7.8 & $-$1.3 & $-$8.7\\
  105531 & 84.30980 & $-$66.34316 & 26.341$\pm$0.095 & 25.137$\pm$0.042 & 24.867$\pm$0.177 & 31.8$\pm$2.5 & 4606$\pm$120& 0.10$\pm$0.01 & 0.6 &  $>$7.7 & $-$1.2 & $-$8.7\\
  105617 & 84.29692 & $-$66.38106 & 26.390$\pm$0.092 & 24.988$\pm$0.051 & 24.736$\pm$0.157 & 33.8$\pm$2.1 & 4319$\pm$97 & 0.11$\pm$0.01 & 0.6 &  $>$7.8 & $-$1.1 & $-$8.5\\
  105618 & 84.33716 & $-$66.37704 & 26.390$\pm$0.094 & 25.030$\pm$0.043 & 24.871$\pm$0.186 & 29.5$\pm$2.7 & 4375$\pm$97 & 0.11$\pm$0.01 & 0.6 &  $>$7.8 & $-$1.2 & $-$8.6\\
  105629 & 84.29288 & $-$66.36293 & 26.394$\pm$0.096 & 24.703$\pm$0.042 & 23.127$\pm$0.109 & 64.1$\pm$0.4 & 3948$\pm$95 & 0.16$\pm$0.01 & 0.7 &     7.3 & $-$0.5 & $-$7.8\\
  105914 & 84.28570 & $-$66.35744 & 26.538$\pm$0.089 & 25.092$\pm$0.049 & 24.764$\pm$0.201 & 37.2$\pm$2.3 & 4260$\pm$91 & 0.10$\pm$0.01 & 0.6 &  $>$7.7 & $-$1.1 & $-$8.6\\
  105940 & 84.30993 & $-$66.34090 & 26.547$\pm$0.095 & 25.391$\pm$0.052 & 24.695$\pm$0.064 & 45.4$\pm$1.0 & 4687$\pm$129& 0.08$\pm$0.01 & 0.6 &  $>$7.8 & $-$1.1 & $-$8.7\\
  201295 & 84.22925 & $-$66.35225 & 22.757$\pm$0.013 & 22.165$\pm$0.014 & 21.964$\pm$0.060 & 18.8$\pm$1.0 & 6408$\pm$67 & 1.71$\pm$0.06 & 1.1 &  $>$7.6 & $-$0.0 & $-$7.4\\
  201370 & 84.25741 & $-$66.35658 & 22.840$\pm$0.014 & 22.237$\pm$0.015 & 22.110$\pm$0.060 & 15.0$\pm$1.1 & 6353$\pm$74 & 1.59$\pm$0.05 & 1.1 &     7.4 & $-$0.1 & $-$7.5\\
  201488 & 84.20598 & $-$66.36387 & 22.955$\pm$0.014 & 22.451$\pm$0.014 & 22.178$\pm$0.097 & 21.1$\pm$1.5 & 6759$\pm$51 & 1.40$\pm$0.05 & 1.1 &  $>$7.6 & $-$0.1 & $-$7.6\\
  201790 & 84.24566 & $-$66.36048 & 23.256$\pm$0.018 & 22.730$\pm$0.014 & 22.033$\pm$0.211 & 38.7$\pm$2.2 & 6676$\pm$61 & 1.07$\pm$0.03 & 1.0 &     7.4 & $-$0.1 & $-$7.6\\
  201932 & 84.24048 & $-$66.36494 & 23.396$\pm$0.018 & 22.664$\pm$0.015 & 22.271$\pm$0.141 & 29.9$\pm$1.8 & 5824$\pm$63 & 1.00$\pm$0.03 & 1.0 &     7.4 & $-$0.2 & $-$7.5\\
  201954 & 84.25608 & $-$66.35287 & 23.420$\pm$0.020 & 23.044$\pm$0.016 & 20.726$\pm$0.180 & 66.5$\pm$0.4 & 7323$\pm$79 & 0.90$\pm$0.03 & 1.0 &     7.5 &    0.5 & $-$7.2\\
  202333 & 84.26053 & $-$66.36747 & 23.759$\pm$0.023 & 22.743$\pm$0.015 & 22.240$\pm$0.244 & 37.9$\pm$2.6 & 4968$\pm$40 & 0.89$\pm$0.02 & 1.1 &     7.0 & $-$0.1 & $-$7.5\\
  202599 & 84.25690 & $-$66.36718 & 23.990$\pm$0.029 & 23.351$\pm$0.021 & 23.235$\pm$0.070 & 15.0$\pm$1.3 & 6187$\pm$100& 0.56$\pm$0.01 & 0.9 &  $>$7.7 & $-$0.5 & $-$8.1\\
  202644 & 84.24506 & $-$66.36098 & 24.025$\pm$0.029 & 23.551$\pm$0.025 & 22.090$\pm$0.110 & 56.8$\pm$0.6 & 6890$\pm$117& 0.52$\pm$0.01 & 0.9 &  $>$7.7 & $-$0.1 & $-$7.7\\
  202652 & 84.25620 & $-$66.37402 & 24.030$\pm$0.026 & 23.406$\pm$0.021 & 23.203$\pm$0.077 & 19.4$\pm$1.3 & 6247$\pm$96 & 0.54$\pm$0.01 & 0.9 &  $>$7.7 & $-$0.5 & $-$8.1\\
  202708 & 84.23724 & $-$66.34436 & 24.074$\pm$0.025 & 23.287$\pm$0.019 & 23.203$\pm$0.078 & 16.0$\pm$1.4 & 5628$\pm$76 & 0.55$\pm$0.01 & 0.8 &     7.6 & $-$0.5 & $-$8.0\\
  202776 & 84.24561 & $-$66.35986 & 24.135$\pm$0.029 & 23.375$\pm$0.030 & 22.167$\pm$0.108 & 54.0$\pm$0.7 & 5721$\pm$101& 0.51$\pm$0.01 & 0.9 &  $>$7.7 & $-$0.1 & $-$7.6\\
  202830 & 84.24810 & $-$66.37537 & 24.183$\pm$0.029 & 23.475$\pm$0.022 & 23.293$\pm$0.083 & 19.7$\pm$1.4 & 5915$\pm$97 & 0.48$\pm$0.01 & 0.9 &  $>$7.7 & $-$0.6 & $-$8.1\\
  203320 & 84.23020 & $-$66.37553 & 24.555$\pm$0.034 & 23.460$\pm$0.025 & 23.504$\pm$0.067 & 15.3$\pm$1.3 & 4800$\pm$60 & 0.47$\pm$0.01 & 0.9 &     7.2 & $-$0.6 & $-$8.0\\
  203367 & 84.24718 & $-$66.35929 & 24.594$\pm$0.035 & 23.658$\pm$0.026 & 22.882$\pm$0.087 & 45.4$\pm$0.8 & 5170$\pm$80 & 0.38$\pm$0.01 & 0.8 &     7.5 & $-$0.4 & $-$7.8\\
  203375 & 84.27716 & $-$66.39177 & 24.599$\pm$0.048 & 23.815$\pm$0.039 & 23.692$\pm$0.083 & 18.1$\pm$1.5 & 5638$\pm$149& 0.34$\pm$0.01 & 0.8 &     7.6 & $-$0.7 & $-$8.2\\
  203439 & 84.23264 & $-$66.36665 & 24.645$\pm$0.034 & 23.643$\pm$0.026 & 23.602$\pm$0.075 & 18.0$\pm$1.3 & 4998$\pm$65 & 0.39$\pm$0.01 & 0.8 &     7.4 & $-$0.7 & $-$8.1\\
  203478 & 84.26115 & $-$66.36638 & 24.677$\pm$0.039 & 23.795$\pm$0.024 & 23.520$\pm$0.141 & 27.1$\pm$2.0 & 5322$\pm$95 & 0.34$\pm$0.01 & 0.8 &     7.6 & $-$0.7 & $-$8.1\\
  203499 & 84.26345 & $-$66.39032 & 24.689$\pm$0.035 & 23.834$\pm$0.030 & 23.635$\pm$0.065 & 23.2$\pm$1.1 & 5405$\pm$99 & 0.33$\pm$0.01 & 0.8 &     7.7 & $-$0.7 & $-$8.2\\
  203605 & 84.21983 & $-$66.38663 & 24.776$\pm$0.033 & 23.939$\pm$0.026 & 23.786$\pm$0.127 & 20.6$\pm$2.0 & 5460$\pm$91 & 0.30$\pm$0.01 & 0.8 &  $>$7.8 & $-$0.8 & $-$8.3\\
  203619 & 84.25516 & $-$66.37281 & 24.783$\pm$0.038 & 23.916$\pm$0.030 & 23.722$\pm$0.065 & 23.1$\pm$1.1 & 5368$\pm$103& 0.31$\pm$0.01 & 0.8 &  $>$7.7 & $-$0.7 & $-$8.2\\
  203809 & 84.22269 & $-$66.36948 & 24.918$\pm$0.033 & 24.057$\pm$0.029 & 24.021$\pm$0.091 & 14.9$\pm$1.6 & 5387$\pm$94 & 0.27$\pm$0.01 & 0.8 &     7.7 & $-$0.9 & $-$8.4\\
  203860 & 84.25838 & $-$66.37823 & 24.947$\pm$0.044 & 23.730$\pm$0.025 & 23.742$\pm$0.119 & 19.2$\pm$2.0 & 4584$\pm$60 & 0.37$\pm$0.01 & 0.9 &     7.1 & $-$0.7 & $-$8.1\\
  203873 & 84.25300 & $-$66.38037 & 24.959$\pm$0.045 & 23.764$\pm$0.027 & 23.661$\pm$0.117 & 24.5$\pm$1.8 & 4621$\pm$62 & 0.36$\pm$0.01 & 0.8 &     7.3 & $-$0.7 & $-$8.0\\
  203903 & 84.19118 & $-$66.34936 & 24.980$\pm$0.040 & 24.001$\pm$0.030 & 23.774$\pm$0.137 & 26.6$\pm$1.9 & 5058$\pm$86 & 0.28$\pm$0.01 & 0.7 &     7.7 & $-$0.8 & $-$8.2\\
  203930 & 84.24300 & $-$66.36428 & 25.003$\pm$0.041 & 24.073$\pm$0.031 & 23.928$\pm$0.091 & 21.9$\pm$1.5 & 5186$\pm$95 & 0.26$\pm$0.01 & 0.8 &  $>$7.8 & $-$0.8 & $-$8.3\\
  203954 & 84.24202 & $-$66.35572 & 25.018$\pm$0.040 & 24.006$\pm$0.032 & 23.685$\pm$0.163 & 31.1$\pm$2.1 & 4977$\pm$77 & 0.28$\pm$0.01 & 0.7 &     7.7 & $-$0.7 & $-$8.1\\
  203958 & 84.26858 & $-$66.38790 & 25.021$\pm$0.043 & 24.175$\pm$0.036 & 24.132$\pm$0.071 & 15.0$\pm$1.4 & 5433$\pm$120& 0.24$\pm$0.01 & 0.7 &  $>$7.8 & $-$0.9 & $-$8.4\\
  204044 & 84.21298 & $-$66.34674 & 25.082$\pm$0.039 & 24.109$\pm$0.026 & 23.968$\pm$0.124 & 22.5$\pm$1.9 & 5074$\pm$84 & 0.25$\pm$0.01 & 0.7 &  $>$7.7 & $-$0.8 & $-$8.3\\
  204046 & 84.26278 & $-$66.38299 & 25.083$\pm$0.048 & 24.130$\pm$0.032 & 24.149$\pm$0.073 & 13.8$\pm$1.5 & 5126$\pm$107& 0.25$\pm$0.01 & 0.7 &     7.7 & $-$0.9 & $-$8.3\\
  204220 & 84.20907 & $-$66.37826 & 25.201$\pm$0.046 & 24.195$\pm$0.034 & 23.487$\pm$0.092 & 44.2$\pm$0.9 & 4989$\pm$85 & 0.23$\pm$0.01 & 0.7 &     7.7 & $-$0.6 & $-$8.1\\
  204224 & 84.24075 & $-$66.37636 & 25.208$\pm$0.053 & 24.224$\pm$0.031 & 24.182$\pm$0.094 & 17.7$\pm$1.7 & 5045$\pm$100& 0.22$\pm$0.01 & 0.7 &     7.6 & $-$0.9 & $-$8.4\\
  204245 & 84.24843 & $-$66.39069 & 25.220$\pm$0.044 & 24.264$\pm$0.034 & 24.062$\pm$0.111 & 25.1$\pm$1.7 & 5118$\pm$103& 0.22$\pm$0.01 & 0.7 &     7.7 & $-$0.9 & $-$8.3\\
  204248 & 84.22416 & $-$66.38059 & 25.223$\pm$0.052 & 24.221$\pm$0.032 & 24.263$\pm$0.044 & 13.4$\pm$1.2 & 4998$\pm$92 & 0.23$\pm$0.01 & 0.7 &     7.6 & $-$0.9 & $-$8.4\\
  204500 & 84.23834 & $-$66.38866 & 25.382$\pm$0.062 & 24.160$\pm$0.031 & 24.269$\pm$0.040 & 14.2$\pm$1.3 & 4575$\pm$81 & 0.25$\pm$0.01 & 0.7 &     7.5 & $-$1.0 & $-$8.3\\
  204524 & 84.27125 & $-$66.37828 & 25.402$\pm$0.081 & 24.259$\pm$0.049 & 24.210$\pm$0.067 & 21.0$\pm$1.6 & 4709$\pm$112& 0.23$\pm$0.01 & 0.7 &     7.6 & $-$0.9 & $-$8.3\\
  204575 & 84.26226 & $-$66.36685 & 25.437$\pm$0.060 & 24.208$\pm$0.036 & 23.720$\pm$0.272 & 40.0$\pm$2.8 & 4564$\pm$79 & 0.24$\pm$0.01 & 0.8 &     7.4 & $-$0.7 & $-$8.1\\
  204668 & 84.24878 & $-$66.35863 & 25.502$\pm$0.058 & 24.380$\pm$0.038 & 24.069$\pm$0.170 & 32.3$\pm$2.2 & 4744$\pm$82 & 0.20$\pm$0.01 & 0.7 &     7.7 & $-$0.9 & $-$8.3\\
  204698 & 84.25701 & $-$66.36720 & 25.523$\pm$0.072 & 24.462$\pm$0.039 & 23.726$\pm$0.204 & 45.5$\pm$1.8 & 4872$\pm$123& 0.18$\pm$0.01 & 0.7 &  $>$7.8 & $-$0.7 & $-$8.2\\
  204728 & 84.20835 & $-$66.36141 & 25.544$\pm$0.051 & 24.566$\pm$0.043 & 24.117$\pm$0.162 & 35.5$\pm$1.9 & 5060$\pm$111& 0.16$\pm$0.01 & 0.7 &     7.7 & $-$0.9 & $-$8.4\\
  204734 & 84.25307 & $-$66.38390 & 25.545$\pm$0.058 & 24.527$\pm$0.040 & 24.104$\pm$0.240 & 35.1$\pm$2.8 & 4964$\pm$106& 0.17$\pm$0.01 & 0.7 &     7.7 & $-$0.9 & $-$8.4\\
  204736 & 84.21478 & $-$66.35324 & 25.545$\pm$0.056 & 24.478$\pm$0.035 & 24.238$\pm$0.071 & 28.6$\pm$1.2 & 4860$\pm$100& 0.18$\pm$0.01 & 0.7 &  $>$7.8 & $-$0.9 & $-$8.4\\
  204762 & 84.24497 & $-$66.37016 & 25.562$\pm$0.066 & 24.441$\pm$0.033 & 23.977$\pm$0.102 & 37.9$\pm$1.3 & 4746$\pm$88 & 0.19$\pm$0.01 & 0.7 &     7.7 & $-$0.8 & $-$8.3\\
  204788 & 84.19713 & $-$66.36378 & 25.582$\pm$0.076 & 24.443$\pm$0.039 & 24.018$\pm$0.130 & 36.8$\pm$1.6 & 4715$\pm$101& 0.19$\pm$0.01 & 0.7 &     7.7 & $-$0.8 & $-$8.3\\
  204806 & 84.23247 & $-$66.36692 & 25.593$\pm$0.058 & 24.491$\pm$0.037 & 24.400$\pm$0.114 & 22.4$\pm$1.9 & 4785$\pm$89 & 0.18$\pm$0.01 & 0.7 &  $>$7.8 & $-$1.0 & $-$8.4\\
  204817 & 84.25305 & $-$66.35604 & 25.598$\pm$0.052 & 24.538$\pm$0.037 & 24.468$\pm$0.055 & 20.6$\pm$1.2 & 4875$\pm$95 & 0.17$\pm$0.01 & 0.7 &     7.1 & $-$1.0 & $-$8.5\\
  204835 & 84.21503 & $-$66.37206 & 25.608$\pm$0.059 & 24.688$\pm$0.040 & 24.491$\pm$0.146 & 24.2$\pm$2.3 & 5212$\pm$131& 0.15$\pm$0.01 & 0.6 &  $>$7.8 & $-$1.0 & $-$8.6\\
  204965 & 84.23400 & $-$66.36196 & 25.695$\pm$0.062 & 24.606$\pm$0.043 & 24.393$\pm$0.130 & 27.7$\pm$1.9 & 4813$\pm$103& 0.16$\pm$0.01 & 0.7 &     7.7 & $-$1.0 & $-$8.5\\
  205004 & 84.25818 & $-$66.36052 & 25.730$\pm$0.068 & 24.374$\pm$0.033 & 22.897$\pm$0.185 & 61.5$\pm$0.8 & 4381$\pm$71 & 0.20$\pm$0.01 & 0.7 &     7.5 & $-$0.4 & $-$7.8\\
  205037 & 84.20301 & $-$66.35402 & 25.750$\pm$0.069 & 24.749$\pm$0.035 & 24.586$\pm$0.150 & 24.0$\pm$2.4 & 5000$\pm$117& 0.14$\pm$0.01 & 0.6 &  $>$7.8 & $-$1.1 & $-$8.6\\
  205062 & 84.23379 & $-$66.35196 & 25.770$\pm$0.058 & 24.471$\pm$0.036 & 24.171$\pm$0.276 & 34.3$\pm$3.3 & 4457$\pm$64 & 0.19$\pm$0.01 & 0.7 &     7.6 & $-$0.9 & $-$8.3\\
  205181 & 84.20958 & $-$66.36611 & 25.850$\pm$0.069 & 24.459$\pm$0.032 & 24.412$\pm$0.082 & 25.1$\pm$1.5 & 4334$\pm$71 & 0.19$\pm$0.01 & 0.7 &     7.5 & $-$1.0 & $-$8.4\\
  205191 & 84.20804 & $-$66.37716 & 25.860$\pm$0.064 & 24.565$\pm$0.034 & 24.000$\pm$0.182 & 43.1$\pm$1.7 & 4462$\pm$68 & 0.17$\pm$0.01 & 0.7 &     7.7 & $-$0.8 & $-$8.2\\
  205209 & 84.21629 & $-$66.35535 & 25.869$\pm$0.068 & 24.660$\pm$0.049 & 24.362$\pm$0.161 & 33.0$\pm$2.1 & 4597$\pm$99 & 0.16$\pm$0.01 & 0.7 &     7.7 & $-$1.0 & $-$8.4\\
  205324 & 84.22818 & $-$66.37878 & 25.943$\pm$0.071 & 24.794$\pm$0.041 & 24.309$\pm$0.266 & 38.9$\pm$2.8 & 4699$\pm$98 & 0.14$\pm$0.01 & 0.6 &  $>$7.8 & $-$1.0 & $-$8.4\\
  205416 & 84.24755 & $-$66.36277 & 26.004$\pm$0.080 & 24.765$\pm$0.048 & 24.108$\pm$0.275 & 45.1$\pm$2.4 & 4547$\pm$98 & 0.14$\pm$0.01 & 0.6 &  $>$7.8 & $-$0.9 & $-$8.3\\
  205422 & 84.25428 & $-$66.37085 & 26.006$\pm$0.090 & 24.737$\pm$0.040 & 24.649$\pm$0.119 & 25.0$\pm$2.1 & 4497$\pm$92 & 0.15$\pm$0.01 & 0.6 &  $>$7.8 & $-$1.1 & $-$8.5\\
  205542 & 84.23473 & $-$66.39123 & 26.073$\pm$0.061 & 24.823$\pm$0.041 & 24.657$\pm$0.207 & 28.2$\pm$2.9 & 4528$\pm$75 & 0.14$\pm$0.01 & 0.6 &     7.7 & $-$1.1 & $-$8.5\\
  205676 & 84.24683 & $-$66.39611 & 26.156$\pm$0.071 & 24.656$\pm$0.039 & 23.940$\pm$0.240 & 49.1$\pm$1.8 & 4191$\pm$72 & 0.16$\pm$0.01 & 0.7 &     7.5 & $-$0.8 & $-$8.2\\
  205677 & 84.21790 & $-$66.35429 & 26.157$\pm$0.084 & 24.922$\pm$0.037 & 24.896$\pm$0.004 & 21.5$\pm$1.3 & 4553$\pm$98 & 0.12$\pm$0.01 & 0.6 &  $>$7.7 & $-$1.2 & $-$8.6\\
  205691 & 84.22599 & $-$66.37111 & 26.163$\pm$0.082 & 24.973$\pm$0.045 & 24.637$\pm$0.045 & 34.2$\pm$1.1 & 4629$\pm$111& 0.12$\pm$0.01 & 0.6 &  $>$7.7 & $-$1.1 & $-$8.6\\
  205771 & 84.23836 & $-$66.37691 & 26.210$\pm$0.094 & 24.992$\pm$0.049 & 24.485$\pm$0.055 & 40.4$\pm$1.1 & 4582$\pm$116& 0.12$\pm$0.01 & 0.6 &  $>$7.7 & $-$1.0 & $-$8.5\\
  205778 & 84.21813 & $-$66.36502 & 26.214$\pm$0.091 & 24.467$\pm$0.061 & 24.191$\pm$0.049 & 39.9$\pm$1.0 & 3876$\pm$100& 0.20$\pm$0.01 & 0.7 &     7.0 & $-$0.9 & $-$8.2\\
  205779 & 84.23670 & $-$66.36692 & 26.214$\pm$0.097 & 24.947$\pm$0.045 & 24.540$\pm$0.168 & 37.7$\pm$2.0 & 4500$\pm$100& 0.12$\pm$0.01 & 0.6 &     7.7 & $-$1.1 & $-$8.5\\
  205812 & 84.26352 & $-$66.39044 & 26.235$\pm$0.090 & 25.258$\pm$0.054 & 24.755$\pm$0.174 & 37.4$\pm$2.1 & 5063$\pm$169& 0.09$\pm$0.01 & 0.6 &  $>$7.7 & $-$1.1 & $-$8.7\\
  205814 & 84.24798 & $-$66.37376 & 26.235$\pm$0.095 & 24.923$\pm$0.052 & 24.546$\pm$0.218 & 37.2$\pm$2.5 & 4439$\pm$102& 0.12$\pm$0.01 & 0.6 &  $>$7.7 & $-$1.1 & $-$8.5\\
  205821 & 84.25049 & $-$66.37525 & 26.238$\pm$0.095 & 24.911$\pm$0.041 & 24.647$\pm$0.205 & 33.3$\pm$2.7 & 4419$\pm$97 & 0.12$\pm$0.01 & 0.6 &  $>$7.7 & $-$1.1 & $-$8.5\\
  205858 & 84.23908 & $-$66.35177 & 26.260$\pm$0.083 & 24.978$\pm$0.048 & 24.832$\pm$0.193 & 27.8$\pm$2.8 & 4479$\pm$90 & 0.12$\pm$0.01 & 0.6 &  $>$7.8 & $-$1.2 & $-$8.6\\
  205908 & 84.22574 & $-$66.38602 & 26.280$\pm$0.086 & 25.256$\pm$0.057 & 24.607$\pm$0.206 & 42.7$\pm$2.0 & 4951$\pm$155& 0.09$\pm$0.01 & 0.6 &  $>$7.7 & $-$1.1 & $-$8.7\\
  205930 & 84.20727 & $-$66.35808 & 26.291$\pm$0.096 & 25.121$\pm$0.046 & 24.852$\pm$0.208 & 31.3$\pm$2.8 & 4663$\pm$127& 0.10$\pm$0.01 & 0.6 &  $>$7.8 & $-$1.2 & $-$8.7\\
  205933 & 84.19588 & $-$66.35396 & 26.293$\pm$0.092 & 24.918$\pm$0.050 & 24.971$\pm$0.083 & 20.0$\pm$1.9 & 4355$\pm$99 & 0.12$\pm$0.01 & 0.6 &  $>$7.8 & $-$1.2 & $-$8.6\\
  205942 & 84.24585 & $-$66.38506 & 26.298$\pm$0.087 & 24.978$\pm$0.063 & 24.783$\pm$0.087 & 30.5$\pm$1.5 & 4429$\pm$101& 0.12$\pm$0.01 & 0.6 &  $>$7.8 & $-$1.2 & $-$8.6\\
  205950 & 84.23575 & $-$66.34881 & 26.302$\pm$0.084 & 24.993$\pm$0.042 & 24.594$\pm$0.224 & 37.9$\pm$2.5 & 4443$\pm$89 & 0.12$\pm$0.01 & 0.6 &  $>$7.8 & $-$1.1 & $-$8.5\\
  205955 & 84.21502 & $-$66.37797 & 26.305$\pm$0.085 & 24.878$\pm$0.045 & 24.216$\pm$0.173 & 47.0$\pm$1.5 & 4286$\pm$87 & 0.13$\pm$0.01 & 0.6 &  $>$7.7 & $-$0.9 & $-$8.3\\
  205959 & 84.20051 & $-$66.35242 & 26.307$\pm$0.079 & 24.705$\pm$0.040 & 24.734$\pm$0.080 & 25.4$\pm$1.6 & 4062$\pm$79 & 0.15$\pm$0.01 & 0.7 &     7.5 & $-$1.1 & $-$8.5\\
  205980 & 84.19942 & $-$66.36190 & 26.318$\pm$0.082 & 24.958$\pm$0.045 & 25.054$\pm$0.043 & 17.5$\pm$1.5 & 4375$\pm$88 & 0.12$\pm$0.01 & 0.6 &  $>$7.7 & $-$1.3 & $-$8.7\\
  206061 & 84.23042 & $-$66.37977 & 26.363$\pm$0.094 & 25.006$\pm$0.045 & 24.925$\pm$0.053 & 26.1$\pm$1.5 & 4379$\pm$98 & 0.11$\pm$0.01 & 0.6 &  $>$7.8 & $-$1.2 & $-$8.6\\
  206092 & 84.20873 & $-$66.36808 & 26.378$\pm$0.097 & 24.865$\pm$0.040 & 24.967$\pm$0.064 & 20.2$\pm$1.8 & 4175$\pm$94 & 0.13$\pm$0.01 & 0.6 &     7.7 & $-$1.2 & $-$8.6\\
  206140 & 84.25296 & $-$66.38738 & 26.404$\pm$0.090 & 25.089$\pm$0.049 & 24.657$\pm$0.158 & 39.1$\pm$1.8 & 4435$\pm$97 & 0.11$\pm$0.01 & 0.6 &  $>$7.8 & $-$1.1 & $-$8.6\\
  206142 & 84.21244 & $-$66.37968 & 26.405$\pm$0.085 & 25.173$\pm$0.051 & 25.045$\pm$0.130 & 26.2$\pm$2.1 & 4559$\pm$105& 0.10$\pm$0.01 & 0.6 &     7.7 & $-$1.3 & $-$8.7\\
  206269 & 84.24240 & $-$66.39419 & 26.470$\pm$0.077 & 25.210$\pm$0.054 & 24.916$\pm$0.214 & 33.5$\pm$2.7 & 4511$\pm$91 & 0.09$\pm$0.01 & 0.6 &  $>$7.7 & $-$1.2 & $-$8.7\\
  206357 & 84.19530 & $-$66.34836 & 26.515$\pm$0.084 & 25.168$\pm$0.043 & 24.915$\pm$0.085 & 33.1$\pm$1.4 & 4393$\pm$89 & 0.10$\pm$0.01 & 0.6 &  $>$7.7 & $-$1.2 & $-$8.7\\
  206419 & 84.23594 & $-$66.39544 & 26.545$\pm$0.098 & 25.081$\pm$0.046 & 24.953$\pm$0.129 & 29.8$\pm$2.1 & 4237$\pm$97 & 0.11$\pm$0.01 & 0.6 &     7.7 & $-$1.2 & $-$8.6\\
\enddata
\end{deluxetable*}
\normalsize

\bibliographystyle{../V62/aastex62}

\end{document}